 \documentclass[preprint2]{aastex}

\slugcomment{Not to appear in Nonlearned J., 45.}

\shorttitle{Spitzer jitter and pixel-ICA}
\shortauthors{Morello}

\begin{document}


\title{A blind method to detrend instrumental systematics in exoplanetary light-curves}


\author{G. Morello}
\affil{Department of Physics \& Astronomy, University College London, Gower Street, WC1E6BT, UK}
\email{giuseppe.morello.11@ucl.ac.uk}



\begin{abstract}
The study of the atmospheres of transiting exoplanets requires a photometric precision, and repeatability, of one part in $\sim$10$^4$. This is beyond the original calibration plans of current observatories, hence the necessity to disentangle the instrumental systematics from the astrophysical signals in raw datasets. Most methods used in the literature are based on an approximate instrument model. The choice of parameters of the model and their functional forms can sometimes be subjective, causing controversies in the literature. Recently, \cite{mor14, mor15} have developed a non-parametric detrending method that gave coherent and repeatable results when applied to Spitzer/IRAC datasets that were debated in the literature. Said method is based on Independent Component Analysis (ICA) of individual pixel time-series, hereafter ``pixel-ICA'. The main purpose of this paper is to investigate the limits and advantages of pixel-ICA on a series of simulated datasets with different instrument properties, and a range of jitter timescales and shapes, non-stationarity, sudden change points, etc. The performances of pixel-ICA are compared against the ones of other methods, in particular polynomial centroid division (PCD), and pixel-level decorrelation (PLD) method \citep{dem14}.\\ We find that in simulated cases pixel-ICA performs as well or better than other methods, and it also guarantees a higher degree of objectivity, because of its purely statistical foundation with no prior information on the instrument systematics. The results of this paper, together with previous analyses of Spitzer/IRAC datasets, suggest that photometric precision and repeatability of one part in 10$^4$ can be achieved with current infrared space instruments.
\end{abstract}





\section{Introduction}

The field of extrasolar planetary transits is one of the most productive and innovative subject in Astrophysics in the last decade. Transit observations can be used to measure the size of planets, their orbital parameters \citep{smo03}, stellar properties \citep{ma02,how11}, to study the atmospheres of planets \citep{bro01,cha02,tin07}, to detect small planets \citep{mir02,agol05}, and exomoons \citep{kip09,kip09b}. In particular, the study of planetary atmospheres requires a high level of photometric precision, i.e. one part in $\sim$10$^4$ in stellar flux \citep{bro01}, which is comparable to the effects of current instrumental systematics and stellar activity \citep{ber11,bal12}, hence the necessity of testable methods for data detrending. In some cases, different assumptions, e.g. using different instrumental information or functional forms to describe them, leaded to controversial results even from the same datasets; examples in the literature are \cite{tin07, ehr07, bea08, des09, des11} for the hot-Jupiter HD189733b, and \cite{ste10, bea11, knu11, knu14} for the warm-Neptune GJ436b. Some of these controversies are based on Spitzer/IRAC datasets at 3.6 and 4.5 $\mu$m. The main systematic effect for these two channels is an almost regular undulation with period $\sim$3000 s, so called pixel-phase effect, as it is correlated with the relative position of the source centroid with respect to a pixel center \citep{faz04, mc06}. Conventional parametric techniques correct for this effect dividing the measured flux by a polynomial function of the coordinates of the photometric centroid; some variants may include time-dependence (e.g. \cite{ste10, bea11}). Newer techniques attempt to map the intra-pixel variability at a fine-scale level, e.g. adopting spatial weighting functions \citep{bal10, cow12, lew13} or interpolating grids \citep{ste12, ste12b}. The results obtained with these methods appear to be strongly dependent on a few assumptions, e.g. the degree of the polynomial adopted, the photometric technique, the centroid determination, calibrating instrument systematics over the out-of-transit only or the whole observation (e.g. \cite{bea11, dl14, zel14}). Also, the very same method, applied to different observations of the same system, often leads to significantly different results. Non-parametric methods have been proposed to guarantee an higher degree of objectivity \citep{cw09, tha10, gib12, wal12, wal13, wal14}. \cite{mor14, mor15} reanalyzed the 3.6 and 4.5 $\mu$m Spitzer/IRAC primary transits of HD189733b and GJ436b obtained during the cryogenic regime, so called ``cold Spitzer'' era, adopting a blind source separation technique, based on an Independent Component Analysis (ICA) of individual pixel time series, in this paper called ``pixel-ICA''. The results obtained with this method are repeatable over different epochs, and a photometric precision of one part in $\sim$10$^4$ in stellar flux is achieved, with no signs of significant stellar variability as suggested in the previous literature \citep{des11, knu11}. The use of ICA to decorrelate the transit signals from astrophysical and instrumental noise, in spectrophotometric observations, has been proposed by \cite{wal12, wal13, wal14}. The reason to prefer such blind detrending methods is twofold: they require very little, if any, prior knowledge of the instrument systematics and astrophysical signals, therefore they also ensure a higher degree of objectivity compared to methods based on approximate instrument systematics models. As an added value, they give stable results over several datasets, also in those cases where more conventional methods have been unsuccessful. Recently, \cite{dem14} proposed a different pixel-level decorrelation method (PLD) that uses pixel time series to correct for the pixel-phase effect, while simultaneously modeling the astrophysical signals and possible detector sensitivity variability in a parametric way. PLD has been applied to some Spitzer/IRAC eclipses and synthetic Spitzer data, showing better performances compared to previously published detrending methods. 

In this paper, we test the pixel-ICA detrending algorithm over simulated datasets, for which instrumental systematic effects are fully under control, to understand better its advantages and limits. In particular, we:
\begin{enumerate}
\item consider some toy-models that can reproduce systematic effects similar to those seen in Spitzer/IRAC data\footnote{we do not mean to emulate the Spitzer/IRAC system, but rather to  study some mechanisms that, in particular configurations, may reproduce in part the effects observed in Spitzer/IRAC datasets.};
\item test the pixel-ICA method on simulated datasets;
\item explore the limits of its applicability;
\item compare its performances with the most common parametric method, based on division by a polynomial function of the centroid (PCD), and the semi-parametric PLD method.
\end{enumerate}

\section{Brief outline of pixel-ICA algorithm}
\label{sec:pixel-ICA}

ICA is a statistical technique that transforms a set of signals into an equivalent set of maximally independent components. It is widely used in a lot of different contexts, e.g. Neuroscience, Econometrics, Photography, and Astrophysics, to separate the source signals present in a set of observations/recordings \citep{hyv01}. The underlying assumption is that real signals are linear mixtures of independent source signals. The validity of this assumption for astrophysical observations has been discussed with more details in \cite{mor15} (App. A). The major strenghts of this approach are:
\begin{enumerate}
\item it requires the minimal amount of prior information;
\item even if the assumption is not valid, the method is able to detrend in part the source signals.
\end{enumerate}
If additional information is available, some variants of ICA can be used to obtain better results \citep{igu02,sto02,bar11}, but these are not considered in this paper. 

Pixel-ICA method uses individual pixel time series from an array to decorrelate the transit signal in photometric observations of stars with a transiting planet. The main steps are:
\begin{enumerate}
\item ICA transformation of the time series to get the independent components;
\item identification of the transit component, by inspection or other methods (e.g. \cite{wal12});
\item fitting of the non-transit components plus a constant on the out-of-transit of the integral light-curve (sum over the pixel array);
\item subtraction of the non-transit components, with coefficients determined by the fitting, from the integral light-curve;
\item normalization of the detrended light-curve.
\end{enumerate}
The normalized, detrended light-curve is model-fitted with \cite{ma02}, which depends on several stellar and orbital parameters. We typically perform a Nelder-Mead optimisation algorithm \citep{lag98} to obtain first estimates of the best parameters of the model, then we generate Monte Carlo chains of 20,000 elements to sample the posterior distributions (approximately gaussians) of the parameters. The updated best parameters are the mean values of the chains, and the zero-order error bars, $\sigma_{par,0}$, are their standard deviations. The zero-order error bars only accounts for the scatter in the detrended light-curve; they must be increased by a factor that includes the uncertainties due to the detrending process:
\begin{equation}
\label{eqn:sigmapar}
\sigma_{par} = \sigma_{par,0} \sqrt{ \frac{ \sigma_{0}^2 + \sigma_{ICA}^2 }{ \sigma_{0}^2}}
\end{equation}
where $\sigma_{par}$ is the final parameter error bar, $\sigma_0$ is the square root of the likelihood's variance (approximately equal to the standard deviation of residuals), and $\sigma_{ICA}$ is a term associated to the detrending process.
\cite{mor14, mor15} suggest the following formula for $\sigma_{ICA}$:
\begin{equation}
\label{eqn:sigmaica}
\sigma_{ICA}^2 = f^2 \left ( \sum_{j} o_j^2 \textbf{ISR}_j + \sigma_{ntc-fit}^2 \right )
\end{equation}
where $\textbf{ISR}$ is the so-called Interference-to-Signal-Ratio matrix, $o_j$ are the coefficients of the non-transit-components, $m$ is their number, $\sigma_{ntc-fit}$ is the standard deviation of residuals from the theoretical raw light-curve, out of the transit, $f$ is the normalising factor for the detrended light-curve. The sum on the left takes into account the precision of the components extracted by the algorithm; $\sigma_{ntc-fit}$ indicates how well the linear combination of components approximates the out-of-transit. The MULTICOMBI code, i.e. the algorithm that we use for the ICA transformation, provides two Interference-to-Signal-Ratio matrices, $\textbf{ISR}^{EF}$ and $\textbf{ISR}^{WA}$, associated to the sub-algorithms EFICA and WASOBI, respectively. Two approaches has been suggested to derive a single Interference-to-Signal-Ratio matrix:
\begin{equation}
\label{eqn:ISRave}
\textbf{ISR} = \frac{ \textbf{ISR}^{EF} + \textbf{ISR}^{WA}}{2}
\end{equation}
\begin{equation}
\label{eqn:ISRmin}
\textbf{ISR}_{i,j} = min \left ( \textbf{ISR}_{i,j}^{EF}, \textbf{ISR}_{i,j}^{WA} \right )
\end{equation}
Eq. \ref{eqn:ISRave} is a worst-case estimate, while Eq. \ref{eqn:ISRmin} takes into account the outperforming separation capabilities of MULTICOMBI compared to EFICA and WASOBI. We adopt Eq. \ref{eqn:ISRmin} throughout this paper, but results obtained with both options are reported in Tab. \ref{tab7}, \ref{tab8}, \ref{tab9}, and \ref{tab10}. In most cases the differences are negligible.

\section{Instrument simulations}

\subsection{Instrument jitter only}
\label{jitter_only}

\begin{figure}[!h]
\epsscale{0.70}
\plotone{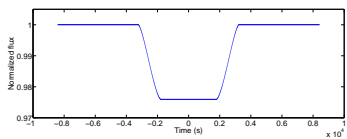}
\caption{Referent transit light-curve adopted in simulations. \label{fig1}}
\end{figure}

We consider an \textit{ideal} transit light-curve with parameters reported in Tab. \ref{tab1}, sampled at 8.4 s over 4$\frac{2}{3}$ hr, totaling 2001 data points, symmetric with respect to the transit minimum.
\begin{table}[!h]
\begin{center}
\caption{Transit parameter values adopted in all simulations: $p = r_p/R_s$ is the ratio of planetary to stellar radii, $a_0 = a/R_s$ is the orbital semimajor axis in units of stellar radius, $i$ is the orbital inclination, $e$ is the eccentricity, $P$ is the orbital period, $\gamma_1$ and $\gamma_2$ are quadratic limb darkening coefficients \citep{how11}. \label{tab1}}
\begin{tabular}{cc}
\tableline\tableline
$p$ & 0.15500 \\
$a_0$ & 9.0 \\
$i$ & 85.80 \\
$e$ & 0.0 \\
$P$ & 2.218573 days\\
$\gamma_1$, $\gamma_2$ & 0.0, 0.0\\
\tableline
\end{tabular}
\end{center}
\end{table}
Fig. \ref{fig1} shows the ideal light-curve. To each data point we associate a number of photons proportional to the expected flux, in particular 50,000 photons in the out-of-transit. We generate random gaussian coordinates for each photon, representing their positions on the plane of the CCD. Finally, we add a grid on this plane: each square of the grid represents a pixel, and the number of photons into a square at a time is the read of an individual pixel in absence of pixel systematics. To simulate the effect of instrumental jitter, we move the gridlines from one data point to another (it is equivalent to shift the coordinates of the photons).

\begin{figure*}
\epsscale{1.10}
\plotone{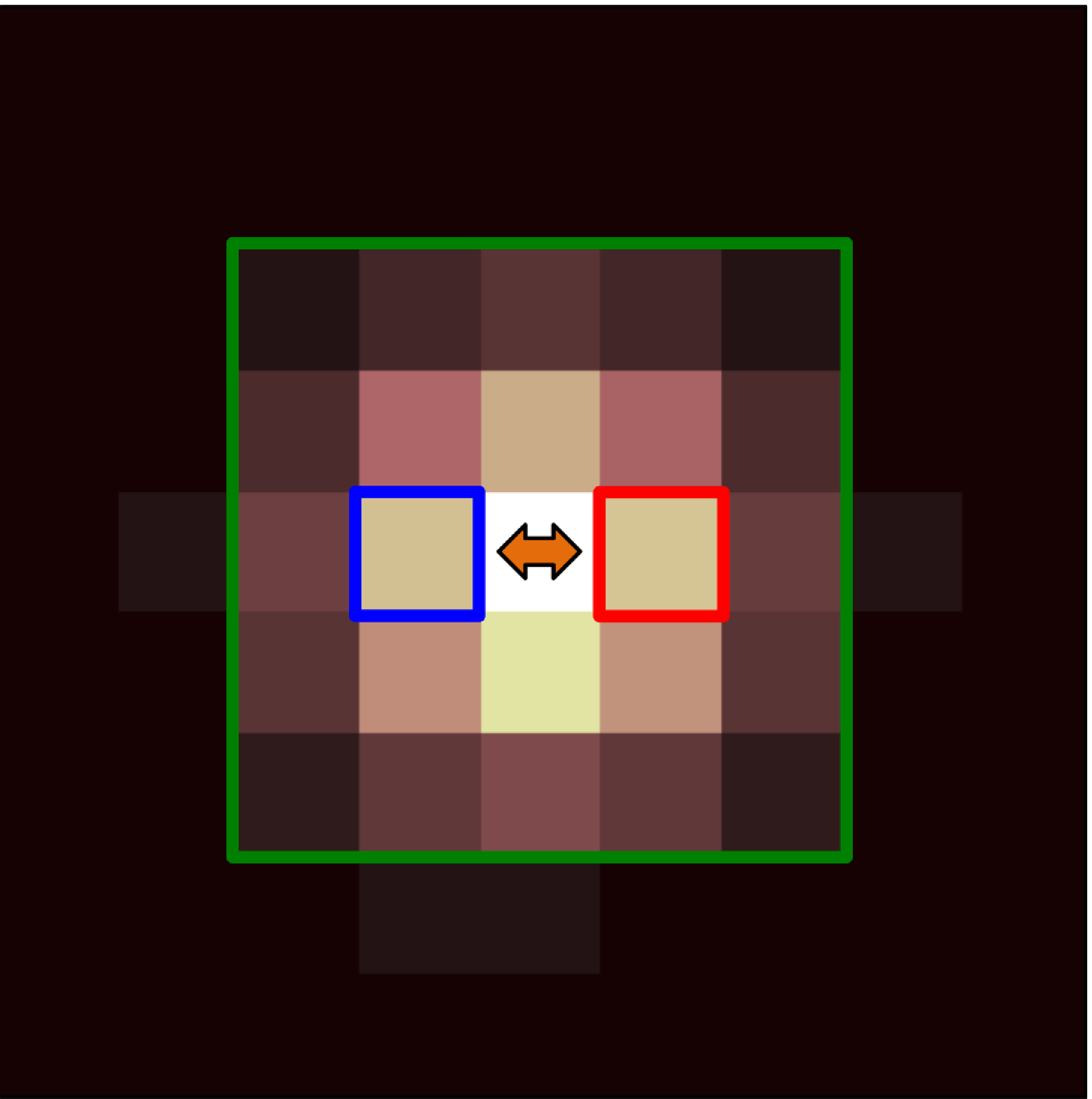}
\epsscale{1.25}
\plotone{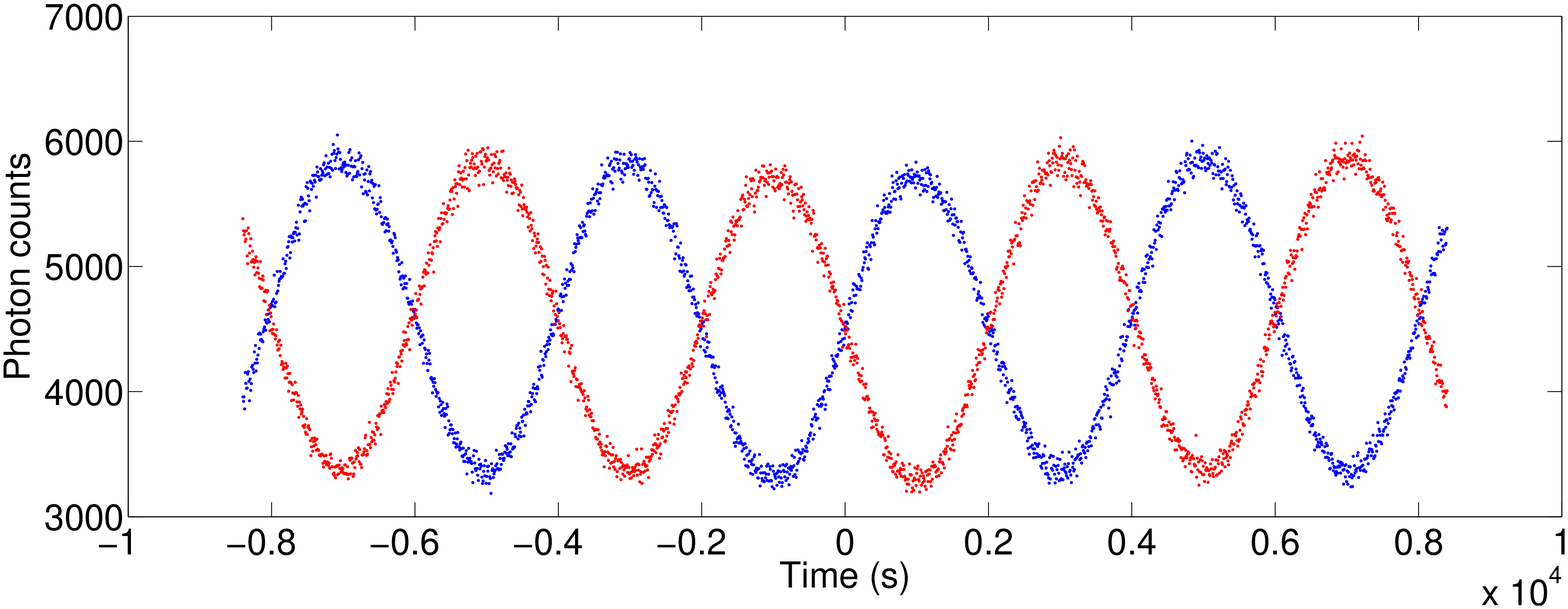}
\plotone{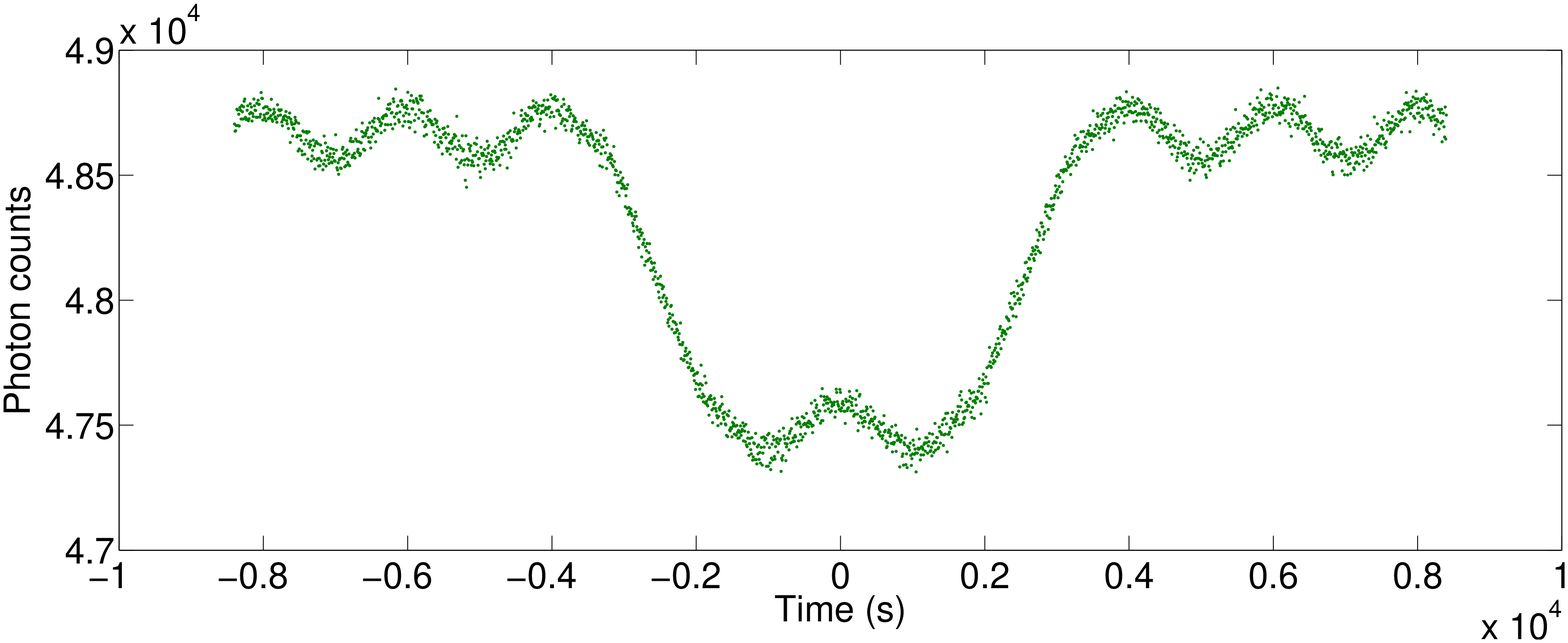}
\plotone{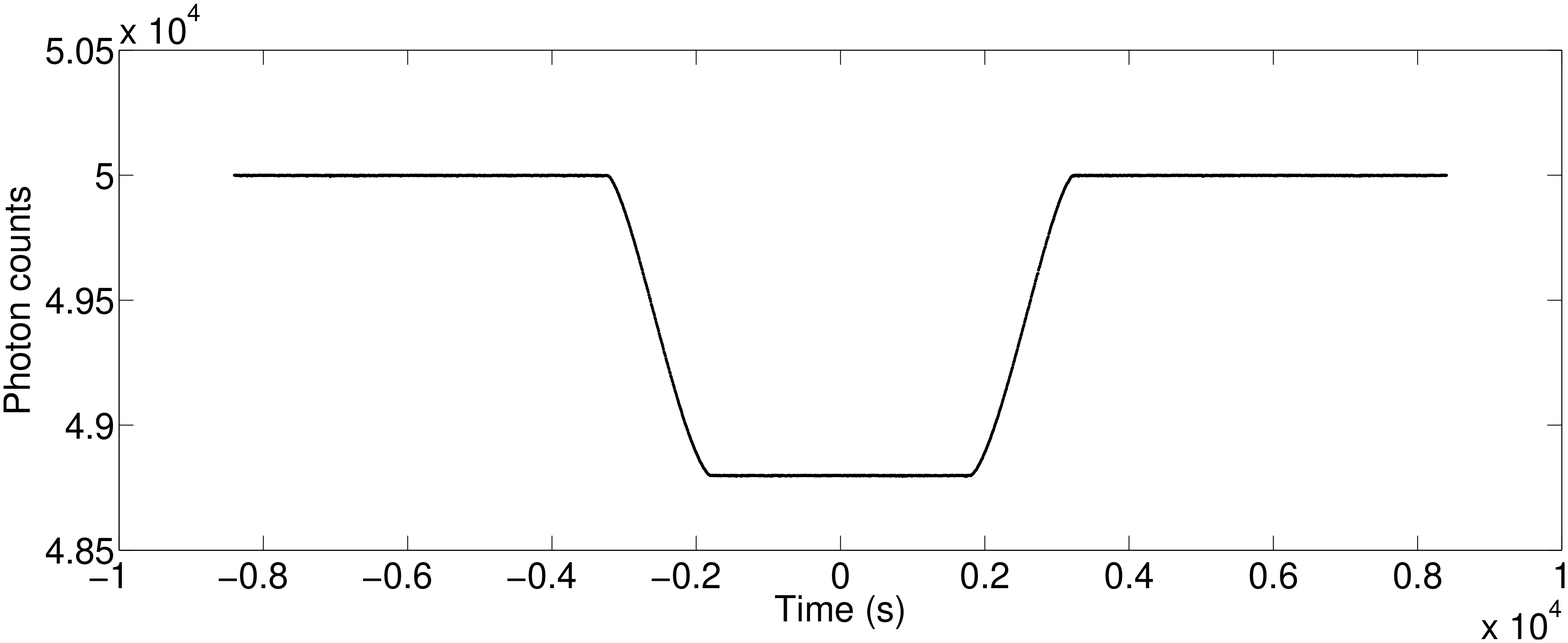}
\caption{Top panel: Representation of one simulated frame onto the focal plane. Bottom panels: simulated time series associated to selected individual pixels (blue and red), a centered 5$\times$5 array (green), and a centered 9$\times$9 array (black). The centroid is assumed oscillating in the direction indicated by the double-headed arrow, with a sinusoidal pattern (sin1, see Tab. \ref{tab2}). \label{fig2}}
\end{figure*}

Fig. \ref{fig2} shows the jitter effect at different levels, i.e. individual pixels, and small and large clusters. Pixel flux variations are related to the changing position of the PSF: the flux is higher when the centre of the PSF is closer to the centre of the pixel. The same is valid for the flux integrated over a small aperture, compared to the width of the PSF. Also note that short time scale fluctuations are present as a sampling effect. Neither of these variations are observed with a large aperture that includes all the photons at any time.


\subsection{Source Poisson noise}

In a real case scenario, the detected photons from the astrophysical source are distributed according to a poissonian function with fluctuations proportional to the square root of the expected number of photons. The value of the proportionality factor is specific to the instrument. Poisson noise in the source is a natural limitation to the photometric precision that can be achieved for any astrophysical target. To include this effect in simulations, we added a random time series to the astrophysical model, multiplied by different scaling factors, then generated frames with total number of photons determined by those noisy models.

\subsection{The effect of pixel systematics}

Spitzer/IRAC datasets for channels 1 and 2 show temporal flux variations correlated to the centroid position, independently from the aperture selected \citep{faz04,mc06}. Here we consider two effects that, combined to the instrument jitter, can produce this phenomenon:
\begin{enumerate}
\item inter-pixel quantum efficiency variations, simulated multiplying the photons in a pixel by a coefficient to get the read, being the coefficients not identical for all the pixels;
\item intra-pixel sensitivity variations, simulated by assigning individual coefficients dependent on the position of the photon into the pixel.
\end{enumerate}

\subsection{Description of simulations}
\label{sec:description}

\begin{figure*}
\epsscale{1.60}
\plotone{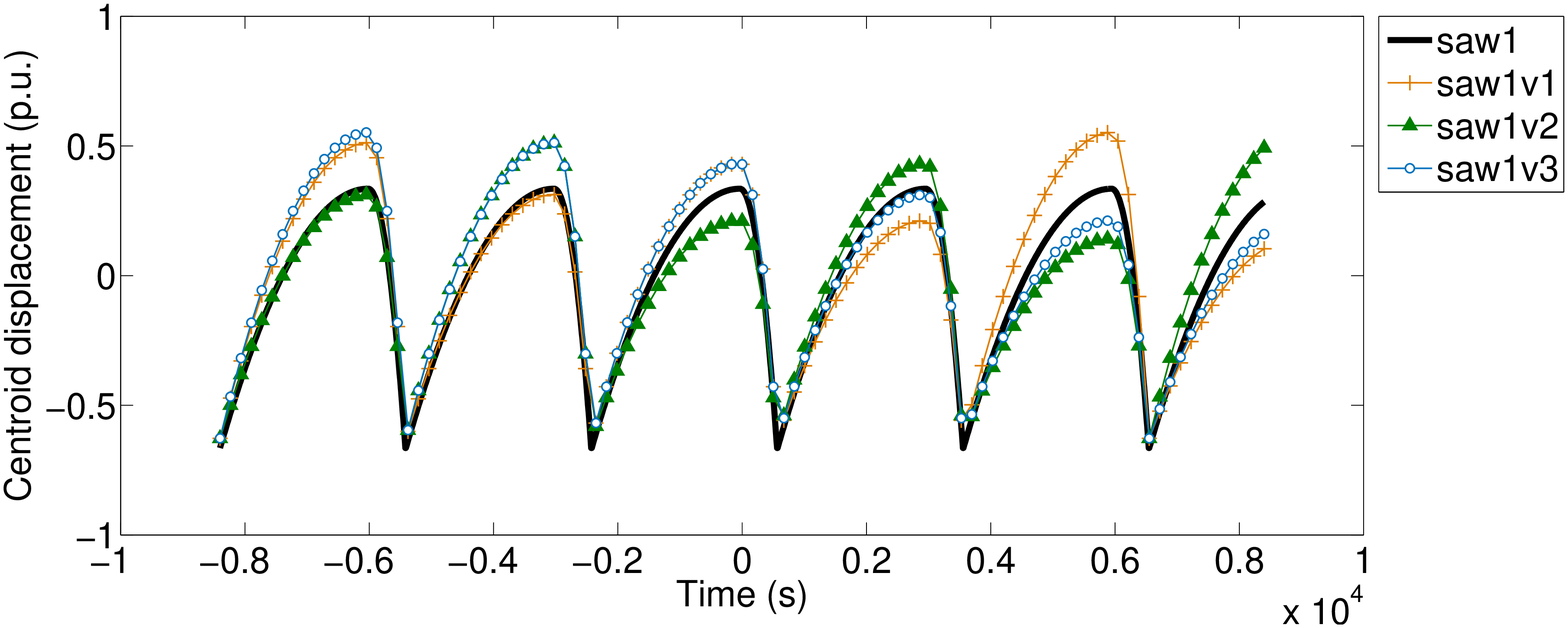}
\plotone{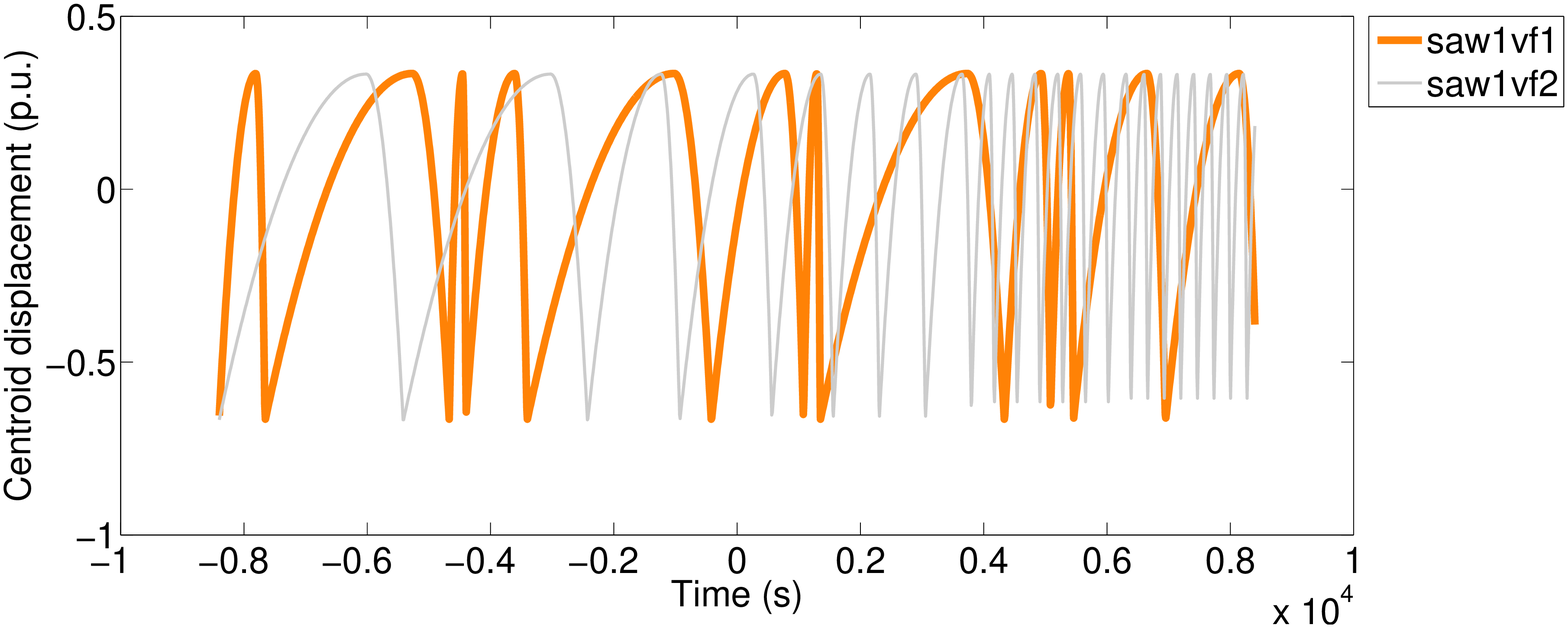}
\caption{Top panel: jitter time series saw1 (black), saw1v1 (ecru, cross markers), saw1v2 (green, triangles), and saw1v3 (cyan, empty circles); markers are represented every 20 data points for reasons of visibility. Bottom panel:  jitter time series saw1vf1 (orange), and saw1vf2 (grey). The other jitter time series are not reported, since their representations are obvious (see Tab. \ref{tab2}).  \label{fig3}}
\end{figure*}

We performed simulations for two values of (gaussian) PSF widths, $\sigma_{PSF}$:
\begin{itemize}
\item $\sigma_{PSF} = $1 p.u. (pixel side units);
\item $\sigma_{PSF} = $0.2 p.u.
\end{itemize}
The two sets of frames differ only in the scaling factor in photon coordinates, therefore no relative differences are attributable to random generation processes. As a comparison, the nominal PSF widths for Spitzer/IRAC channels are in the range 0.6-0.8 p.u. \citep{faz04}. By comparing two more extreme cases, we observe the impact of the PSF width on pixel systematics and their detrending. We also tested eleven Poisson noise amplitudes, linearly spaced between $0$ and $\sqrt{50,000}$. \\
We simulated several jitter time series as detailed in Tab. \ref{tab2} and Fig. \ref{fig3}. For each case, we adopted:
\begin{enumerate}
\item a random generated quantum efficiency map with standard deviations of $\sim$10$^{-2}$, to simulate the inter-pixel variations;
\item a non-uniform response function for each pixel, i.e. 1$-$0.1$d$, where $d$ is the distance from the centre of the pixel, to simulate the intra-pixel variations.
\end{enumerate}
These values are of the same order as the nominal pixel-to-pixel accuracy of flat-fielding and sub-pixel response for Spitzer/IRAC channels 1 and 2 \citep{faz04}, but slightly larger to better visualize their effects.
Finally, we add white noise time series at an arbitrary level of 5 counts/pixel/data point for most cases, to simulate high-frequency pixel noise (HFPN). It is worth to note that the same quantum efficiency maps and noise time series have been adopted for all the simulations with pixel arrays of the same size, to minimize possible aleatory effects when comparing the results. 
\begin{table*}
\begin{center}
\caption{List of jitter time series adopted in simulations and their properties. \label{tab2}}
\begin{tabular}{cccc}
\tableline\tableline
Abbr. & Shape & Peak-to-peak amplitude (p.u.) & Period (s)\\
\tableline
sin1 & sinusoidal & 0.6 & 4014.6\\
cos1 & cosinusoidal & 0.6 & 4014.6\\
sin2 & sinusoidal & 0.6 & 3011.0\\
cos2 & cosinusoidal & 0.6 & 3011.0\\
sin3 & sinusoidal & 0.6 & 2007.3\\
cos3 & cosinusoidal & 0.6 & 2007.3\\
saw1 & $\sim$sawtooth & 0.6 & 2990.4\\
saw1v1 & $\sim$sawtooth & variable & 2990.4\\
saw1v2 & $\sim$sawtooth & variable & 2990.4\\
saw1v3 & $\sim$sawtooth & decreasing & 2990.4\\
saw1vf1 & $\sim$sawtooth & 0.6 & variable\\
saw1vf2 & $\sim$sawtooth & 0.6 & decreasing\\
jump04c & Heaviside step & 0.4 & mid-transit discontinuity\\
\tableline
\end{tabular}
\end{center}
\end{table*}

\clearpage

\section{Results}

After having generated the simulated raw datasets, we applied both pixel-ICA and PCD detrending techniques to evaluate their reliability and robustness in those contexts. More specifically, the polynomial decorrelating function adopted is:
\begin{equation}
p(x,y) = a_1 + a_2(x-\bar{x}) + a_3(y - \bar{y}) + a_4(x-\bar{x})^2 + a_5(y - \bar{y})^2
\end{equation}
where $x$ and $y$ are the centroid coordinates, calculated as weighted means of the counts per pixel, $\bar{x}$ and $\bar{y}$ are the mean centroid coordinates on the out-of-transit, and $a_i$ coefficients are fitted on the out-of-transit. In several cases we tested some variants of PCD, i.e. higher order polynomials, cross terms, etc.

Since the main purpose of this paper is to test the ability of pixel-ICA method to detrend instrument systematics, we first report the detailed results for all the configurations obtained in the limit of zero stellar noise.
We show in particular (Sec. \ref{sec:interPSF1}, \ref{sec:interPSF02}, and \ref{sec:intraPSF02}):
\begin{itemize}
\item the simulated raw light-curves and the corresponding detrended ones;
\item the root mean square (rms) from the theoretical transit light-curve, before and after the detrending processes;
\item the residual systematics in the detrended light-curves;
\item the planetary, orbital, and stellar parameters estimated by fitting the light-curves;
\item for a subsample of cases, the results of full parameter retrieval, including error bars.
\end{itemize}
The impact of stellar poissonian noise at different levels is studied for a few representative cases (Sec. \ref{sec:poisson}). A variant of pixel-ICA algorithm, and PLD detrending method are discussed in Sec. \ref{sec:pixel-ICAvariant} and \ref{sec:PLDdeming}.

\subsection{Case I: inter-pixel effects, large PSF}
\label{sec:interPSF1}

\begin{figure*}
\epsscale{0.92}
\plotone{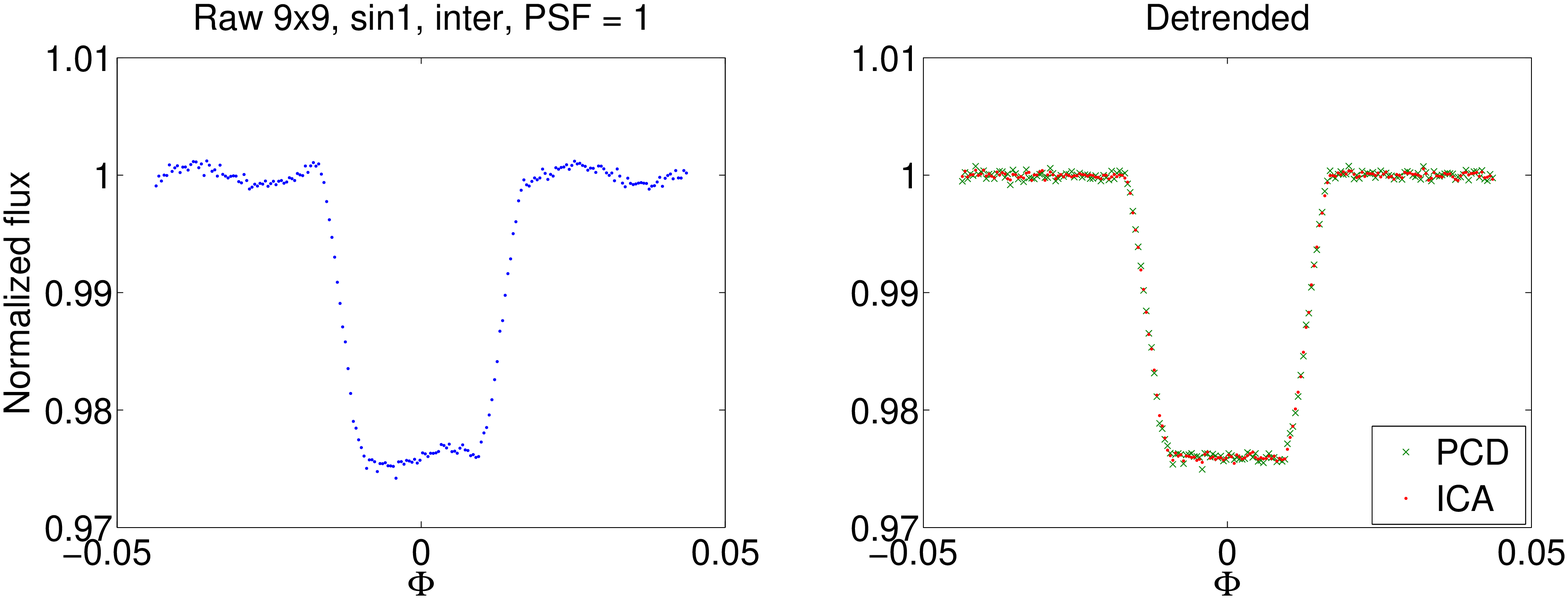}
\plotone{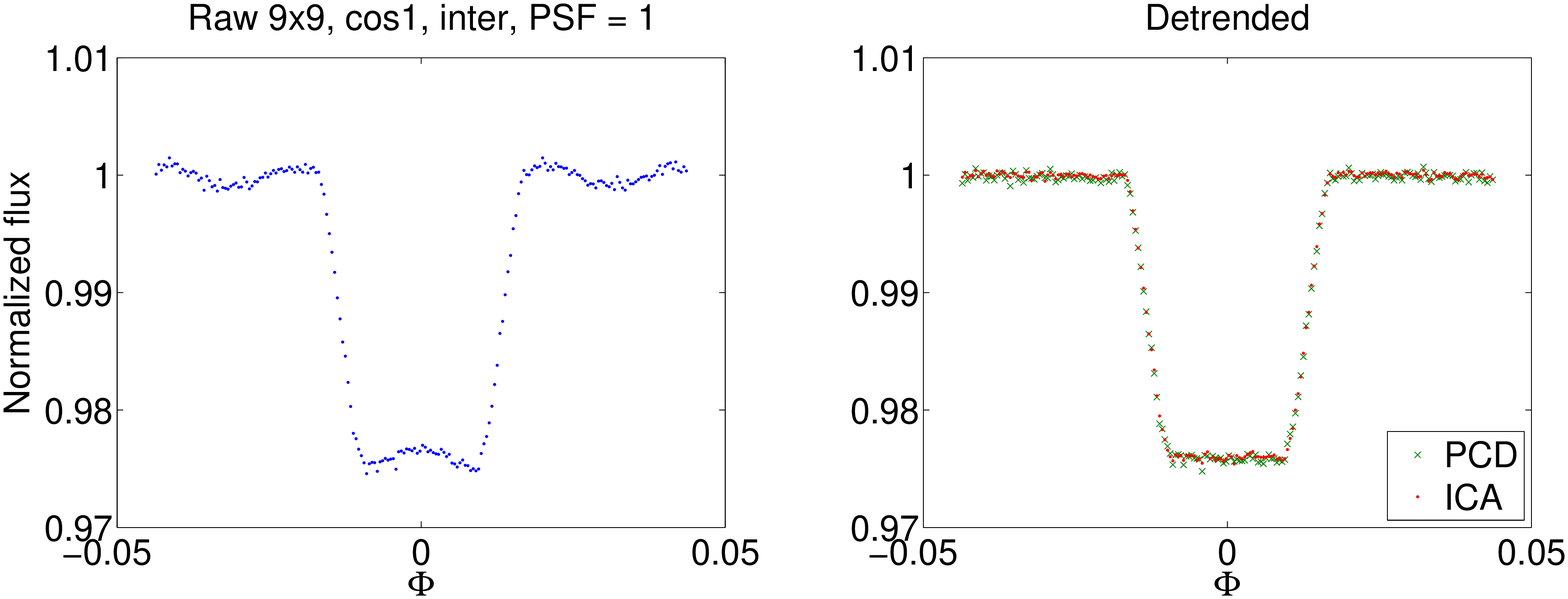}
\plotone{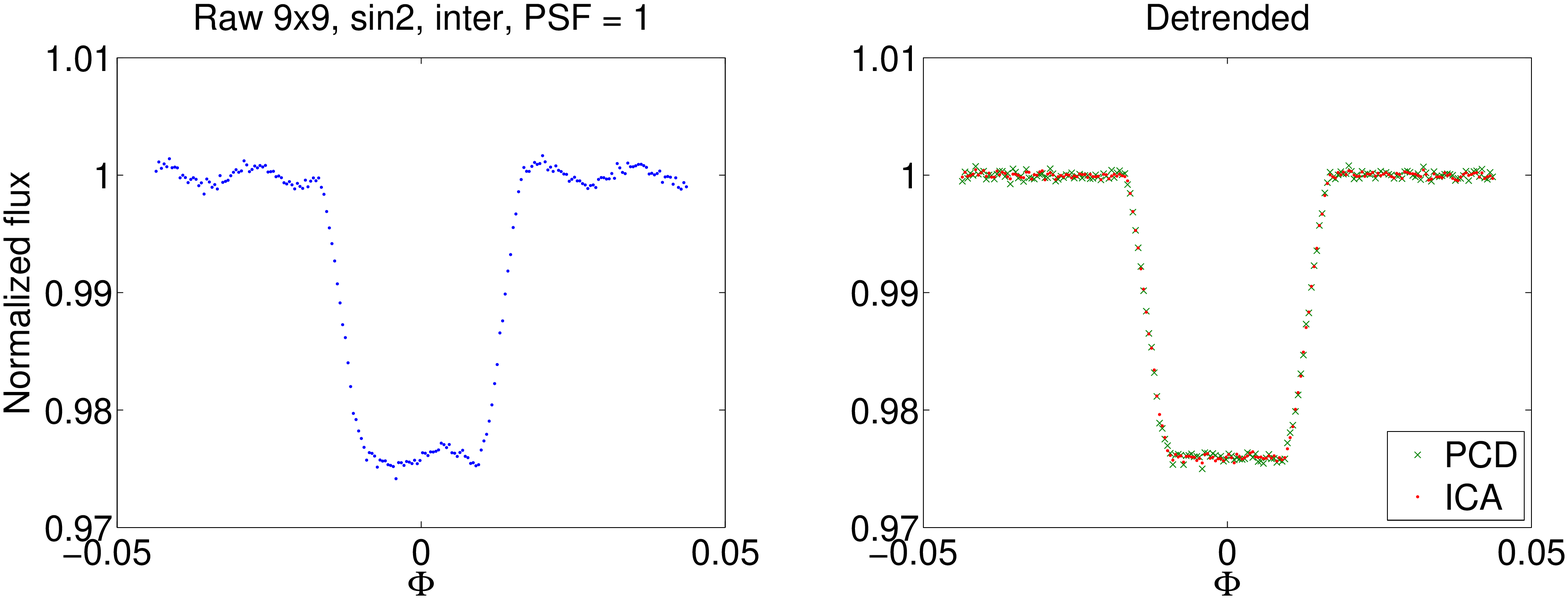}
\plotone{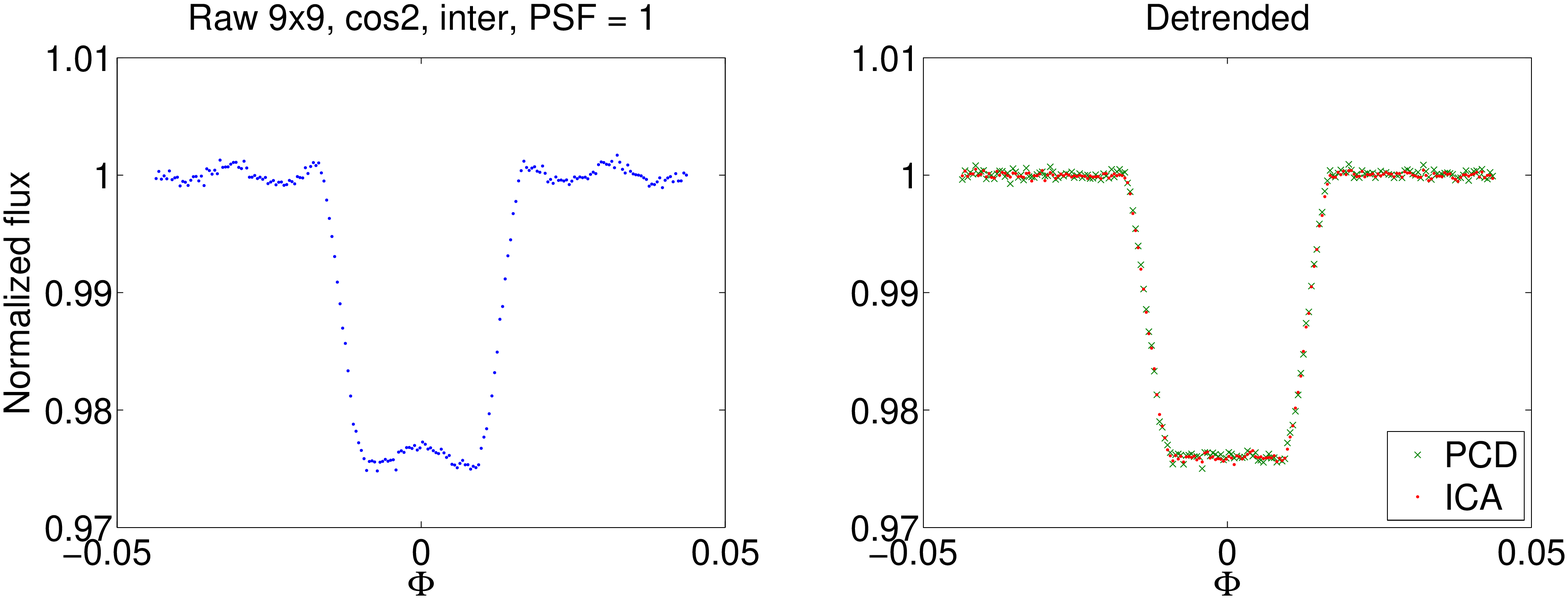}
\plotone{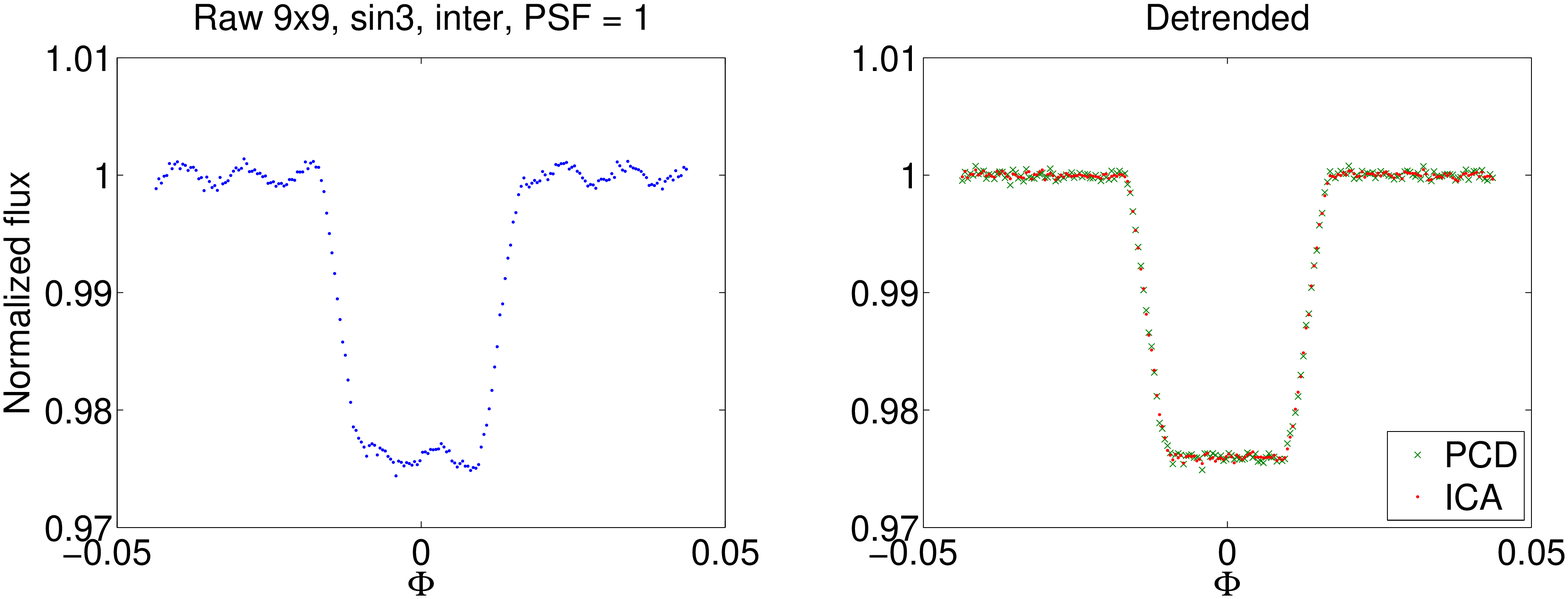}
\plotone{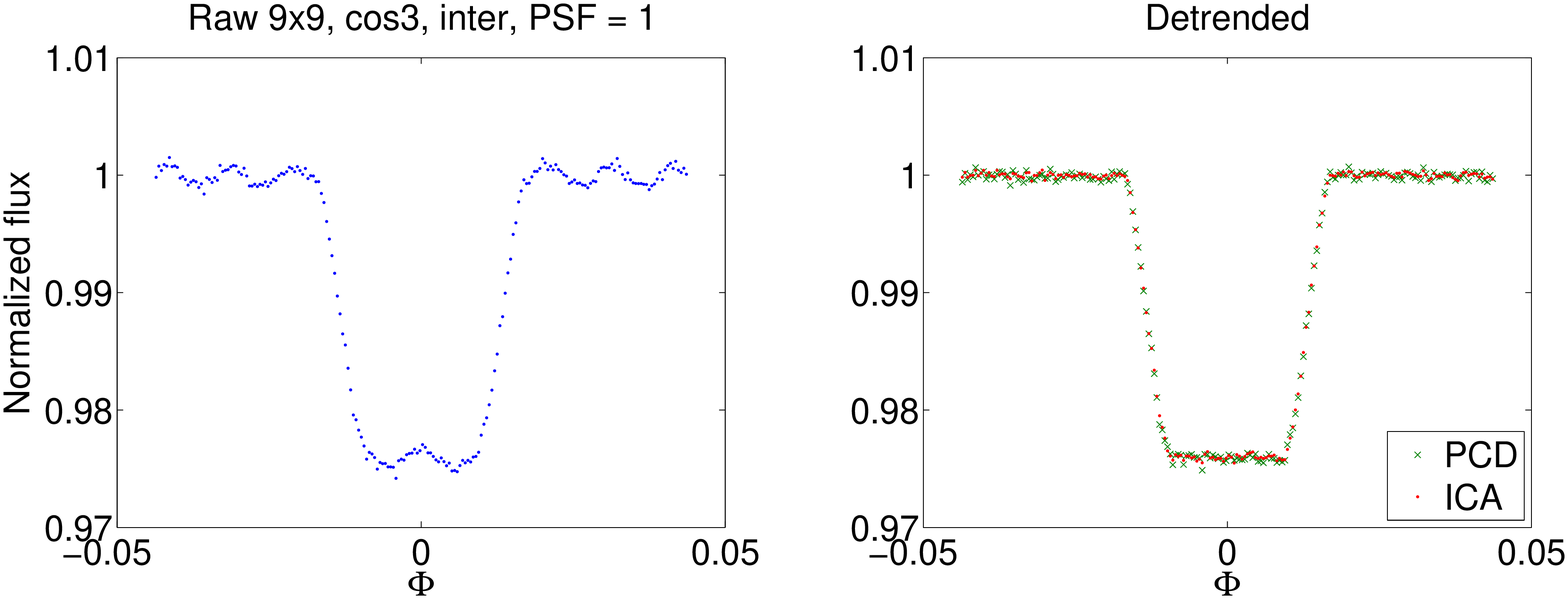}
\plotone{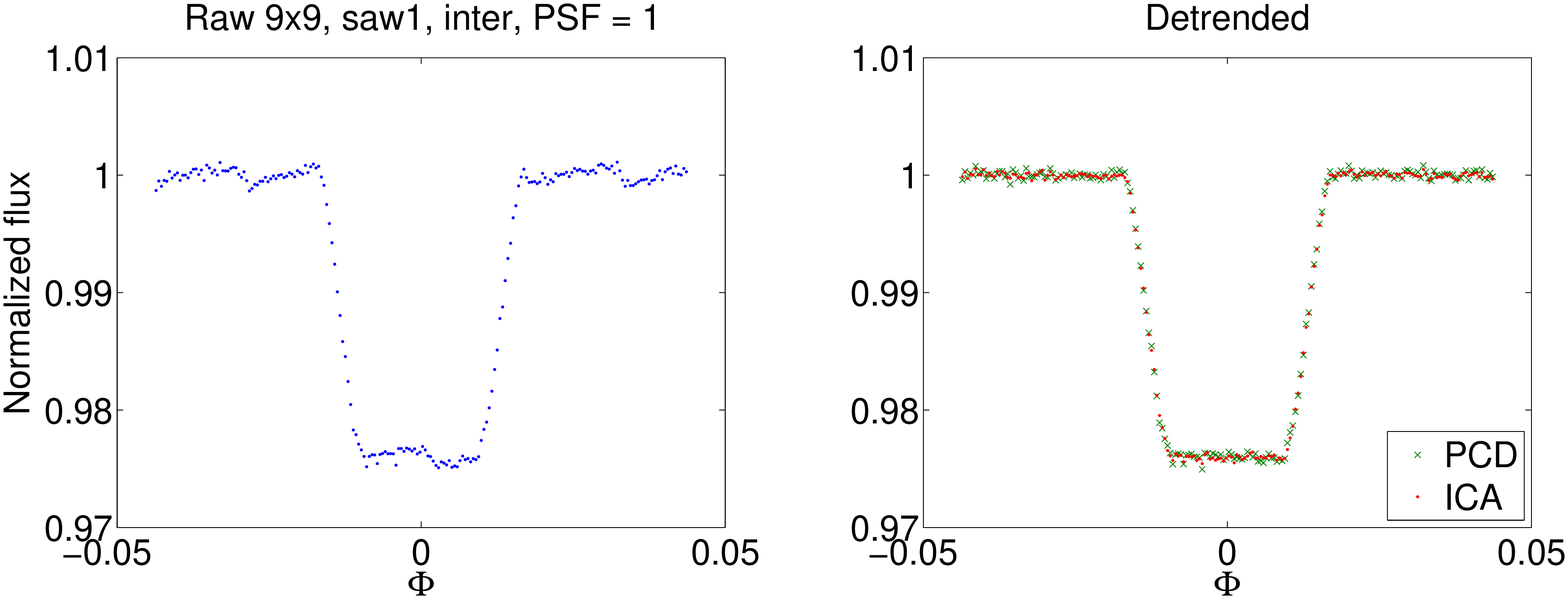}
\plotone{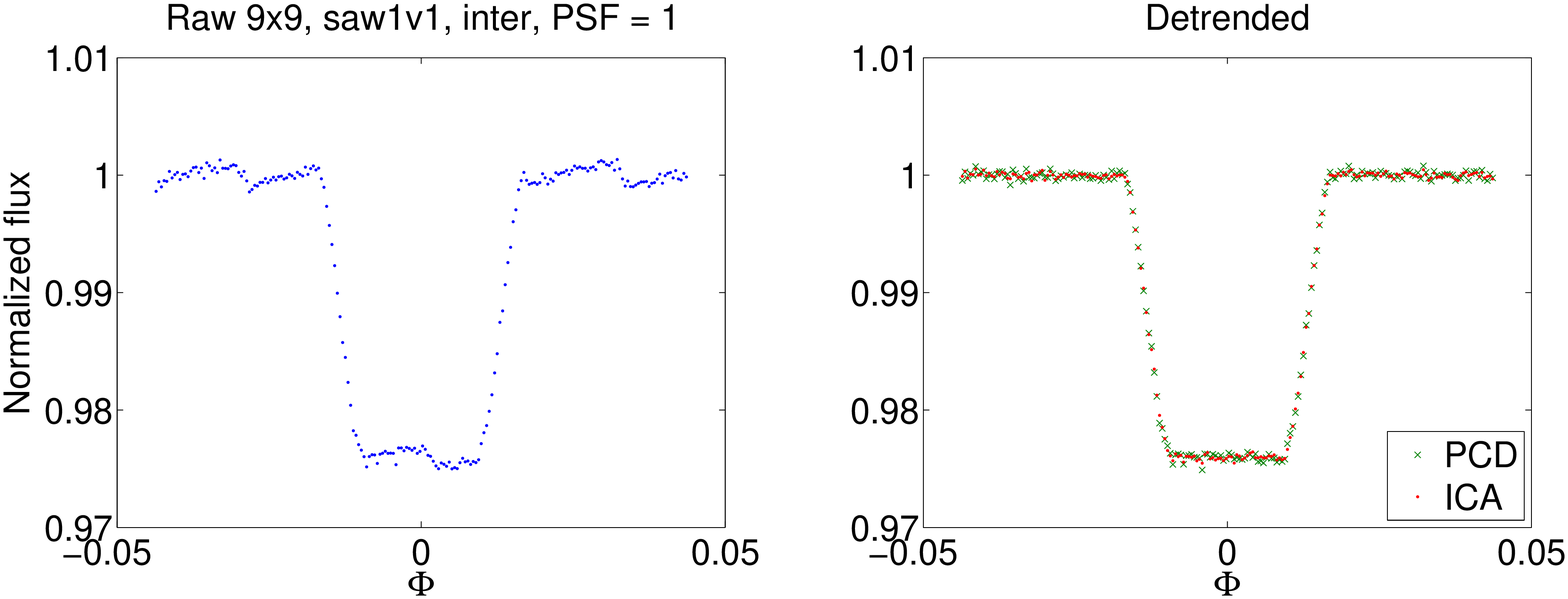}
\plotone{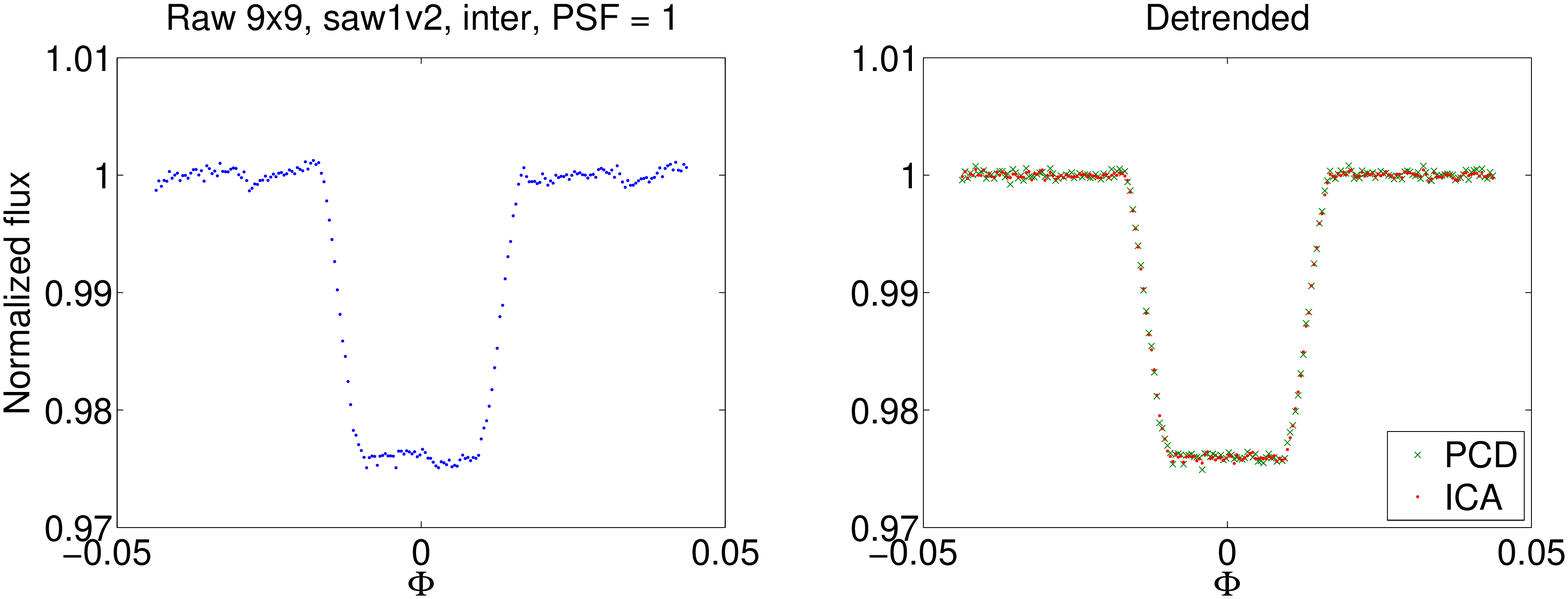}
\plotone{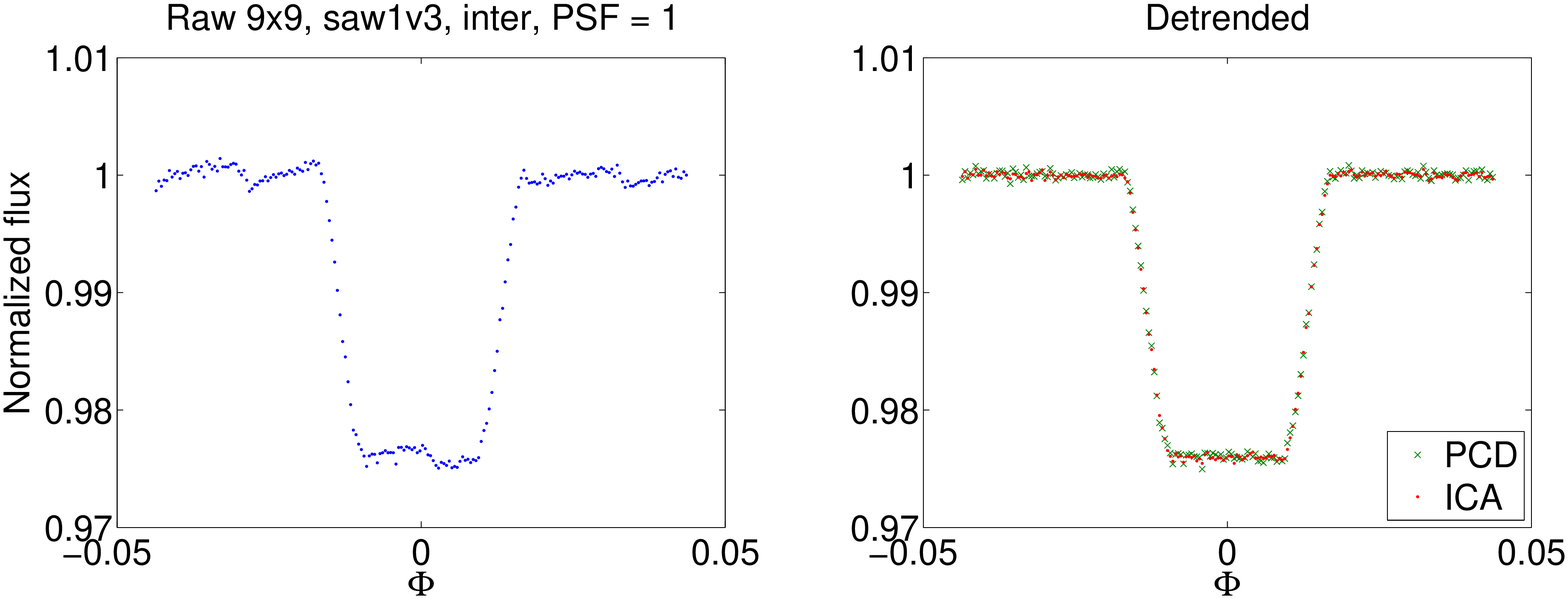}
\plotone{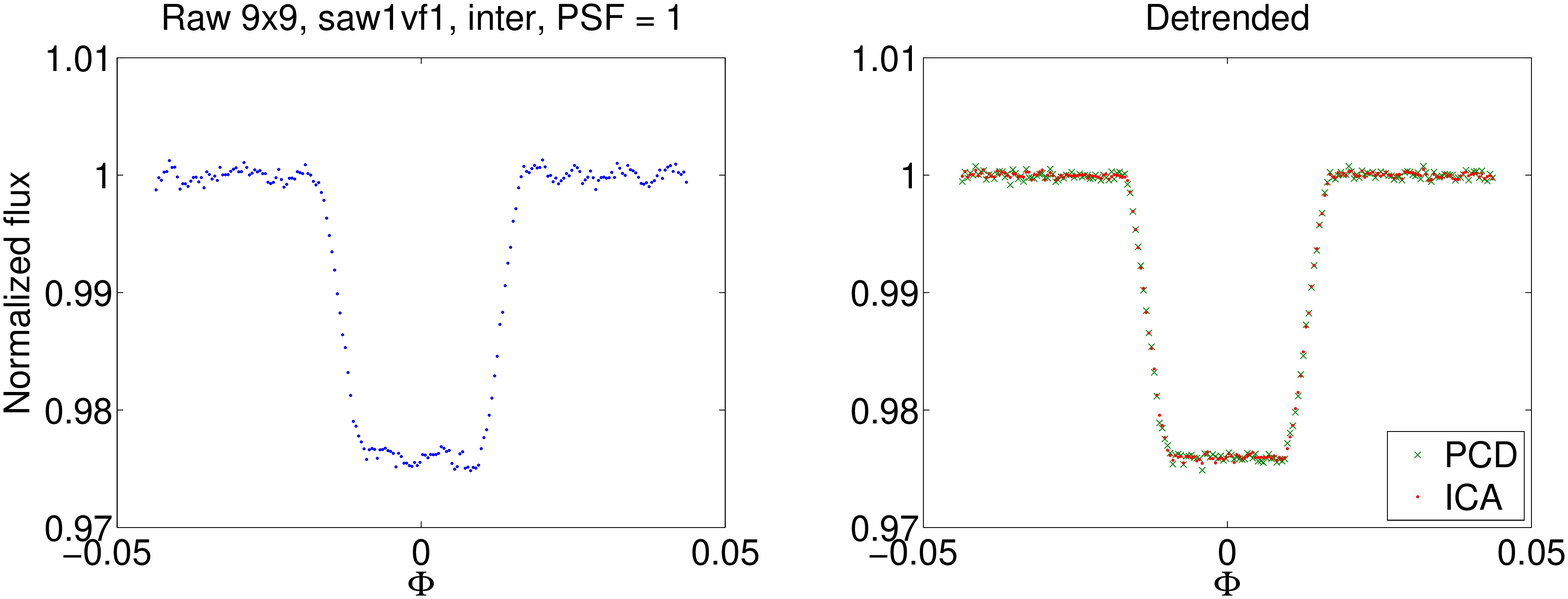}
\plotone{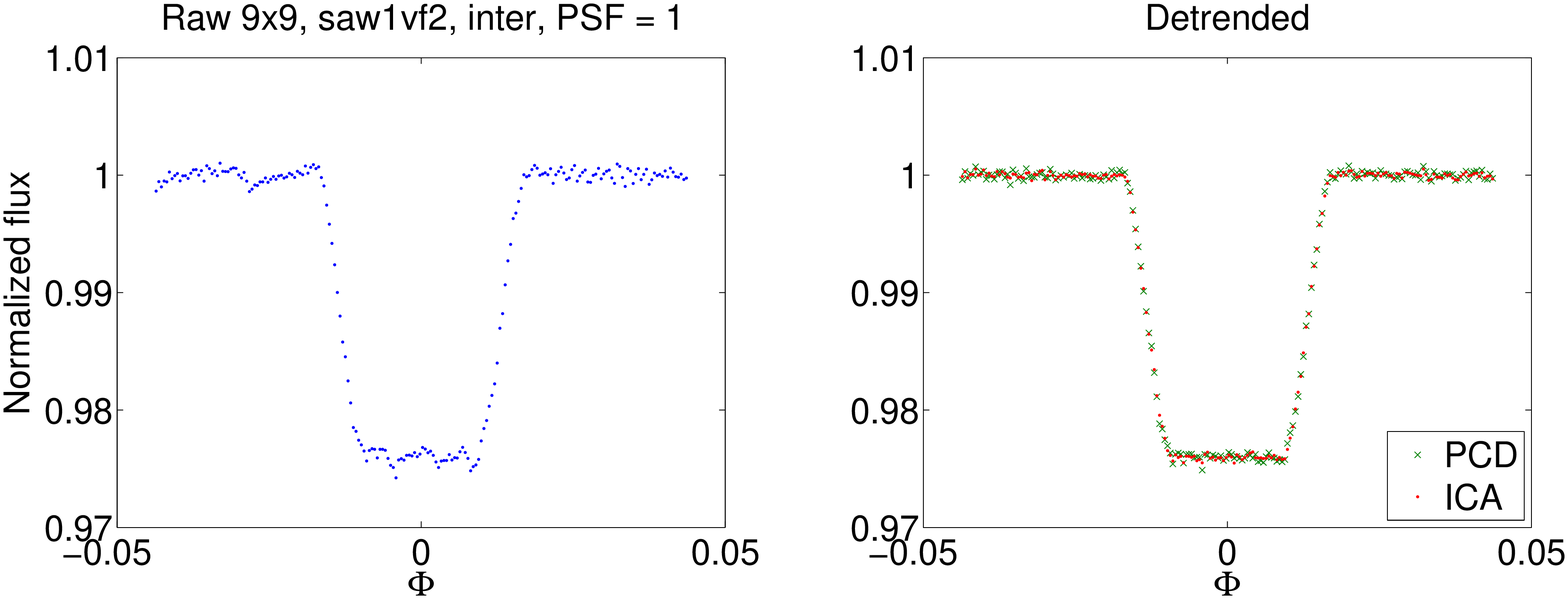}
\plotone{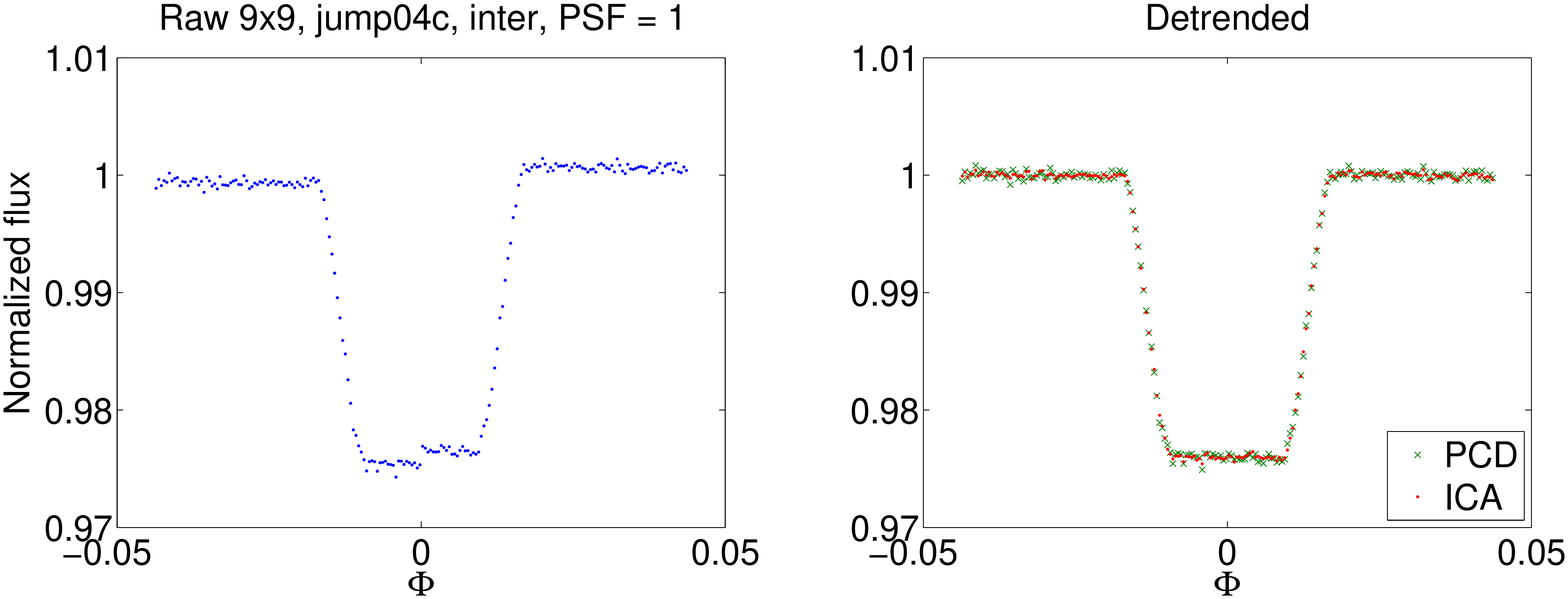}
\plotone{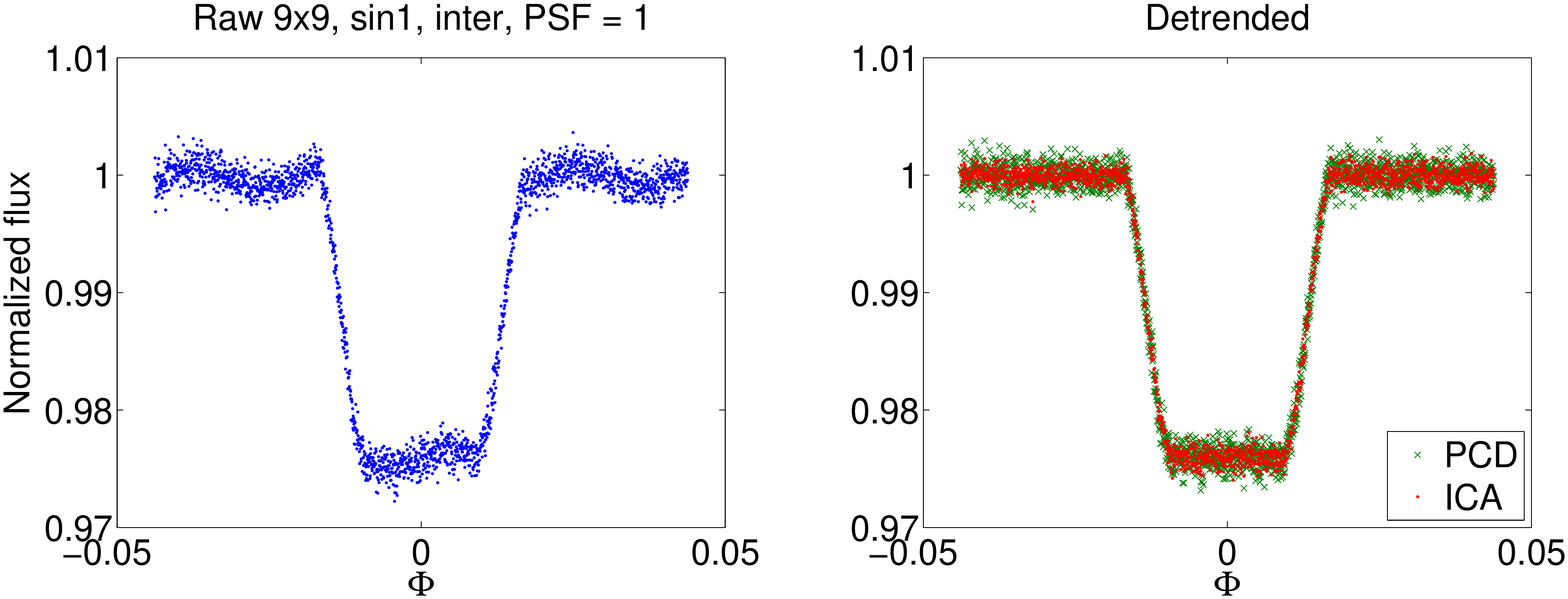}
\caption{Left panels: (blue) raw light-curves simulated with $\sigma_{PSF} = 1 p.u.$, and inter-pixel quantum efficiency variations over 9$\times$9 array of pixels. Right panels: detrended transit light-curves obtained with (green `x') polynomial centroid fitting method, and (red dots) pixel-ICA method. All the light-curves are binned over 10 points, except those in the bottom right, to make clearer visualization of the systematic effects. \label{fig4}}
\end{figure*}

\begin{figure*}
\epsscale{1.60}
\plotone{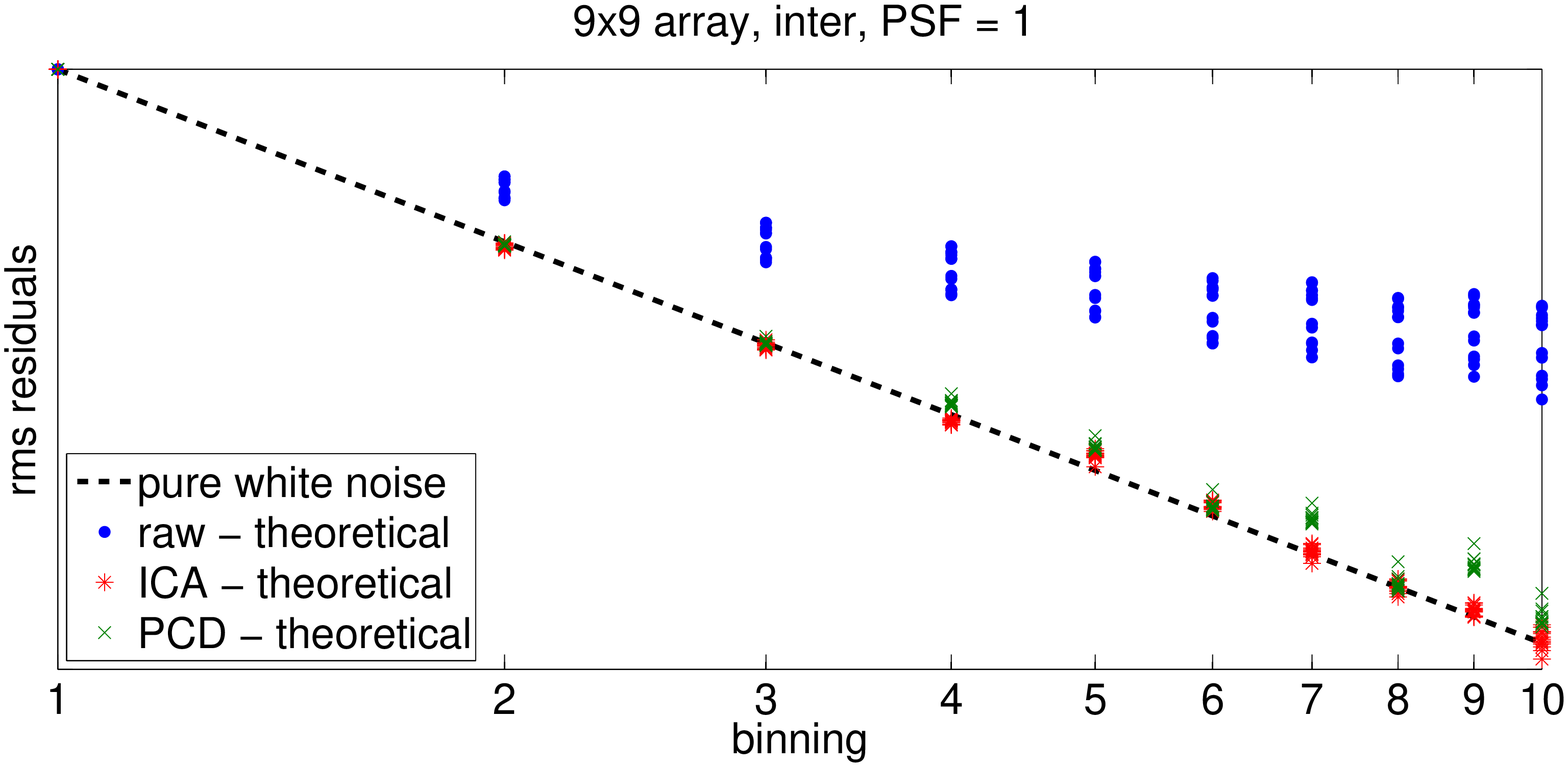}
\caption{Root mean square of residuals for light-curves binned over 1 to 10 points, scaled to their non-binned values. The simulations were obtained with $\sigma_{PSF} =$1 p.u., 9$\times$9 array, and inter-pixel effects. The dashed black line indicates the expected trend for white residuals, blue dots are for normalized raw light-curves, red `$\ast$' are for pixel-ICA detrendend light-curves, and green `x' for PCD detrended light-curves. \label{fig5}}
\end{figure*}

\begin{figure*}
\epsscale{1.50}
\plotone{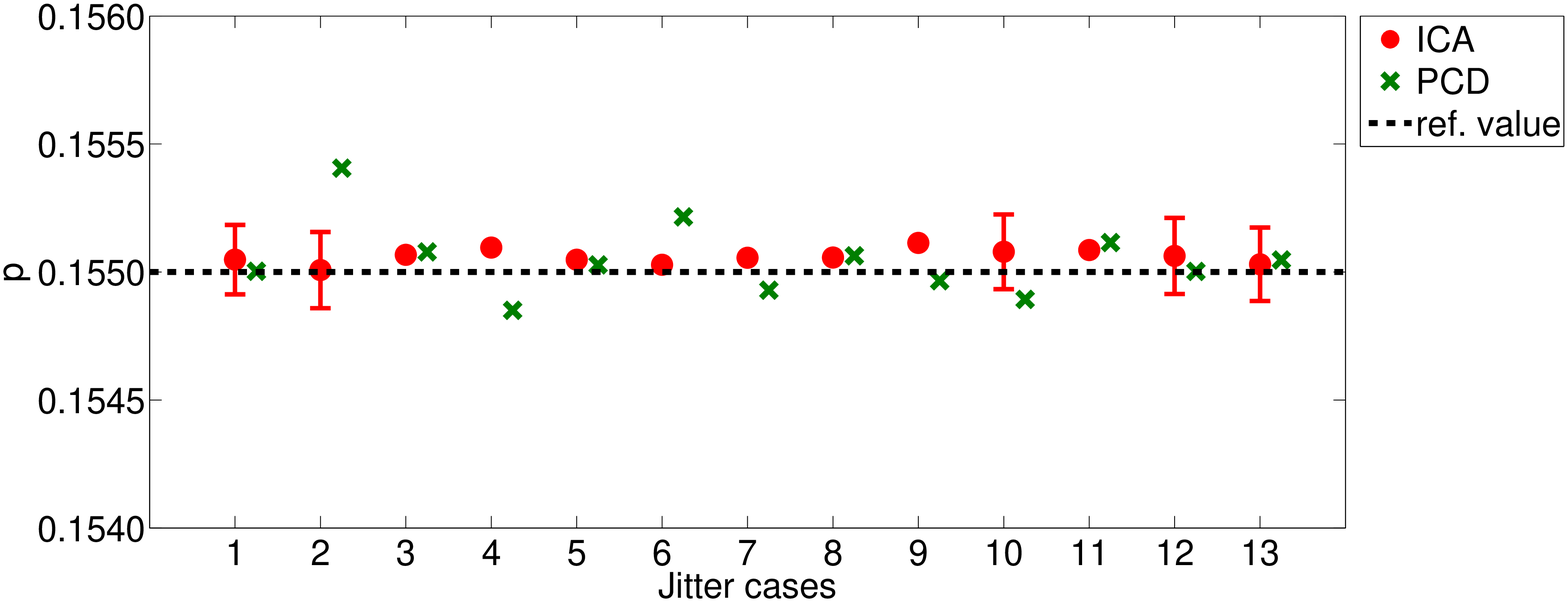}
\plotone{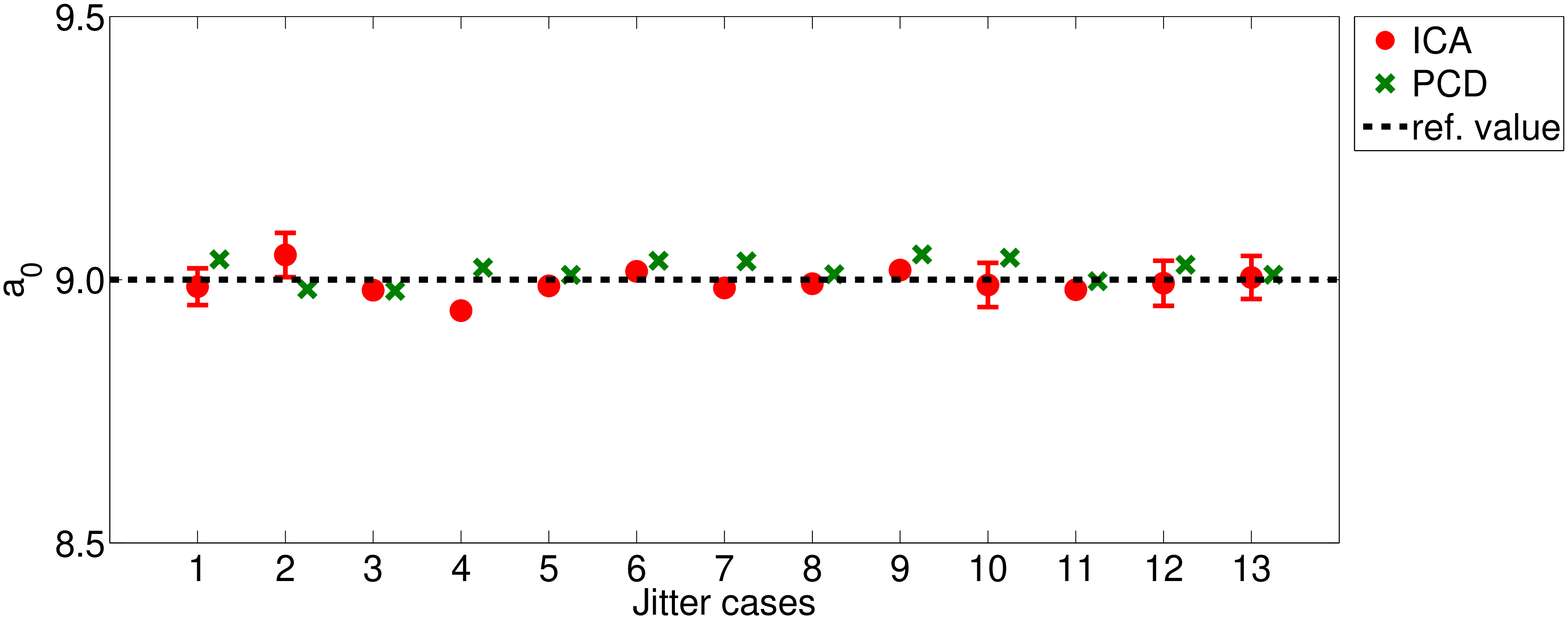}
\plotone{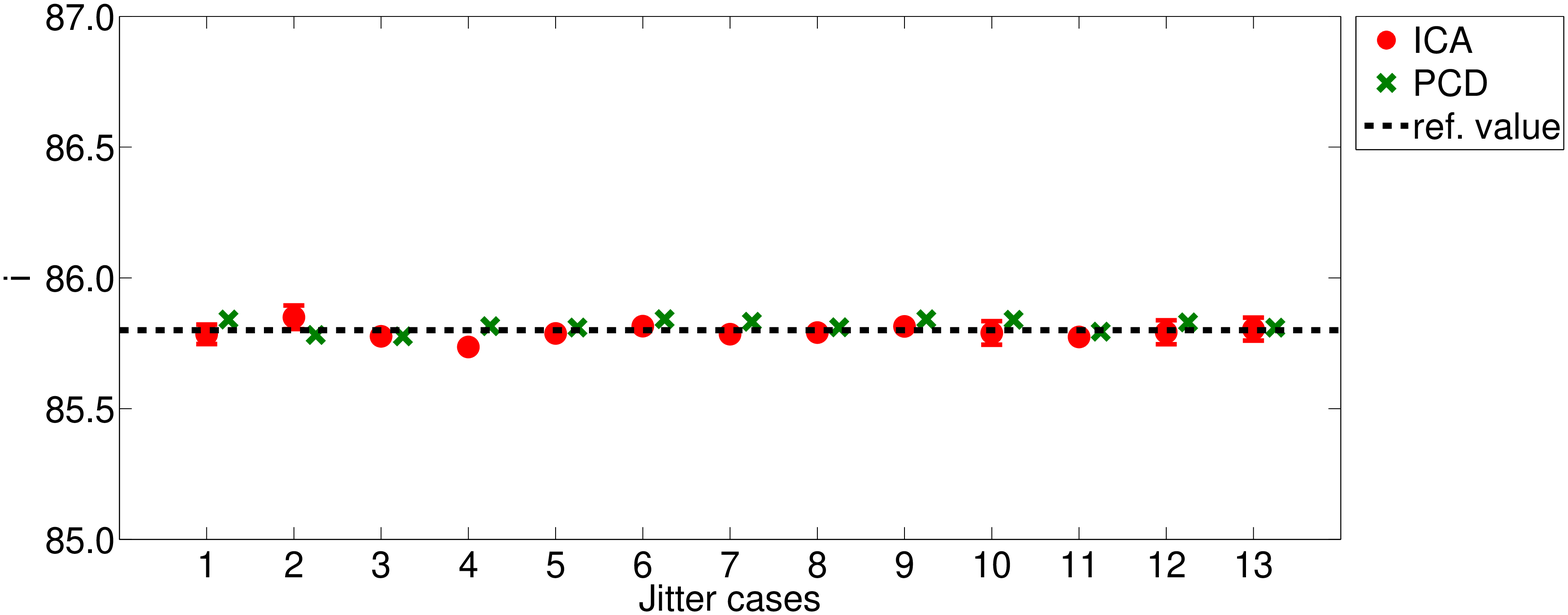}
\caption{Top panel: best estimates of the planet-to-star radii ratio, $p = r_p/R_s$, for detrended light-curves with (red dots) pixel-ICA, and (green x) PCD method ($\sigma_{PSF} = $1 p.u., inter-pixel effects over 9$\times$9 array). Error bars are reported for a few representative cases of jitter signal, i.e. sin1, cos1 (chosen as examples of periodic functions with different phasing), saw1v3 (example with non-stationary amplitude), saw1vf2 (non-stationary frequency), and jump04c (sudden change).  Middle panel: the same for the orbital semimajor axis in units of the stellar radius, $a_0 = a/R_s$. Bottom panel: the same for the orbital inclination, $i$.\label{fig6}}
\end{figure*}

Fig. \ref{fig4} shows the raw light-curves simulated with $\sigma_{PSF} = $1 p.u., and inter-pixel quantum efficiency variations over 9$\times$9 array of pixels, and the correspondent detrended light-curves obtained with pixel-ICA and PCD methods. This array is large enough that the observed modulations are only due to the pixel effects (see Sec. \ref{jitter_only}). Tab. \ref{tab3} reports the discrepancies between the detrended light-curves and the theoretical model. The amplitude of residuals after PCD detrending, i.e. 3.0$\times$10$^{-4}$ for the selected binning, equals the HFPN level (see Sec. \ref{sec:description}), while for pixel-ICA detrended data it is smaller by a factor $\sim$1/3. We note that two independent components are enough to correct for the main instrument systematics within the HFPN level, the other components slightly correct for the residual pixel systematics. Interestingly, a similar behaviour has been observed for real Spitzer datasets \citep{mor14}, but with the best fit model instead of the (unknown) theoretical one. Fig. \ref{fig5} shows how the residuals scale for binning over $n$ points, with $1 \le n \le 10$. This behaviour suggests a high level of temporal structure in raw data, which is not present in ICA-detrended light-curves. Some systematics are still detected in residuals obtained with the parametric method. We checked that the empirical centroid coordinates are accurate within 0.006 p.u. on average, and higher order polynomial corrections lead to identical results. Fig. \ref{fig6} shows the transit parameters retrieved from the detrended light-curves; in a few representative cases, we calculated the error bars as detailed in \cite{mor15}, and Sec. \ref{sec:pixel-ICA} in this paper. Numerical results are reported in Tab. \ref{tab7}. 
\begin{table*}
\begin{center}
\caption{Root mean square of residuals between the light-curves and the theoretical model for simulations with $\sigma_{PSF} = 1$ p.u., and inter-pixel quantum efficiency variations over 9$\times$9 array of pixels; in particular they are calculated for the raw light-curves, light-curves detrended with pixel-ICA, and PCD method, binned over 10 points. \label{tab3}}
\begin{tabular}{cccc}
\tableline\tableline
Jitter & rms (raw $-$ theoretical) & rms (ICA $-$ theoretical) & rms (PCD $-$ theoretical)\\
\tableline
sin1 & 6.5$\times$10$^{-4}$ & 2.0$\times$10$^{-4}$ & 3.0$\times$10$^{-4}$\\
cos1 & 6.9$\times$10$^{-4}$ & 2.1$\times$10$^{-4}$ & 3.2$\times$10$^{-4}$\\
sin2 & 6.7$\times$10$^{-4}$ & 2.0$\times$10$^{-4}$ & 3.0$\times$10$^{-4}$\\
cos2 & 6.5$\times$10$^{-4}$ & 2.1$\times$10$^{-4}$ & 3.1$\times$10$^{-4}$\\
sin3 & 6.6$\times$10$^{-4}$ & 2.0$\times$10$^{-4}$ & 3.0$\times$10$^{-4}$\\
cos3 & 6.6$\times$10$^{-4}$ & 2.0$\times$10$^{-4}$ & 3.1$\times$10$^{-4}$\\
saw1 & 5.4$\times$10$^{-4}$ & 2.1$\times$10$^{-4}$ & 3.1$\times$10$^{-4}$\\
saw1v1 & 6.0$\times$10$^{-4}$ & 2.1$\times$10$^{-4}$ & 3.0$\times$10$^{-4}$\\
saw1v2 & 5.6$\times$10$^{-4}$ & 2.1$\times$10$^{-4}$ & 3.1$\times$10$^{-4}$\\
saw1v3 & 5.9$\times$10$^{-4}$ & 2.1$\times$10$^{-4}$ & 3.1$\times$10$^{-4}$\\
saw1vf1 & 5.5$\times$10$^{-4}$ & 2.1$\times$10$^{-4}$ & 3.0$\times$10$^{-4}$\\
saw1vf2 & 5.3$\times$10$^{-4}$ & 2.0$\times$10$^{-4}$ & 3.0$\times$10$^{-4}$\\
jump04c & 6.9$\times$10$^{-4}$ & 1.9$\times$10$^{-4}$ & 3.0$\times$10$^{-4}$\\
\tableline
\end{tabular}
\end{center}
\end{table*}

\clearpage

\begin{figure*}
\epsscale{0.94}
\plotone{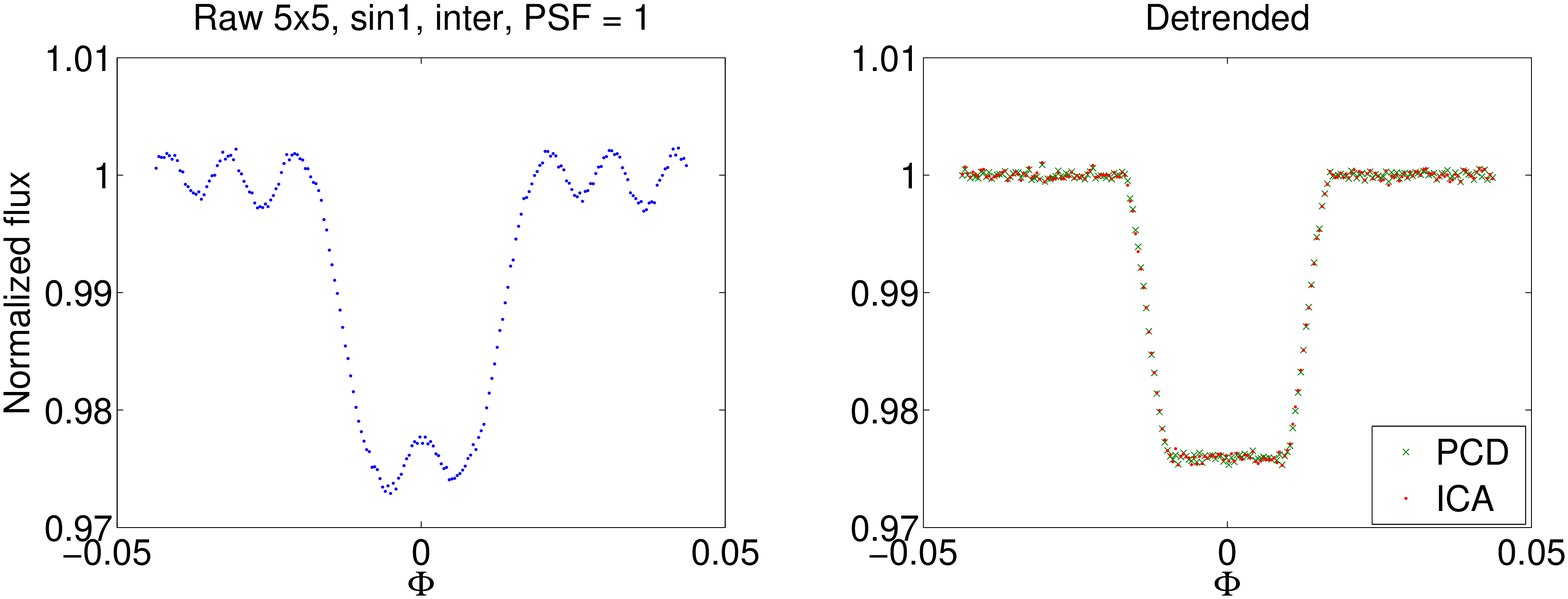}
\plotone{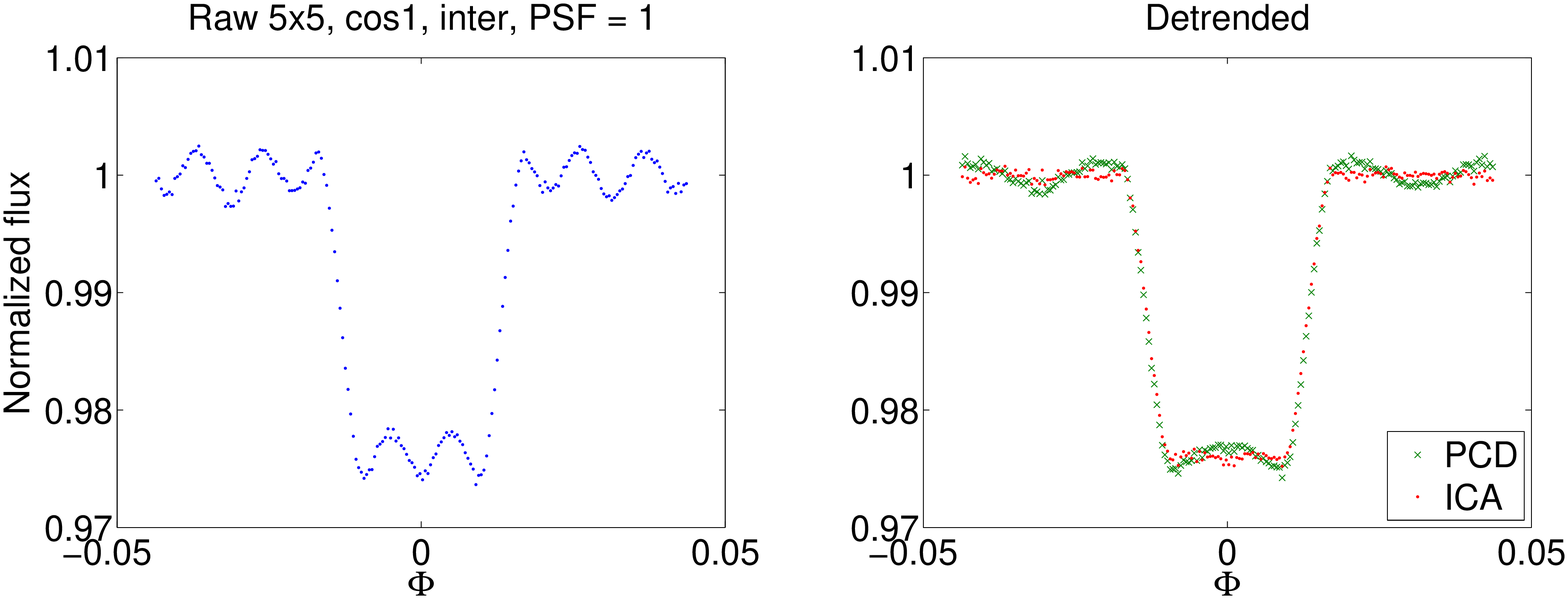}
\plotone{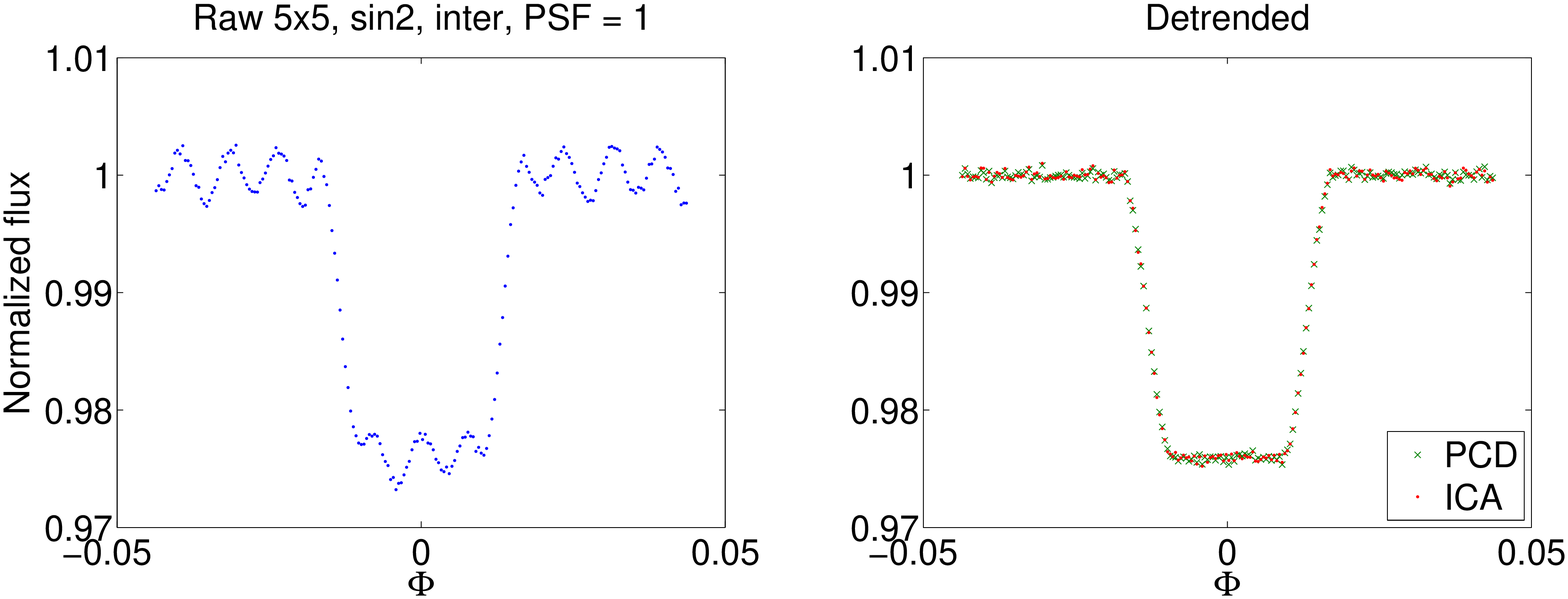}
\plotone{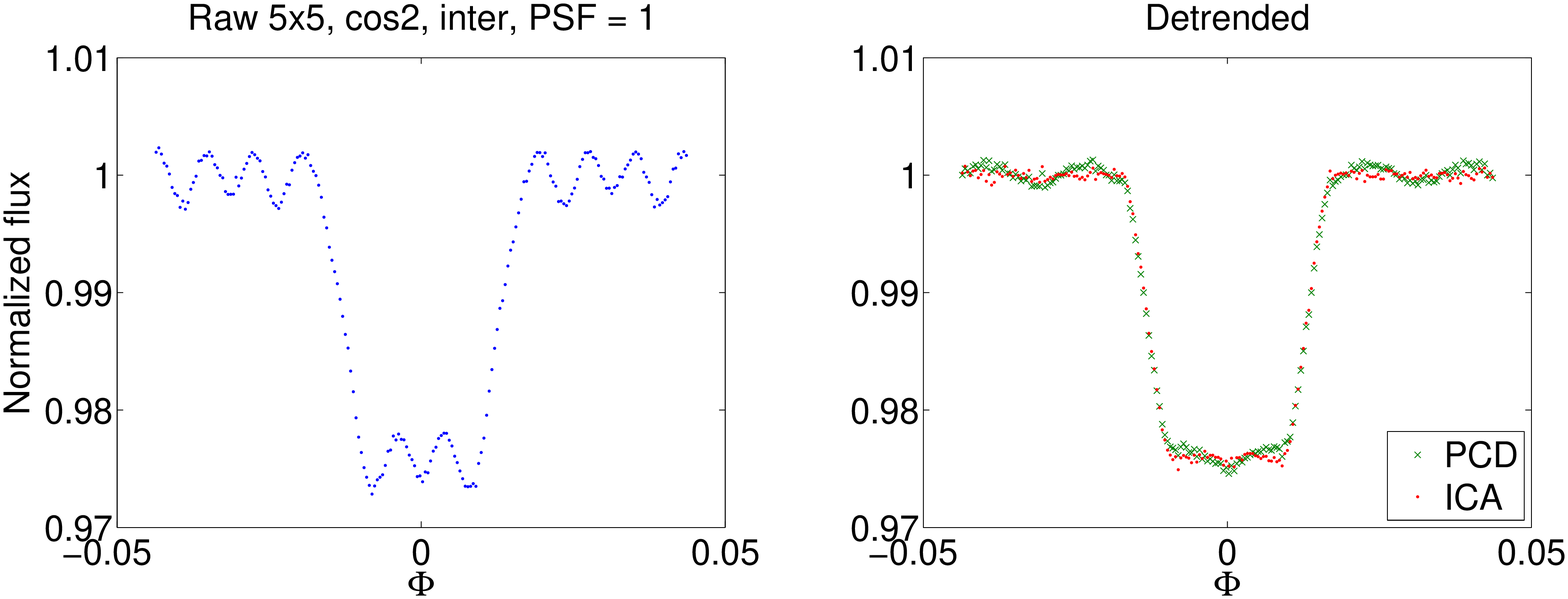}
\plotone{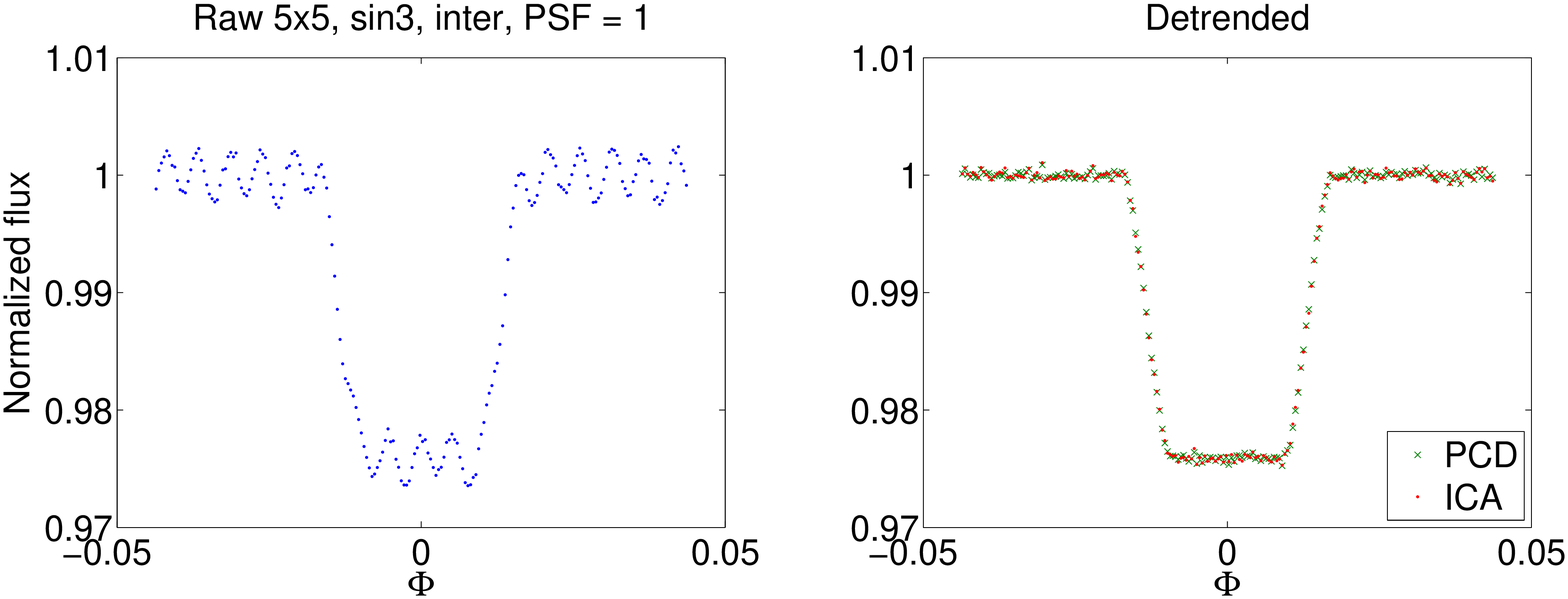}
\plotone{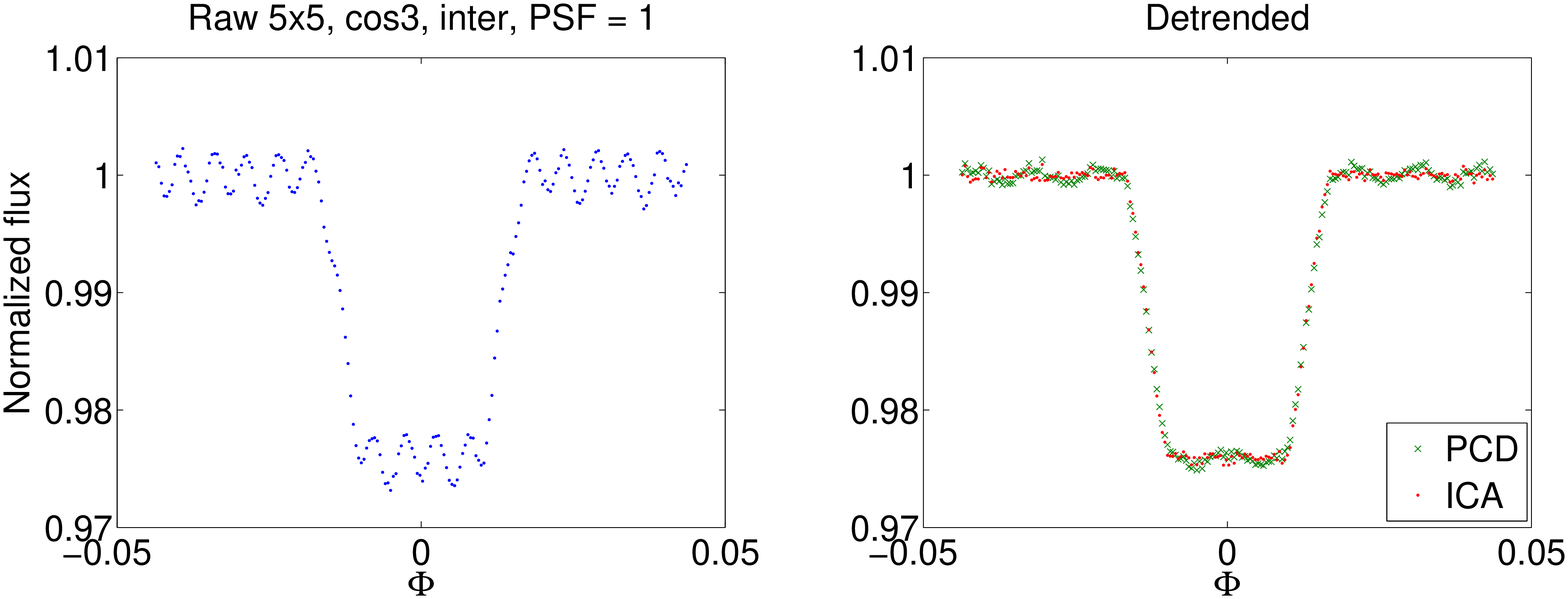}
\plotone{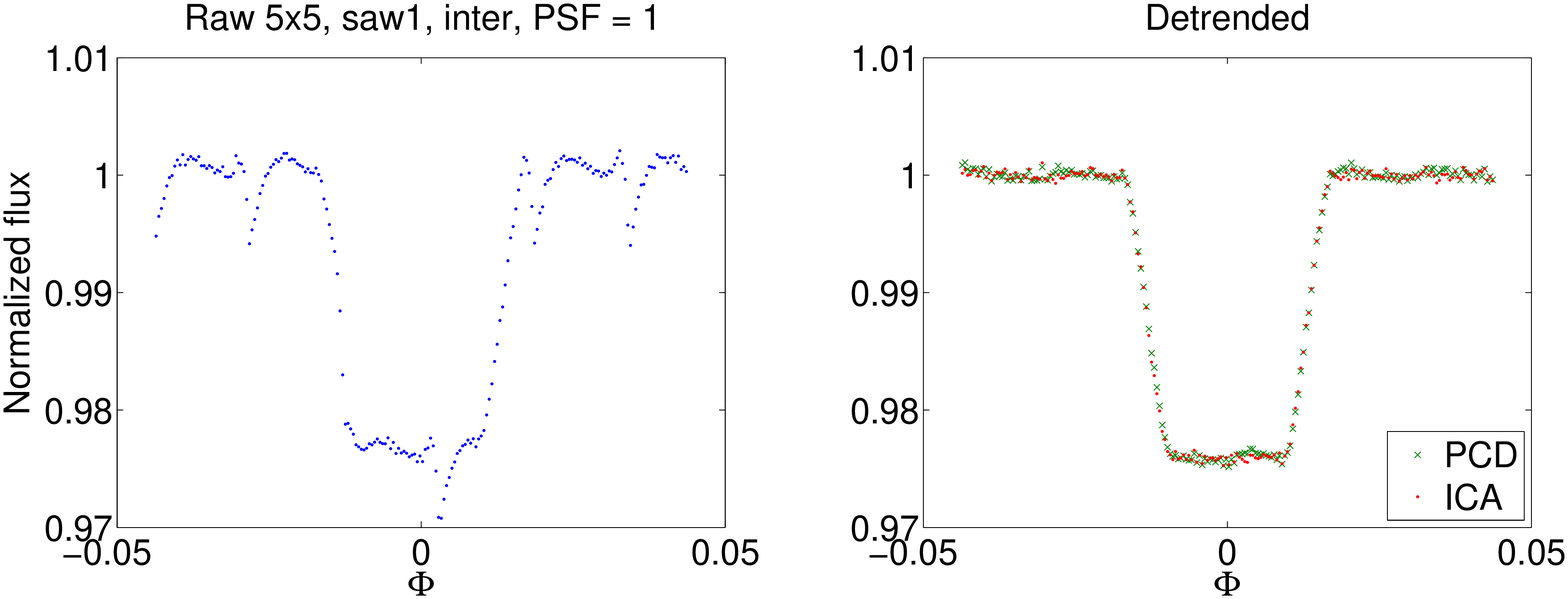}
\plotone{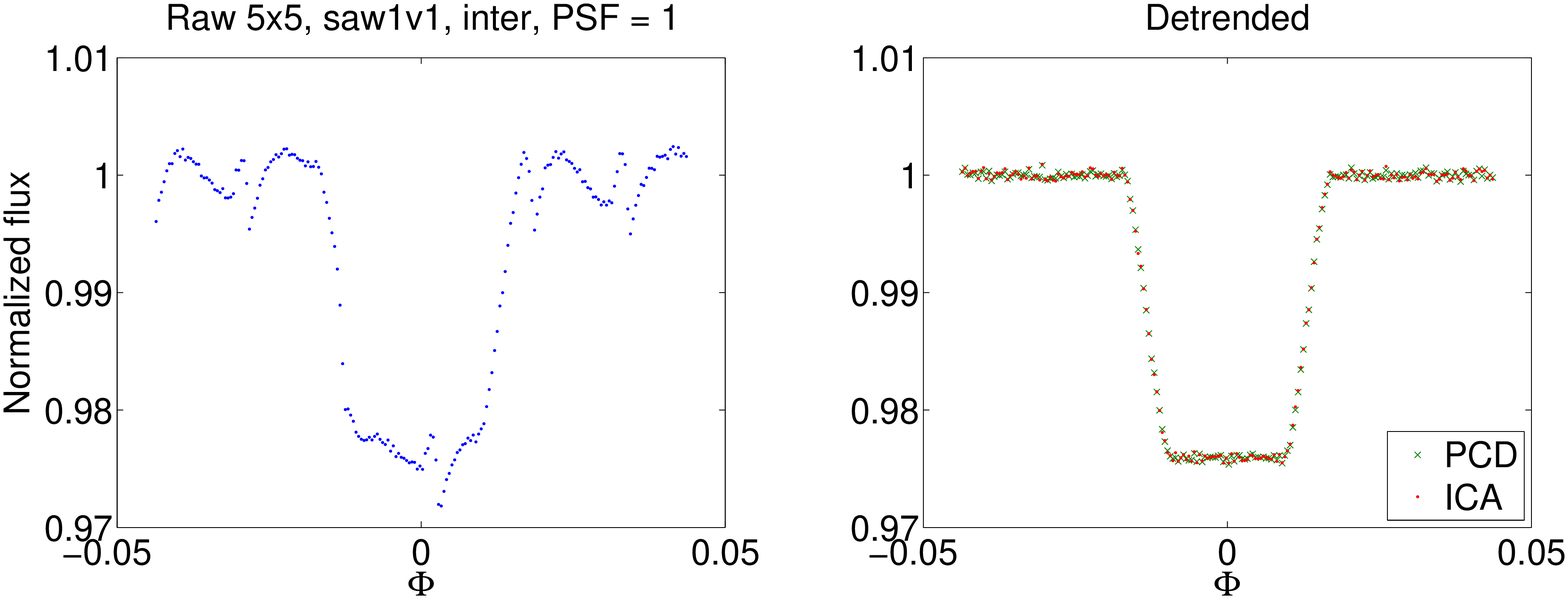}
\plotone{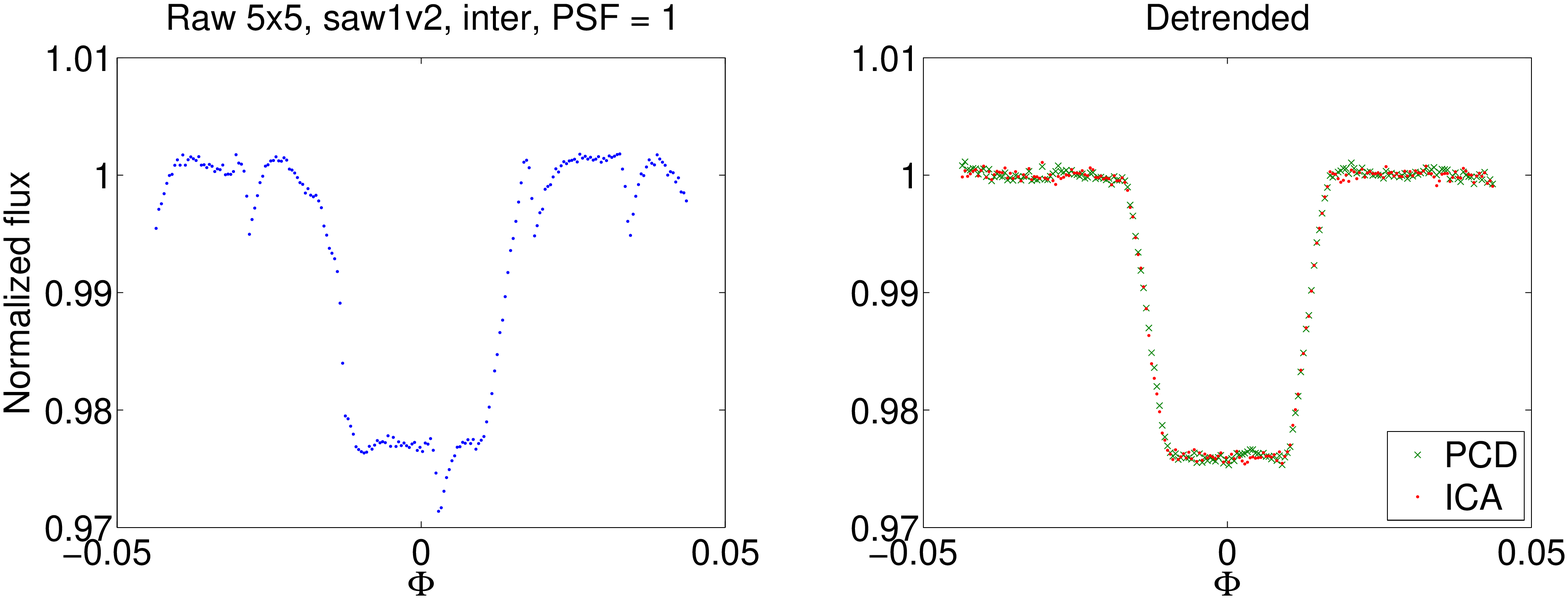}
\plotone{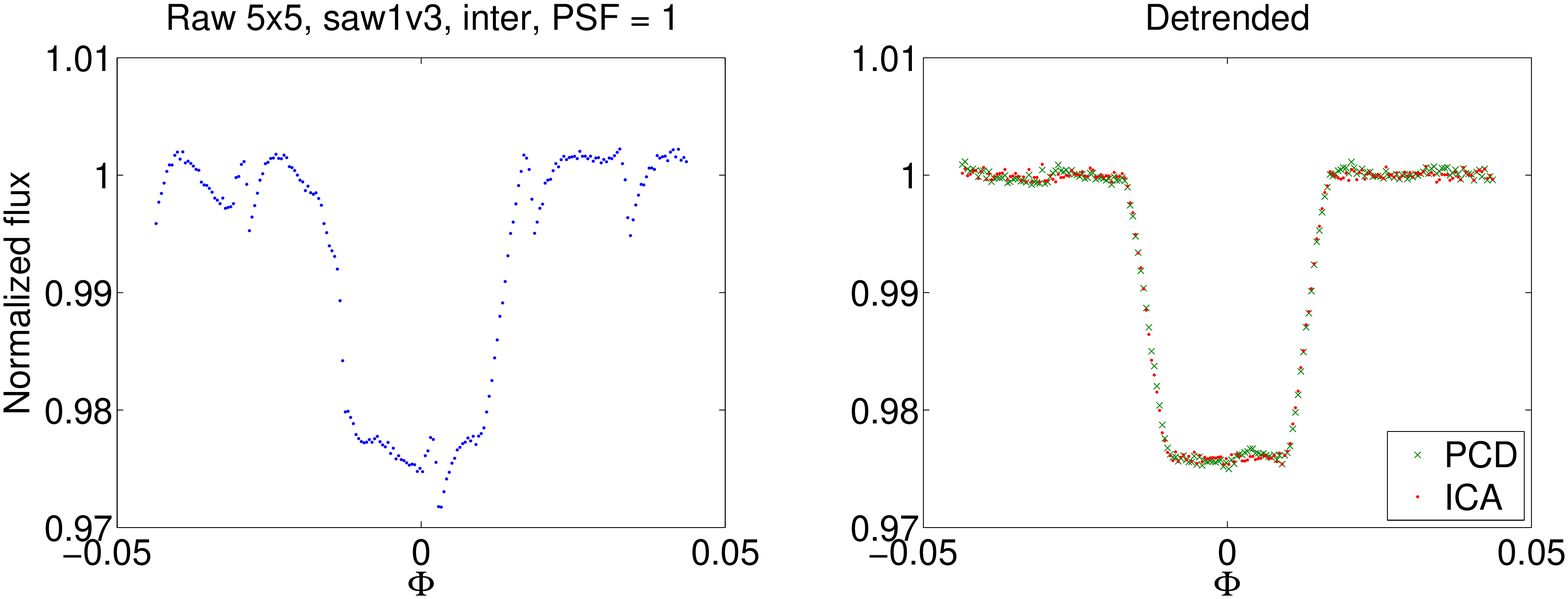}
\plotone{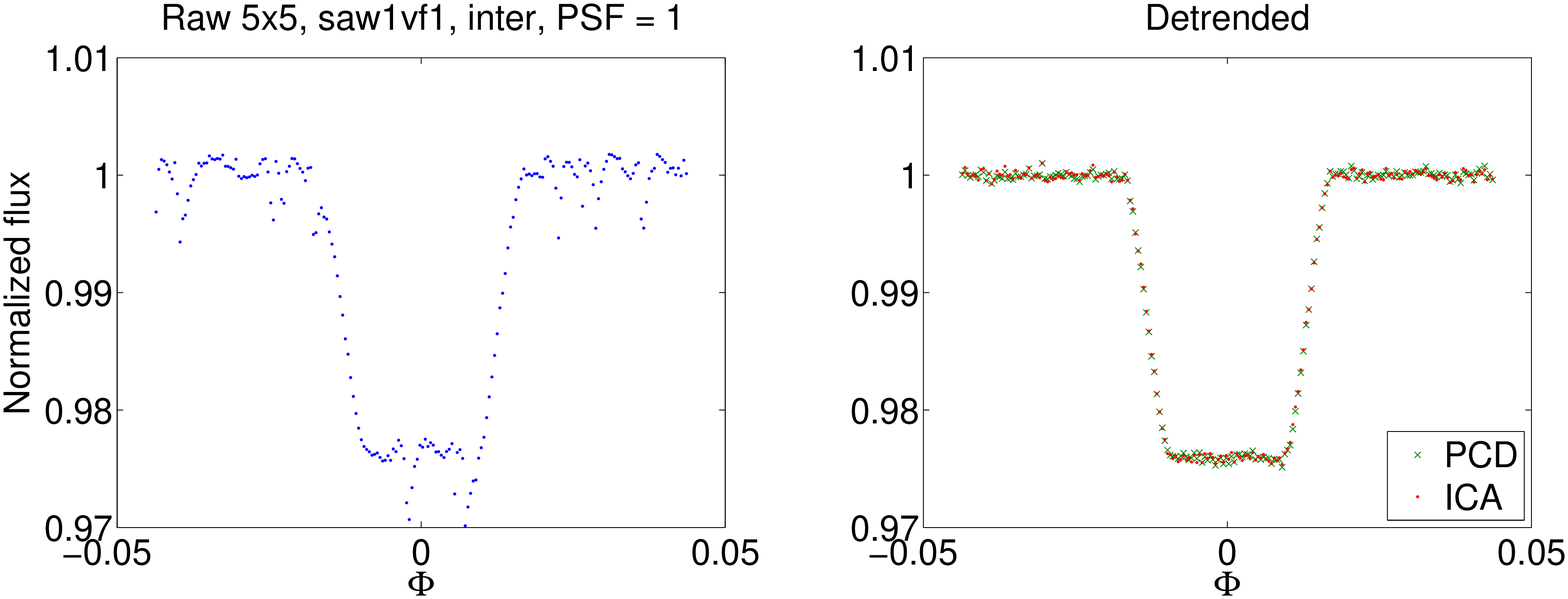}
\plotone{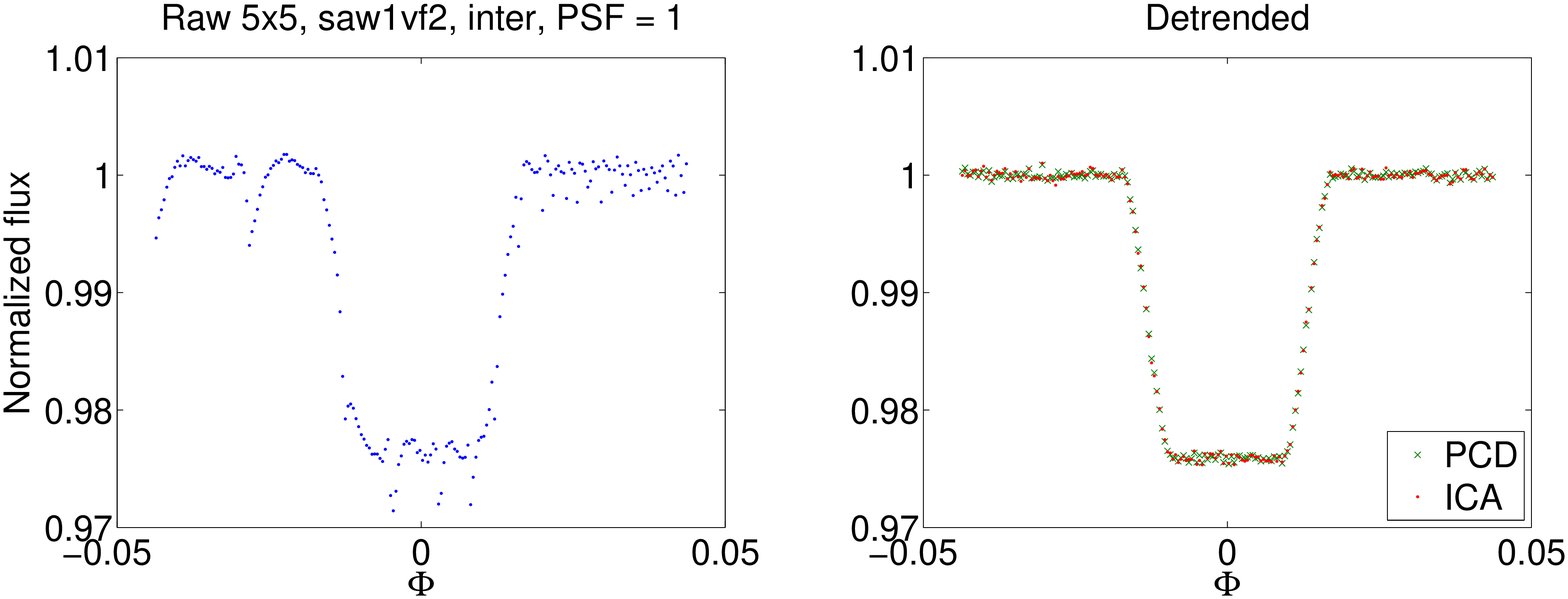}
\plotone{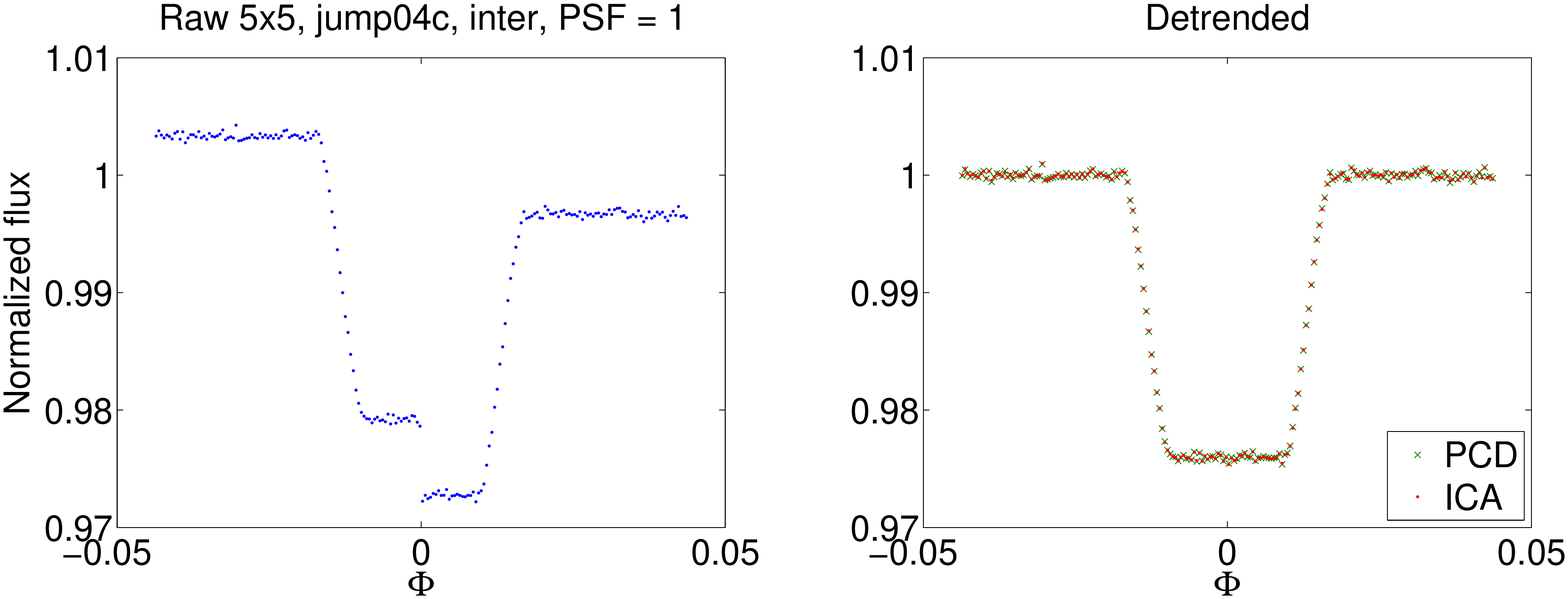}
\caption{Left panels: (blue) raw light-curves simulated with $\sigma_{PSF} = 1 p.u.$, and inter-pixel quantum efficiency variations over 5$\times$5 array of pixels. Right panels: detrended transit light-curves obtained with (green `x') polynomial centroid fitting method, and (red dots) pixel-ICA method. All the light-curves are binned over 10 points to make clearer visualization of the systematic effects. \label{fig7}}
\end{figure*}

\begin{figure*}
\epsscale{1.60}
\plotone{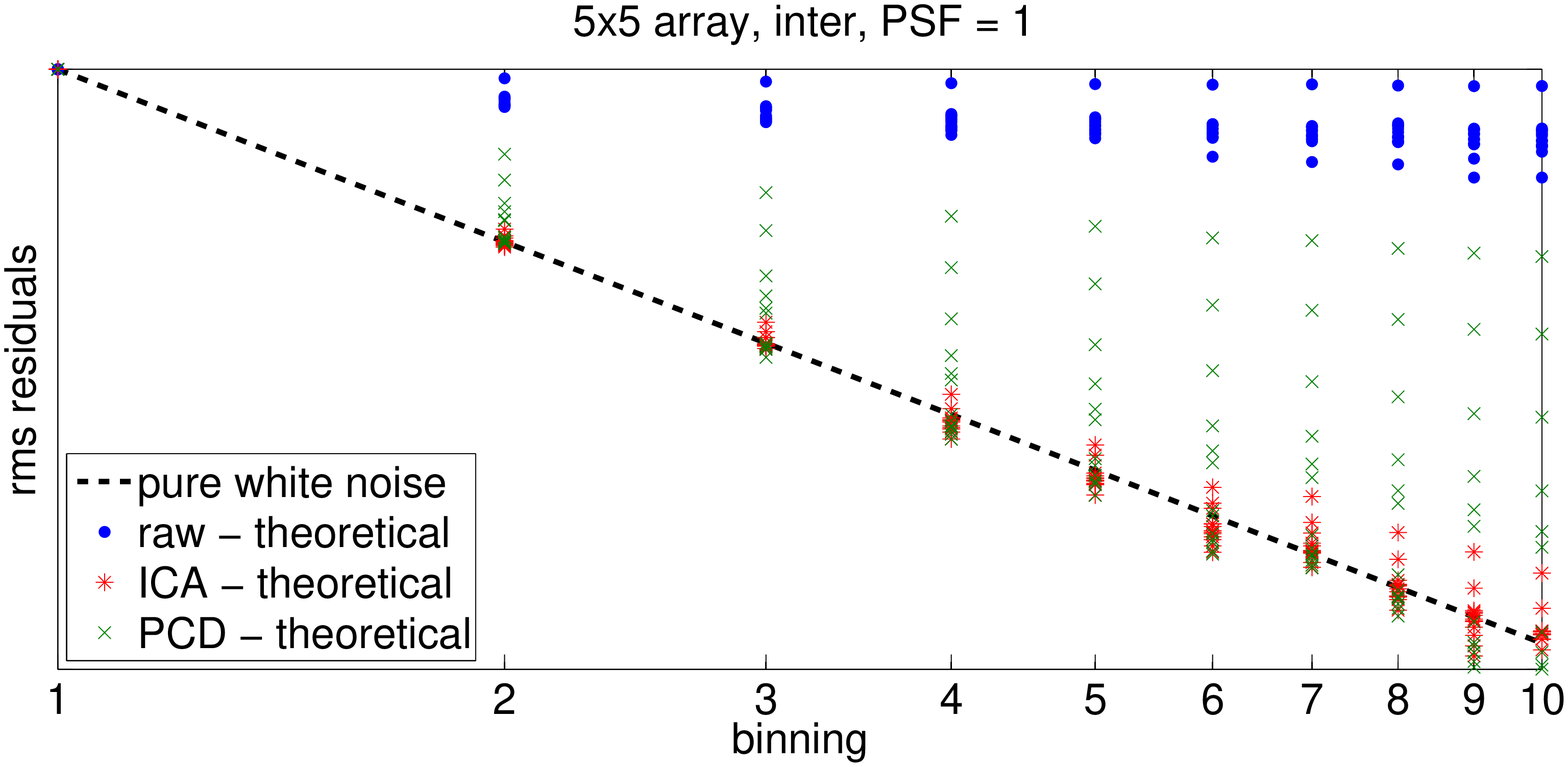}
\caption{Root mean square of residuals for light-curves binned over 1 to 10 points, scaled to their non-binned values. The simulations were obtained with $\sigma_{PSF} =$1 p.u., 5$\times$5 array, and inter-pixel effects. The dashed black line indicates the expected trend for white residuals, blue dots are for normalized raw light-curves, red `$\ast$' are for pixel-ICA detrendend light-curves, and green `x' for PCD detrended light-curves. \label{fig8}}
\end{figure*}

\begin{figure*}
\epsscale{1.60}
\plotone{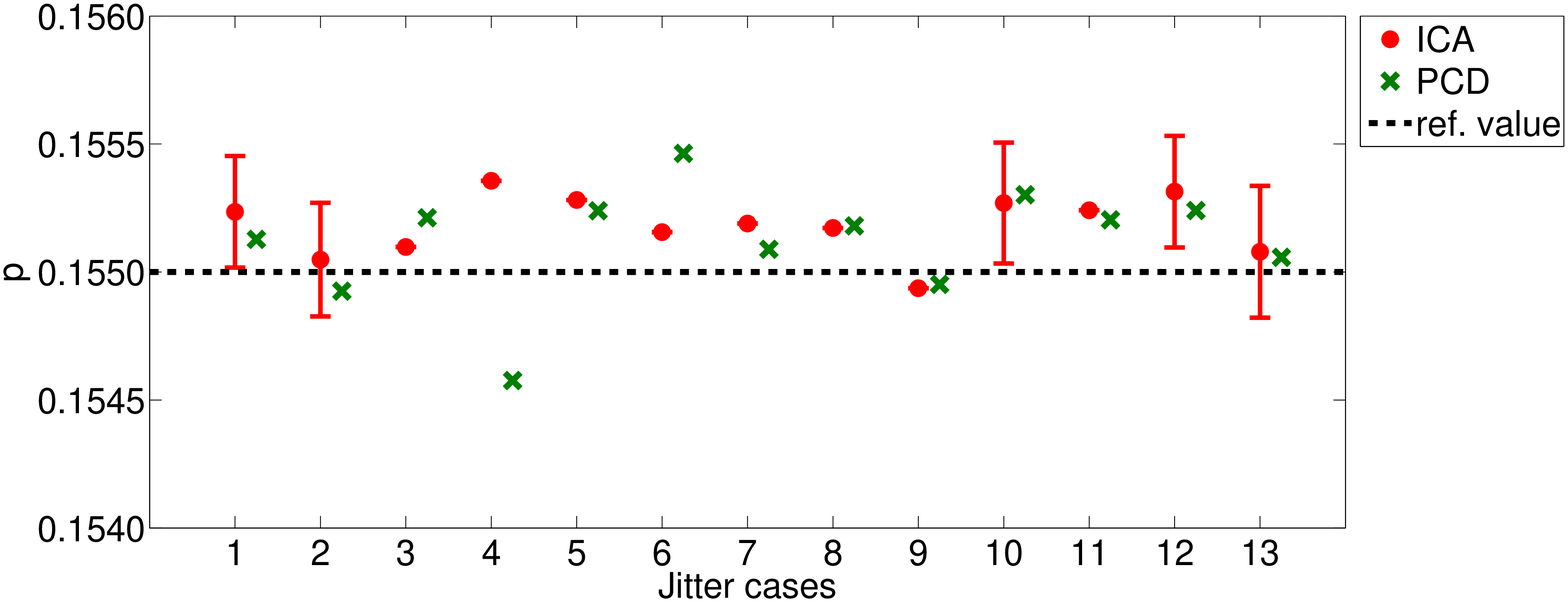}
\plotone{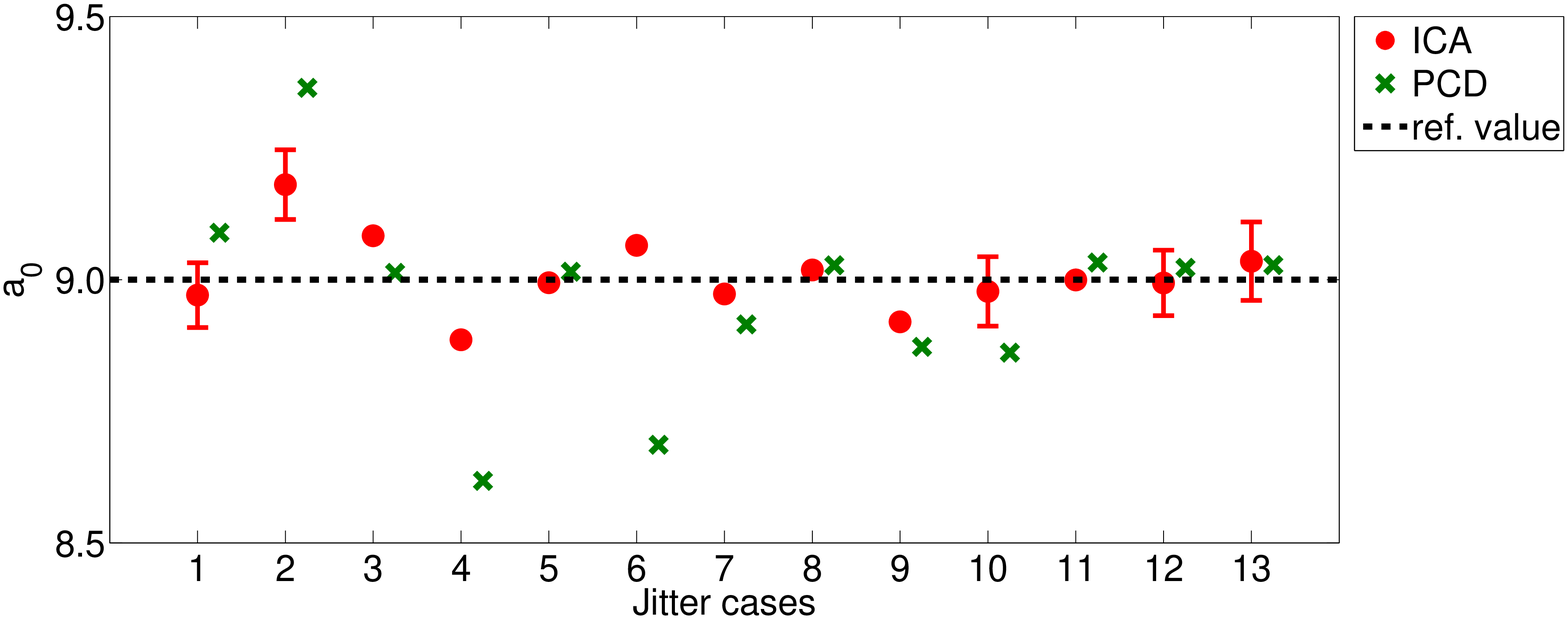}
\plotone{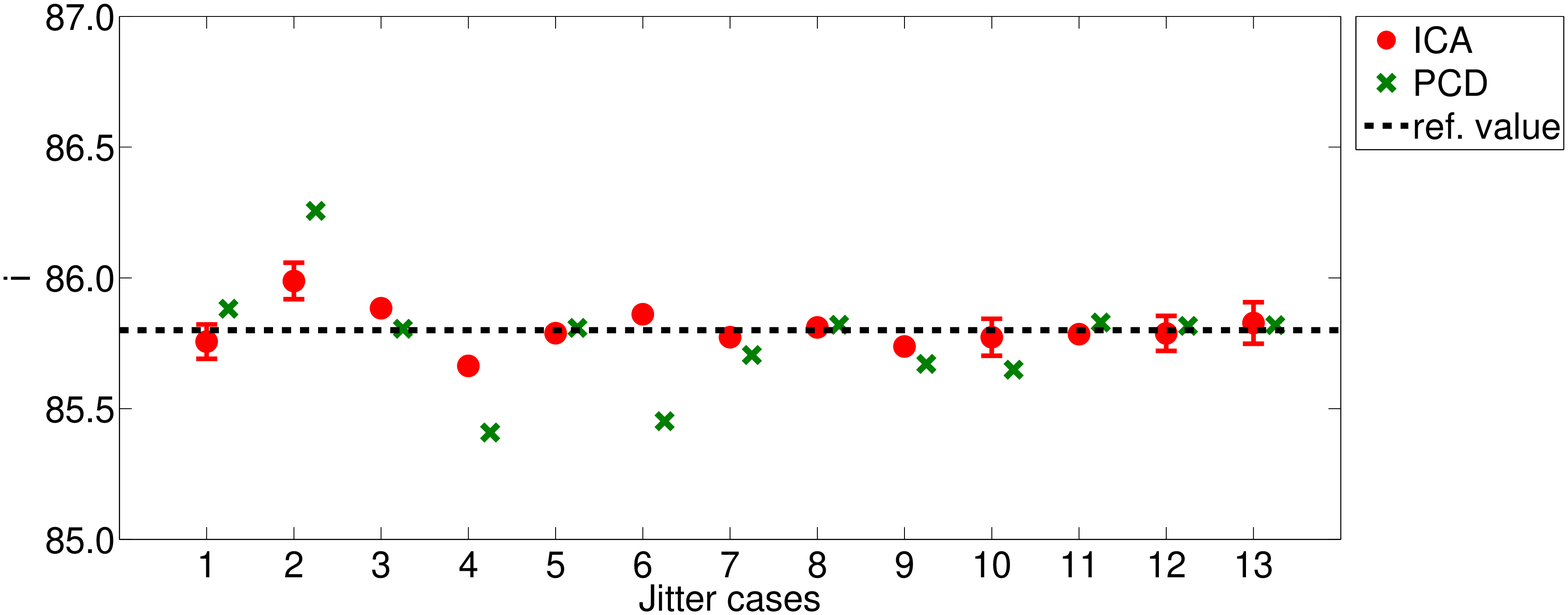}
\caption{Top panel: best estimates of the planet-to-star radii ratio, $p = r_p/R_s$, for detrended light-curves with (red dots) pixel-ICA, and (green `x') PCD method ($\sigma_{PSF} = $1 p.u., inter-pixel effects over 5$\times$5 array). Error bars are reported for representative cases of jitter signal, i.e. sin1, cos1, saw1v3, saw1vf2, and jump04c. Middle panel: the same for the orbital semimajor axis in units of the stellar radius, $a_0 = a/R_s$. Bottom panel: the same for the orbital inclination, $i$. \label{fig9}}
\end{figure*}

For the same configuration, i.e. $\sigma_{PSF} =$1 p.u., inter-pixel effects, we investigated the consequences of considering a smaller array (5$\times$5), which does not include the whole PSF. Fig. \ref{fig7} shows the raw light-curves, and the correspondent detrended ones, obtained with the two methods. Tab. \ref{tab4} reports the discrepancies between those light-curves and the theoretical model. The discrepancies are higher than for the larger pixel-array by a factor $\gtrsim$ 2 (for the raw light-curves), because of the additional effect. After pixel-ICA detrending, the discrepancies are reduced by a factor $\sim$5 (for the selected binning) in most cases, and $\sim$13 for the `jump04c', while the performances of the parametric method are case dependent, and discrepancies are reduced by a factor between 2 and 7 in most cases, and also $\sim$13 for `jump04c'. In all cases, the final discrepancies are higher than the HFPN level, i.e. $\sim$1.7$\times$10$^{-4}$ for the 5$\times$5 array. Fig. \ref{fig8} shows how the residuals scale for binning over $n$ points, with $1 \le n \le 10$. The temporal structure due to jitter effect is dominant in raw data, but little traces of this behaviour are present after pixel-ICA detrending. Even for this aspect, the performances of the parametric method are case dependent. Fig. \ref{fig9} shows the transit parameters retrieved from detrended light-curves; in a few representative cases, we calculated the error bars. Numerical results are reported in Tab. \ref{tab8}. In conclusion, the choice of a non-optimal pixel array introduces additional systematics, that worsen the parameter retrieval, but it is quite remarkable that the pixel-ICA technique gives consistent results in most cases, whereas the parametric technique appears to be less robust.
\begin{table*}
\begin{center}
\caption{Root mean square of residuals between the light-curves and the theoretical model for simulations with $\sigma_{PSF} = 1$ p.u., and inter-pixel quantum efficiency variations over 5$\times$5 array of pixels; in particular they are calculated for the raw light-curves, light-curves detrended with pixel-ICA, and PCD method, binned over 10 points. \label{tab4}}
\begin{tabular}{cccc}
\tableline\tableline
Jitter & rms (raw $-$ theoretical) & rms (ICA $-$ theoretical) & rms (PCD $-$ theoretical)\\
\tableline
sin1 & 1.5$\times$10$^{-3}$ & 3.1$\times$10$^{-4}$ & 2.6$\times$10$^{-4}$\\
cos1 & 1.4$\times$10$^{-3}$ & 3.4$\times$10$^{-4}$ & 8.1$\times$10$^{-4}$\\
sin2 & 1.5$\times$10$^{-3}$ & 3.2$\times$10$^{-4}$ & 2.8$\times$10$^{-4}$\\
cos2 & 1.5$\times$10$^{-3}$ & 3.2$\times$10$^{-4}$ & 6.2$\times$10$^{-4}$\\
sin3 & 1.5$\times$10$^{-3}$ & 3.1$\times$10$^{-4}$ & 2.7$\times$10$^{-4}$\\
cos3 & 1.4$\times$10$^{-3}$ & 3.2$\times$10$^{-4}$ & 4.9$\times$10$^{-4}$\\
saw1 & 1.7$\times$10$^{-3}$ & 3.1$\times$10$^{-4}$ & 3.5$\times$10$^{-4}$\\
saw1v1 & 1.7$\times$10$^{-3}$ & 2.9$\times$10$^{-4}$ & 2.5$\times$10$^{-4}$\\
saw1v2 & 1.6$\times$10$^{-3}$ & 3.6$\times$10$^{-4}$ & 3.7$\times$10$^{-4}$\\
saw1v3 & 1.7$\times$10$^{-3}$ & 3.1$\times$10$^{-4}$ & 4.0$\times$10$^{-4}$\\
saw1vf1 & 1.7$\times$10$^{-3}$ & 3.1$\times$10$^{-4}$ & 2.9$\times$10$^{-4}$\\
saw1vf2 & 1.6$\times$10$^{-3}$ & 3.2$\times$10$^{-4}$ & 2.6$\times$10$^{-4}$\\
jump04c & 3.3$\times$10$^{-3}$ & 2.6$\times$10$^{-4}$ & 2.6$\times$10$^{-4}$\\
\tableline
\end{tabular}
\end{center}
\end{table*}

\clearpage

\subsection{Case II: inter-pixel effects, narrow PSF}
\label{sec:interPSF02}

\begin{figure*}
\epsscale{0.94}
\plotone{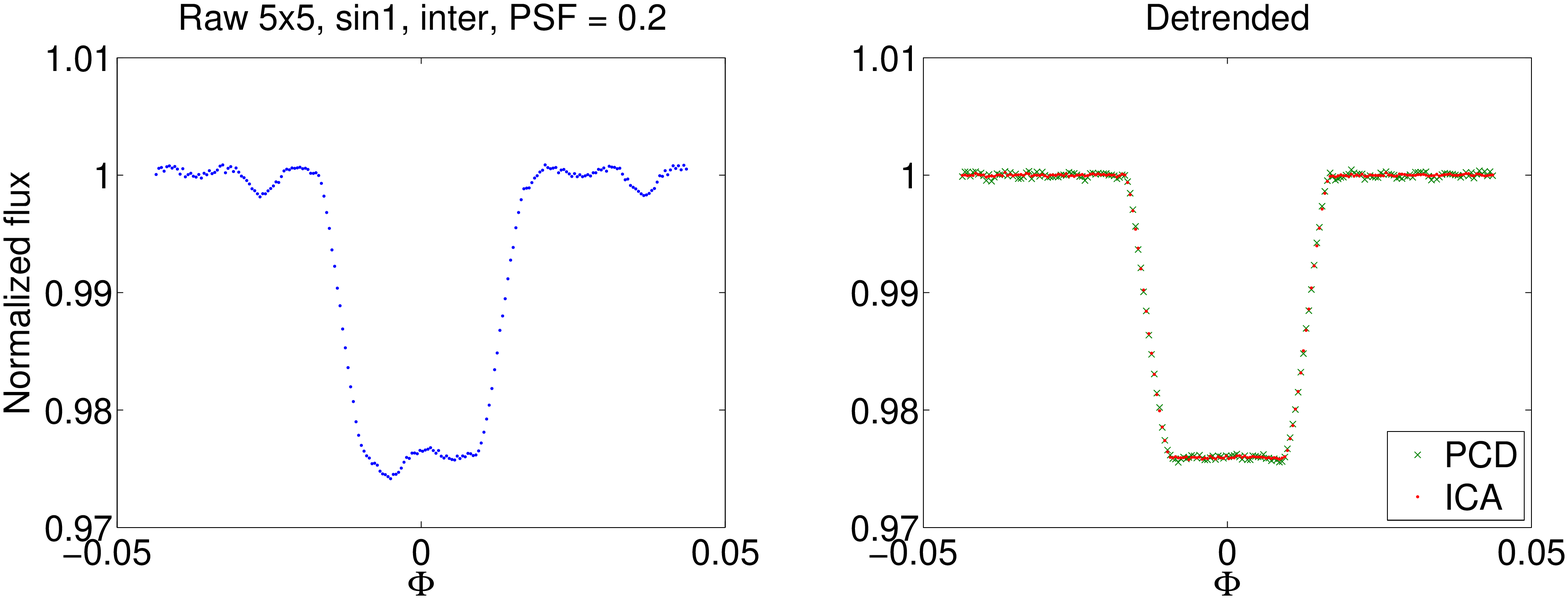}
\plotone{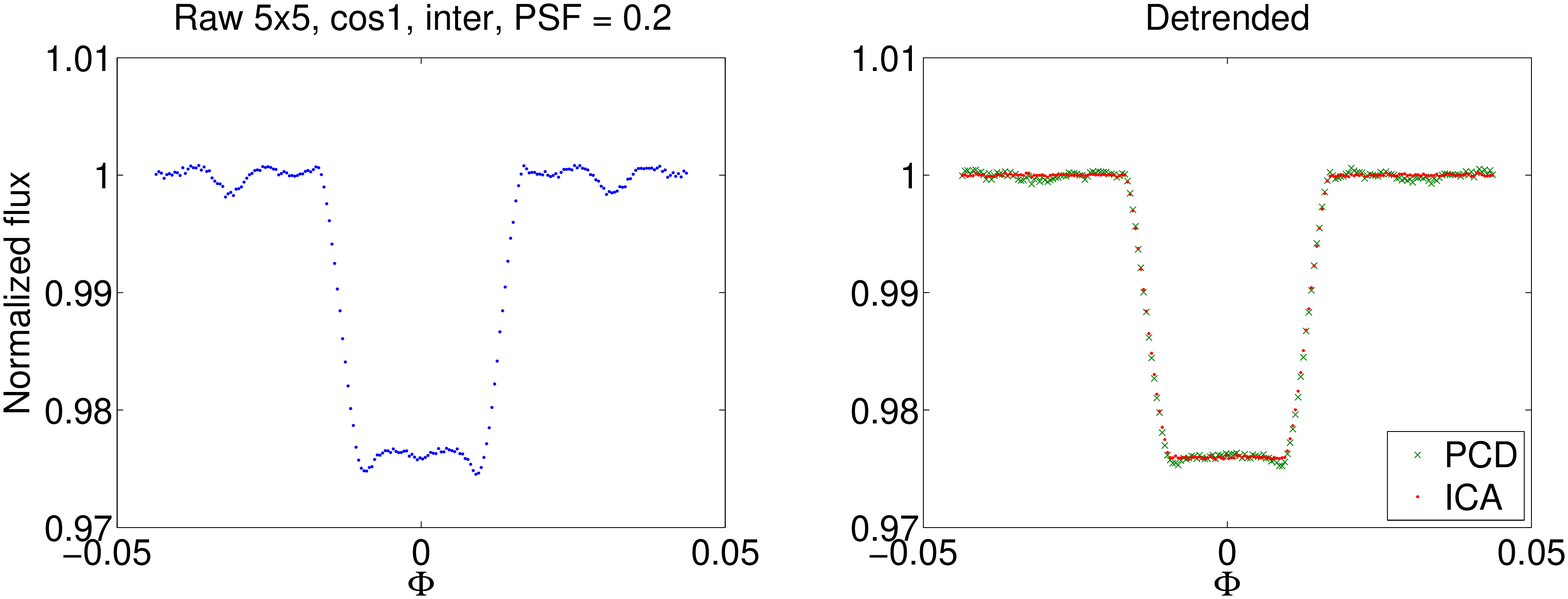}
\plotone{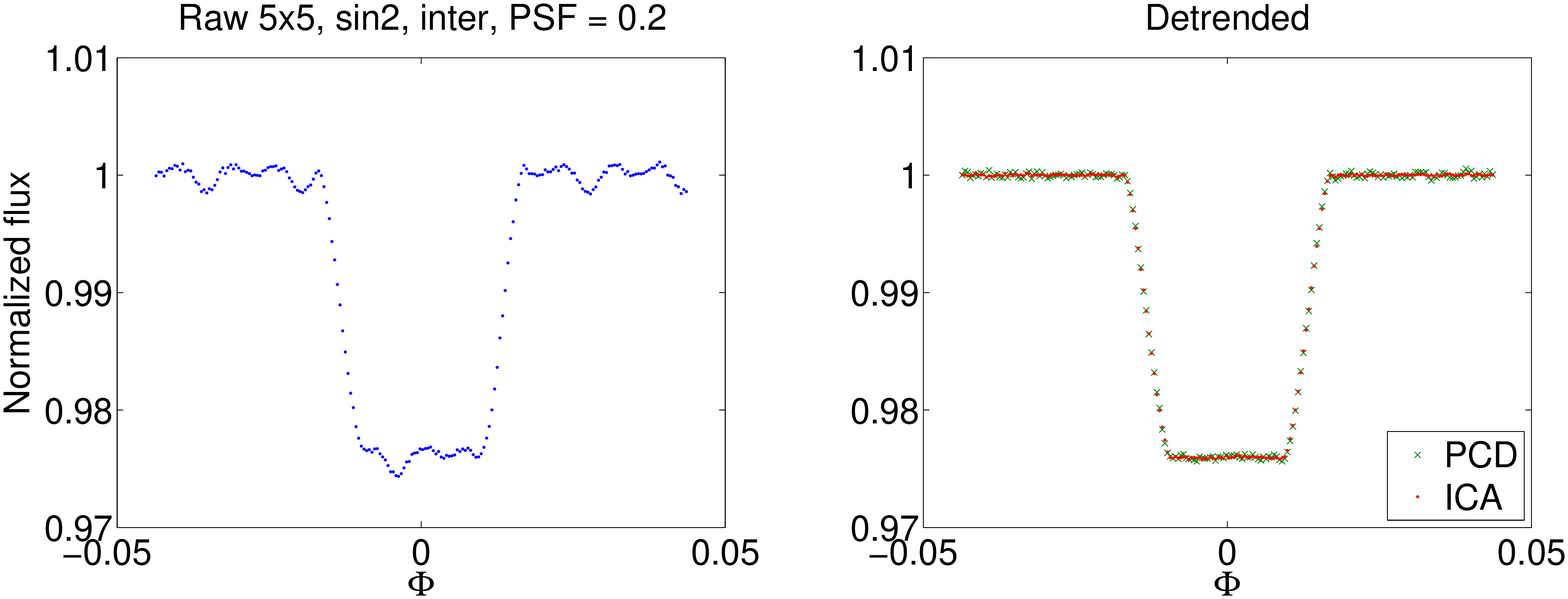}
\plotone{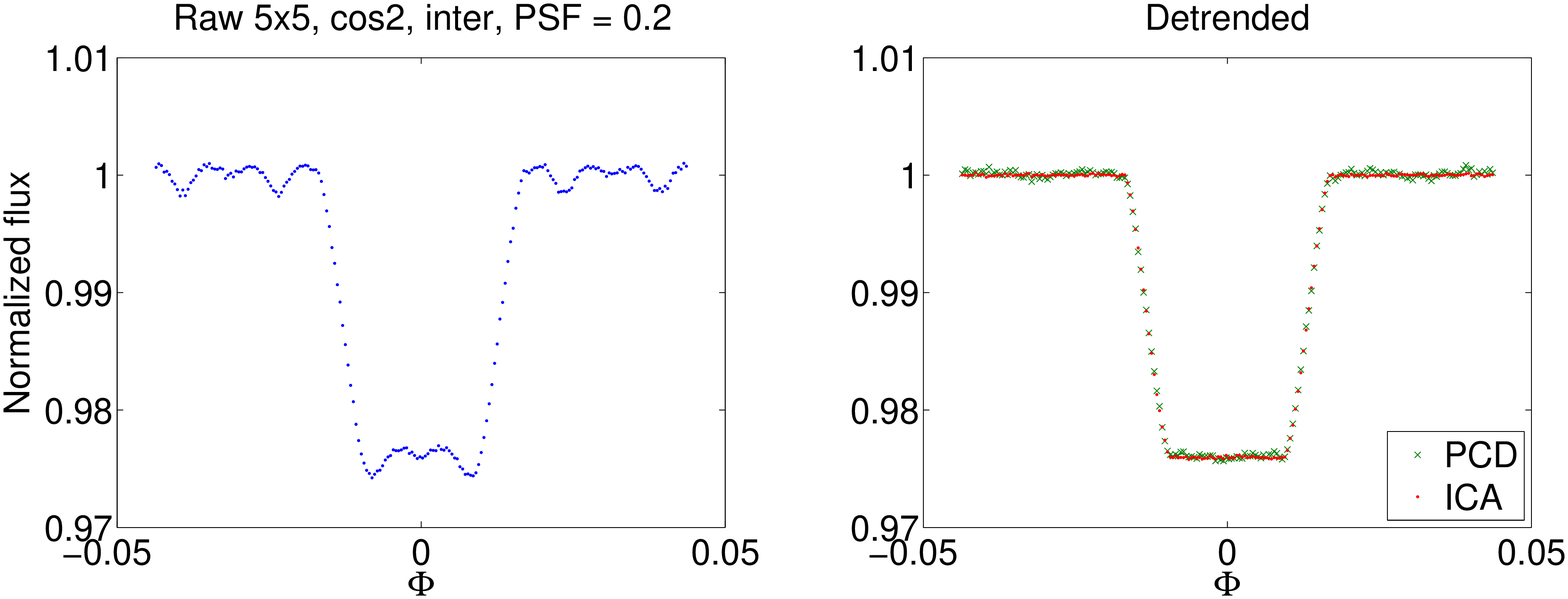}
\plotone{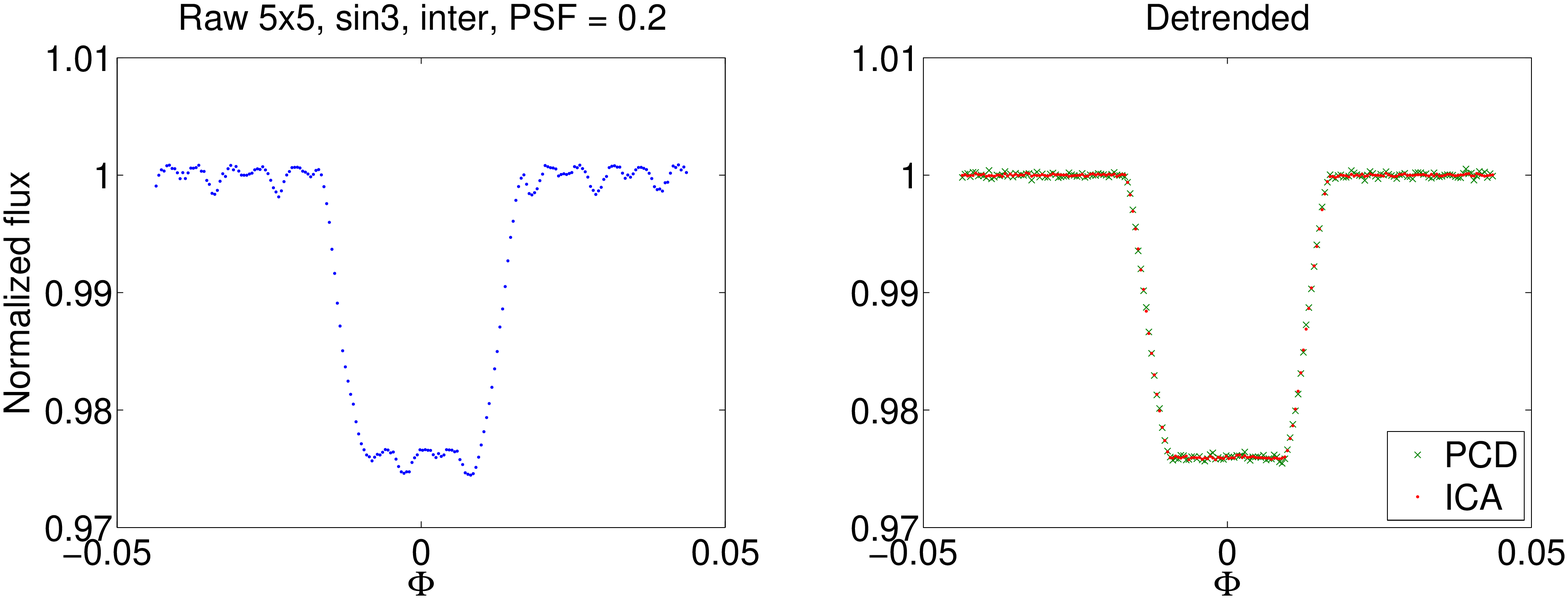}
\plotone{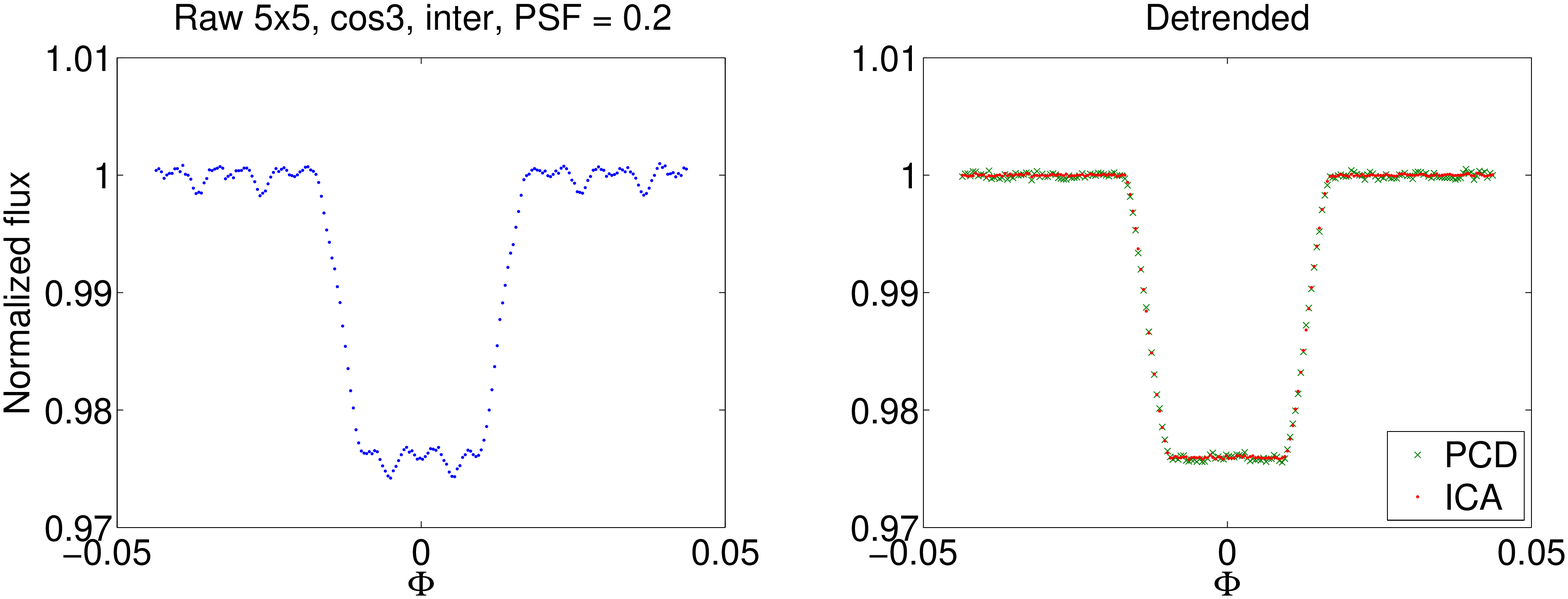}
\plotone{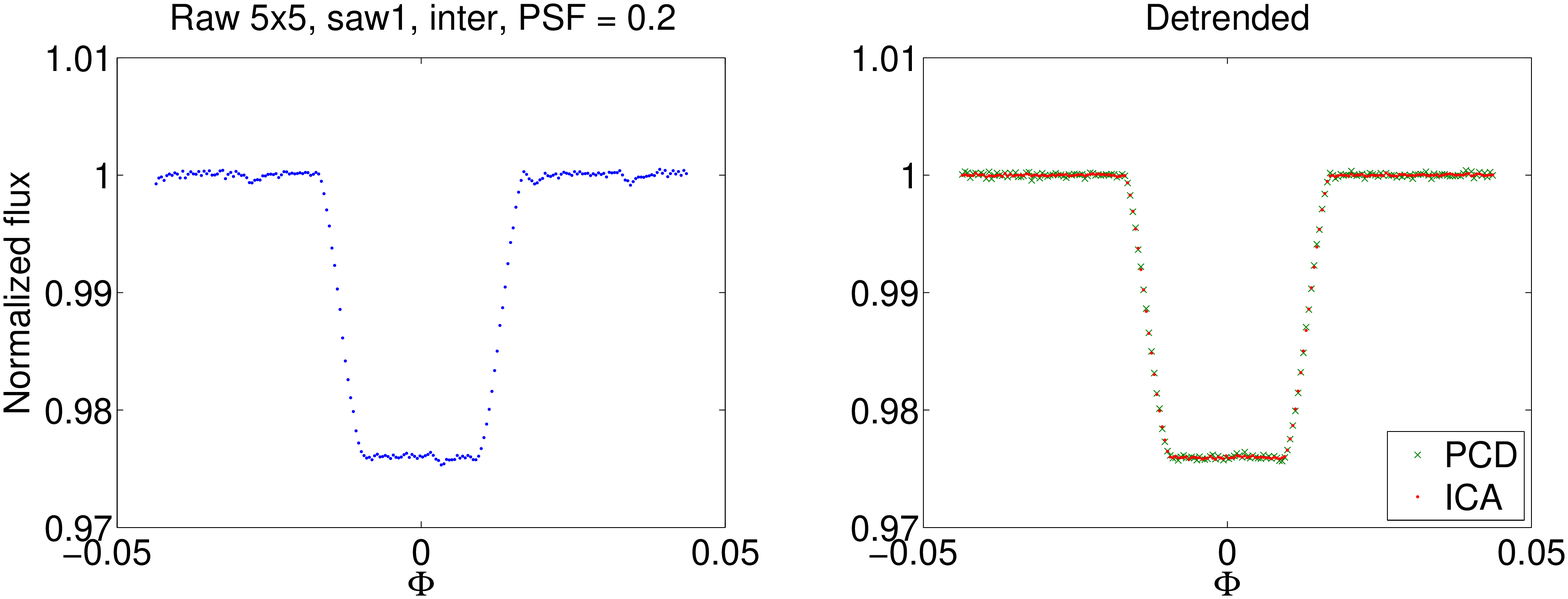}
\plotone{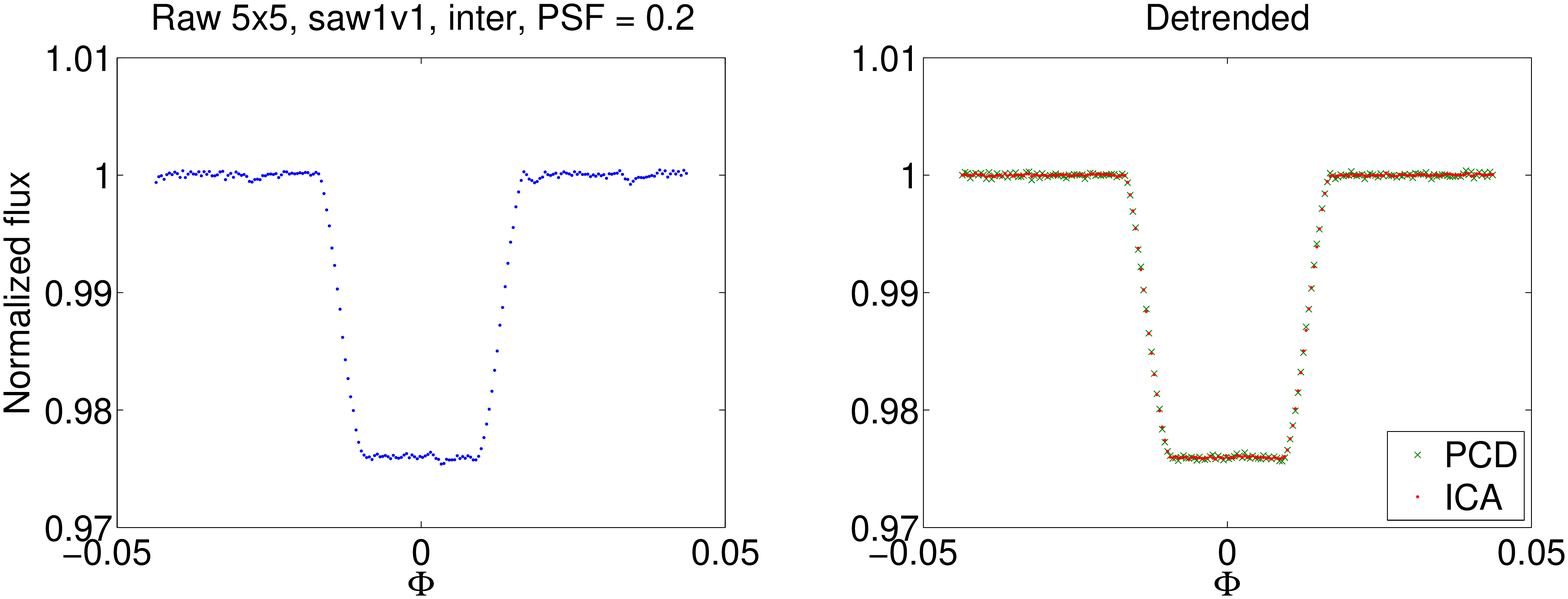}
\plotone{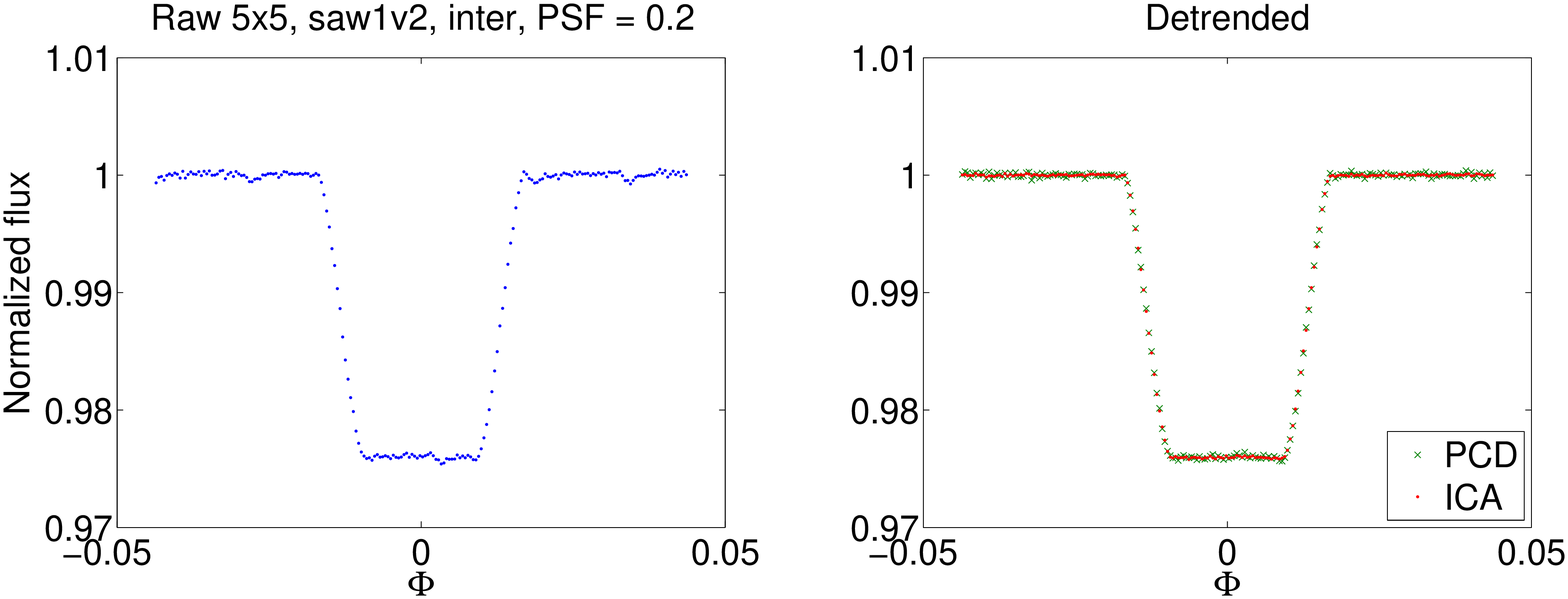}
\plotone{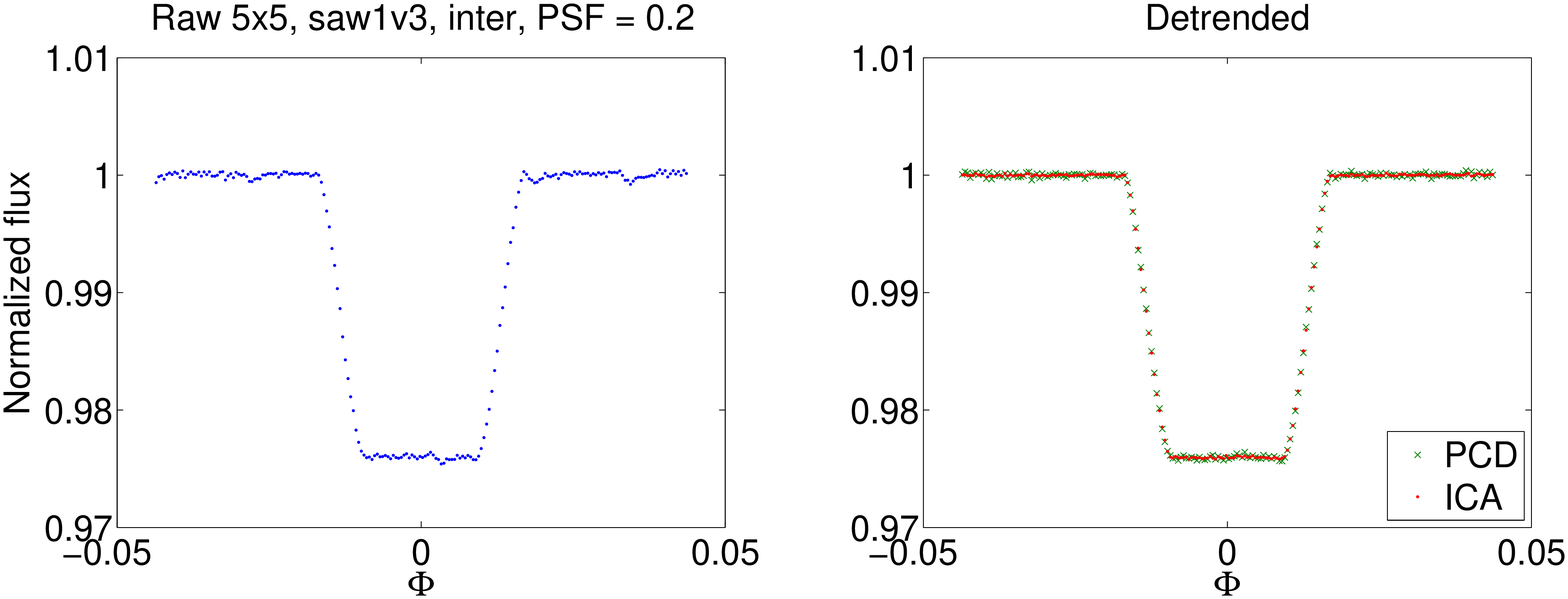}
\plotone{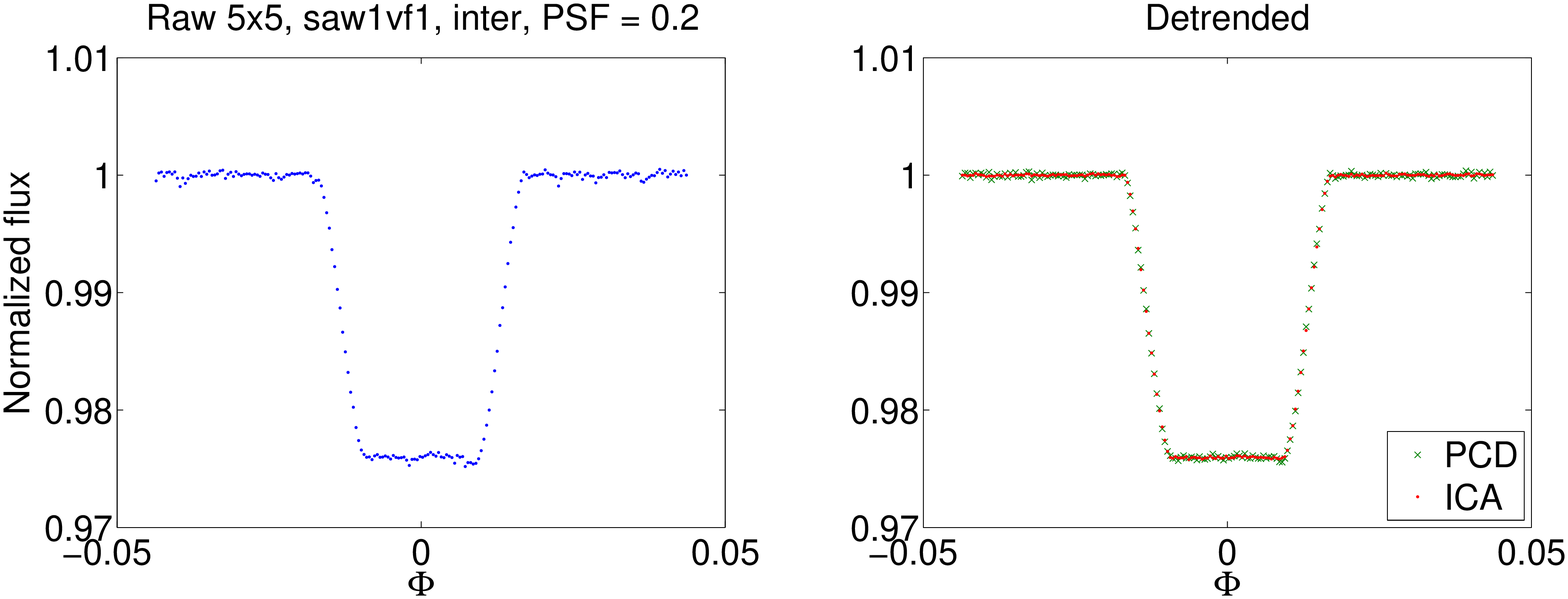}
\plotone{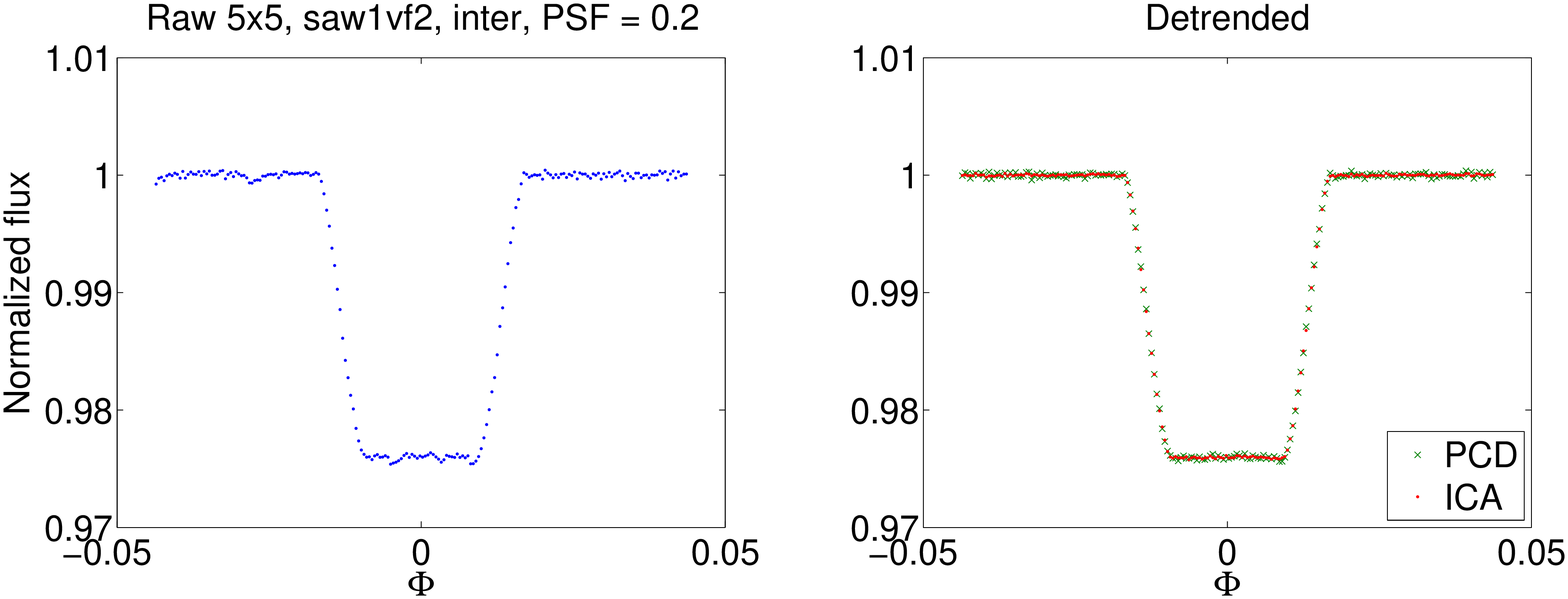}
\plotone{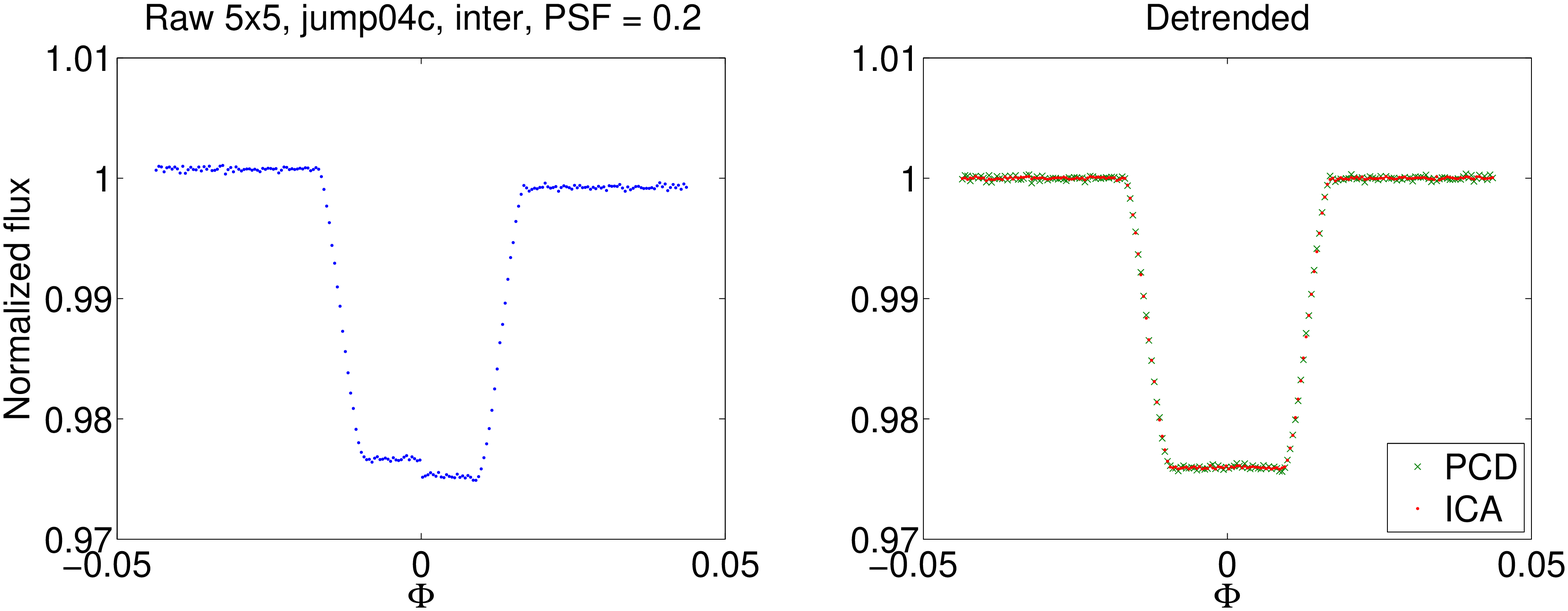}
\caption{Left panels: (blue) raw light-curves simulated with $\sigma_{PSF} = $0.2 p.u., and inter-pixel quantum efficiency variations over 5$\times$5 array of pixels. Right panels: detrended transit light-curves obtained with (green `x') polynomial centroid fitting method, and (red dots) pixel-ICA method. All the light-curves are binned over 10 points to make clearer visualization of the systematic effects. \label{fig10}}
\end{figure*}

\begin{figure*}
\epsscale{1.60}
\plotone{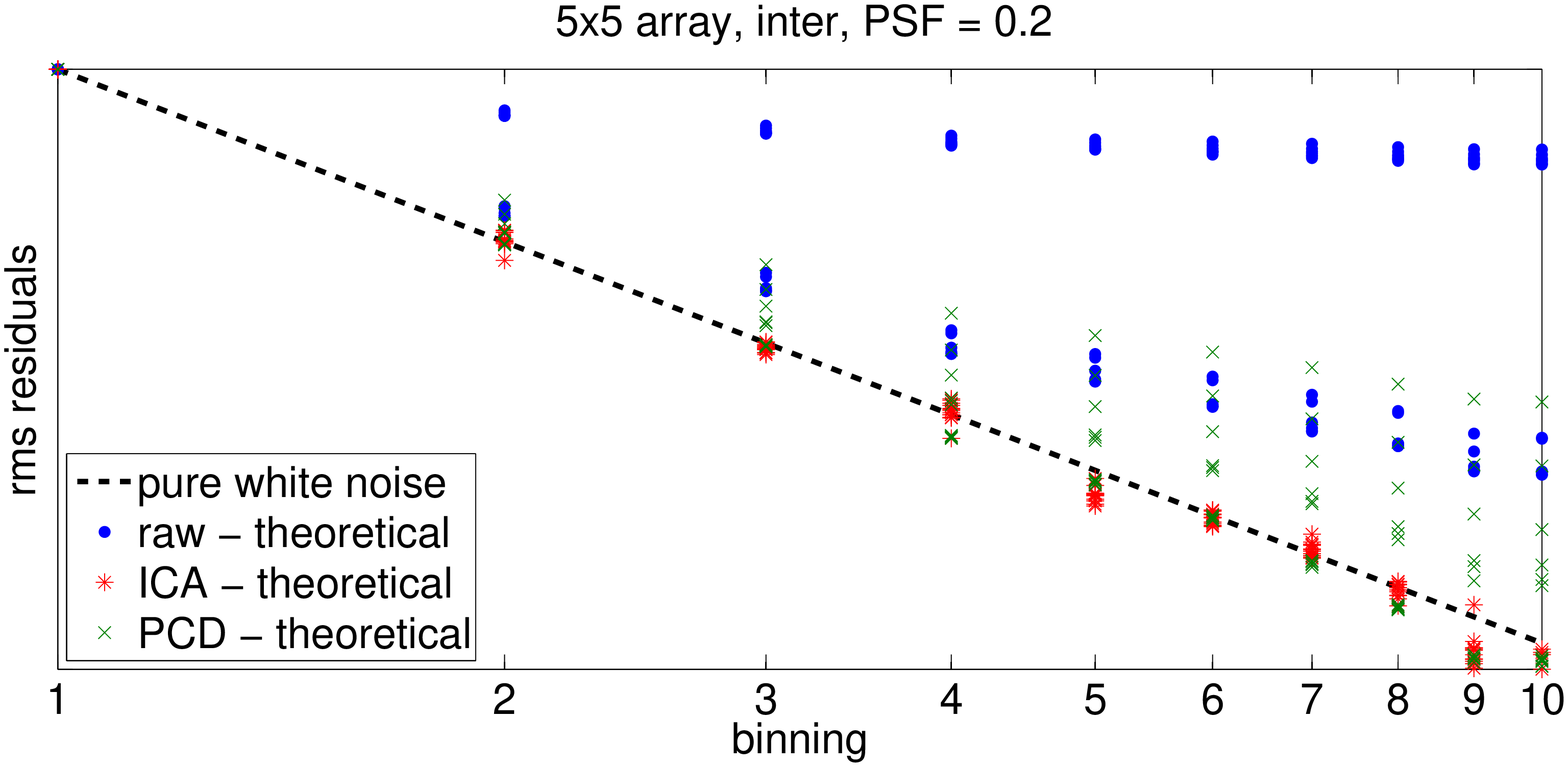}
\caption{Root mean square of residuals for light-curves binned over 1 to 10 points, scaled to their non-binned values. The simulations were obtained with $\sigma_{PSF} =$0.2 p.u., 5$\times$5 array, and inter-pixel effects. The dashed black line indicates the expected trend for white residuals, blue dots are for normalized raw light-curves, red `$\ast$' are for pixel-ICA detrendend light-curves, and green `x' for PCD detrended light-curves. \label{fig11}}
\end{figure*}

\begin{figure*}
\epsscale{1.60}
\plotone{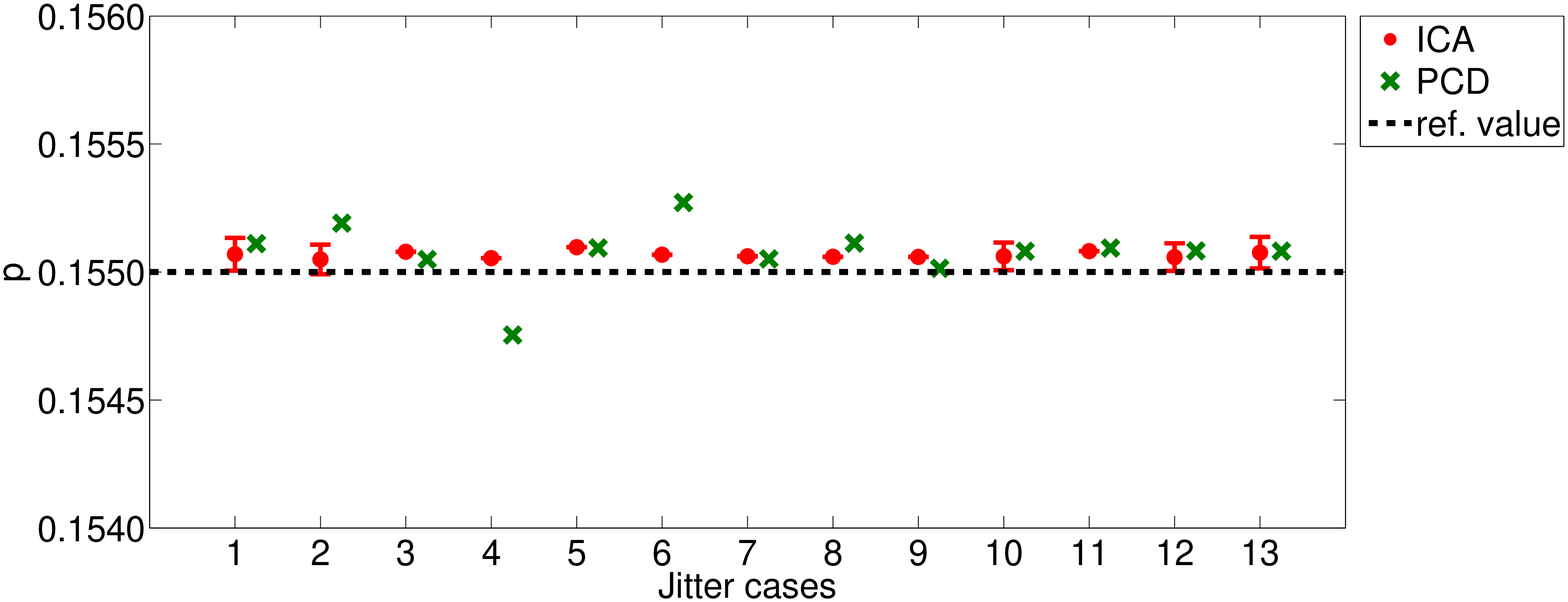}
\plotone{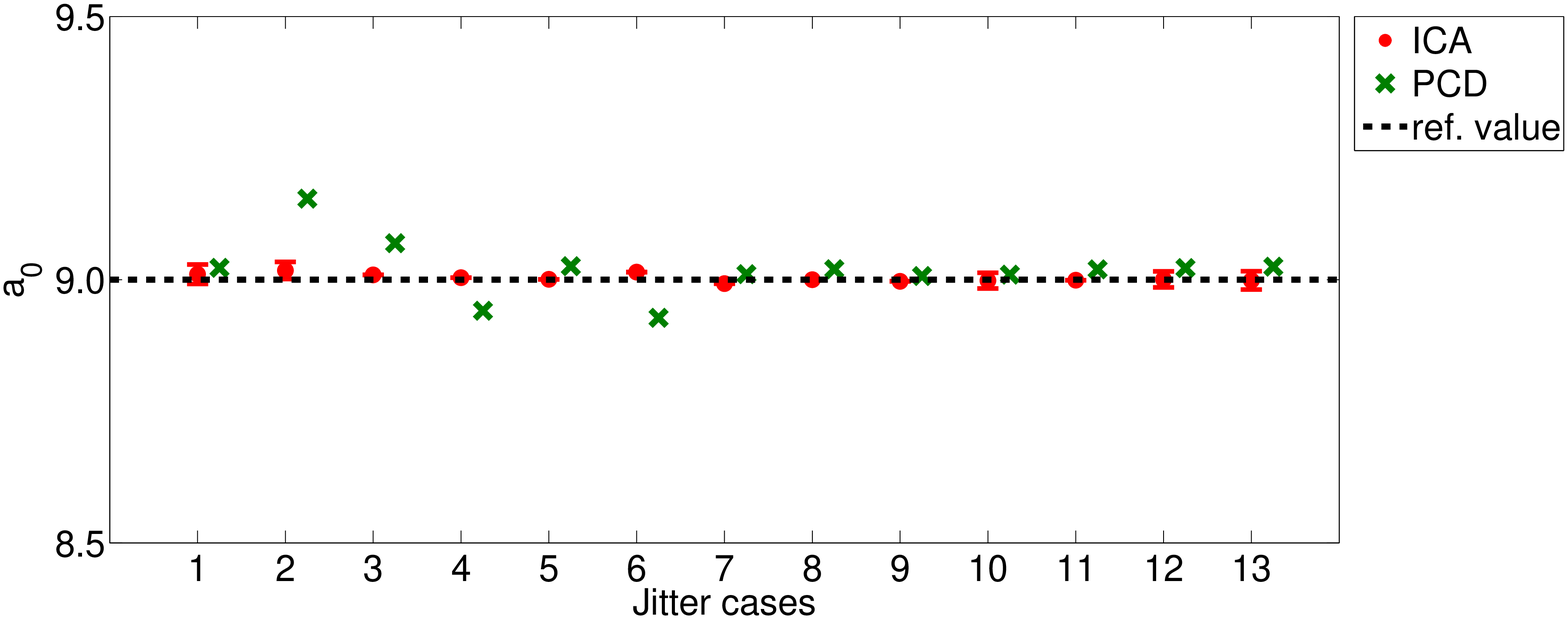}
\plotone{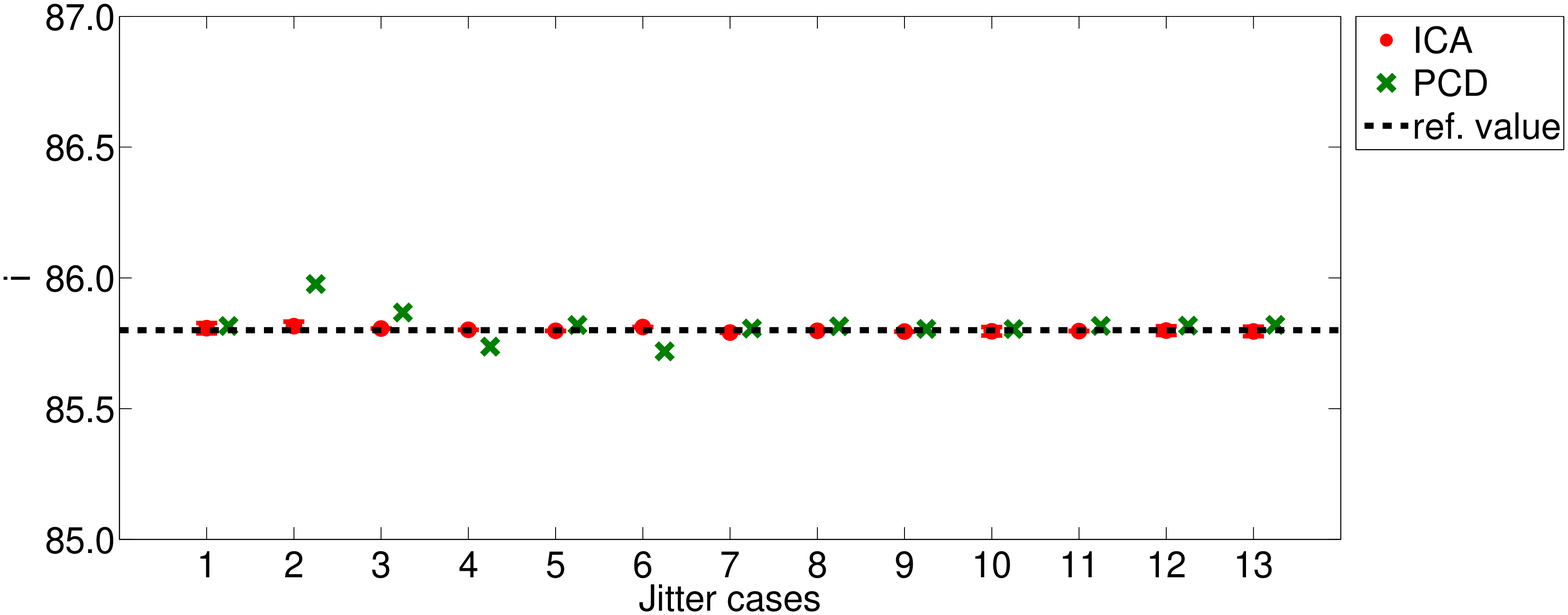}
\caption{Top panel: best estimates of the planet-to-star radii ratio, $p = r_p/R_s$, for detrended light-curves with (red dots) pixel-ICA, and (green `x') PCD method ($\sigma_{PSF} = $0.2 p.u., inter-pixel effects over 5$\times$5 array). Error bars are reported for representative cases of jitter signal, i.e. sin1, cos1, saw1v3, saw1vf2, and jump04c. Middle panel: the same for the orbital semimajor axis in units of the stellar radius, $a_0 = a/R_s$. Bottom panel: the same for the orbital inclination, $i$. \label{fig12}}
\end{figure*}

Fig. \ref{fig10} shows the raw light-curves simulated with $\sigma_{PSF} = $0.2 p.u., and inter-pixel quantum efficiency variations over 5$\times$5 array of pixels, and the correspondent detrended light-curves, obtained with the two methods considered in this paper. The array is large enough that observed modulations are due only to the pixel effects. Tab. \ref{tab5} reports the discrepancies between those light-curves and the theoretical model. Again, the discrepancies after pixel-ICA detrending are below the HFPN level, i.e. 1.5$\times$10$^{-4}$ for the selected binning, outperforming the parametric method by a factor 2-3. We also noted that the empirical centroid coordinates may differ from the ``true centroid coordinates'' up to 0.15-0.33 p.u., which is an error comparable with the jitter amplitude. This is not surprising, given that the PSF is undersampled, being much narrower than the pixel size. Despite the large errors in empirical centroid coordinates, in some cases, the discrepancies between PCD detrended light-curves and the theoretical model are at the HFPN level, and slightly larger in other cases. Fig. \ref{fig11} shows how the residuals scale for binning over $n$ points, with $1 \le n \le 10$. A significant temporal structure is present in the raw data, but not in the pixel-ICA detrended light-curves, while the performances of the parametric method are case dependent. Fig. \ref{fig12} shows the transit parameters retrieved from detrended light-curves; in representative cases, we calculated the error bars. Numerical results are reported in Tab. \ref{tab9}.
\begin{table*}
\begin{center}
\caption{Root mean square of residuals between the light-curves and the theoretical model for simulations with $\sigma_{PSF} = 0.2$ p.u., and inter-pixel quantum efficiency variations over 5$\times$5 array of pixels; in particular they are calculated for the raw light-curves, light-curves detrended with pixel-ICA, and PCD method, binned over 10 points. \label{tab5}}
\begin{tabular}{cccc}
\tableline\tableline
Jitter & rms (raw $-$ theoretical) & rms (ICA $-$ theoretical) & rms (PCD $-$ theoretical)\\
\tableline
sin1 & 7.2$\times$10$^{-4}$ & 7.2$\times$10$^{-5}$ & 1.9$\times$10$^{-4}$\\
cos1 & 7.0$\times$10$^{-4}$ & 7.7$\times$10$^{-5}$ & 2.8$\times$10$^{-4}$\\
sin2 & 7.1$\times$10$^{-4}$ & 7.4$\times$10$^{-5}$ & 1.8$\times$10$^{-4}$\\
cos2 & 7.4$\times$10$^{-4}$ & 7.4$\times$10$^{-5}$ & 2.4$\times$10$^{-4}$\\
sin3 & 7.1$\times$10$^{-4}$ & 7.8$\times$10$^{-5}$ & 1.8$\times$10$^{-4}$\\
cos3 & 7.0$\times$10$^{-4}$ & 7.5$\times$10$^{-5}$ & 2.0$\times$10$^{-4}$\\
saw1 & 2.6$\times$10$^{-4}$ & 6.7$\times$10$^{-5}$ & 1.5$\times$10$^{-4}$\\
saw1v1 & 2.3$\times$10$^{-4}$ & 6.8$\times$10$^{-5}$ & 1.5$\times$10$^{-4}$\\
saw1v2 & 2.3$\times$10$^{-4}$ & 6.6$\times$10$^{-5}$ & 1.5$\times$10$^{-4}$\\
saw1v3 & 2.3$\times$10$^{-4}$ & 6.8$\times$10$^{-5}$ & 1.5$\times$10$^{-4}$\\
saw1vf1 & 2.3$\times$10$^{-4}$ & 6.6$\times$10$^{-5}$ & 1.5$\times$10$^{-4}$\\
saw1vf2 & 2.4$\times$10$^{-4}$ & 6.8$\times$10$^{-5}$ & 1.5$\times$10$^{-4}$\\
jump04c & 7.6$\times$10$^{-4}$ & 7.4$\times$10$^{-5}$ & 1.5$\times$10$^{-4}$\\
\tableline
\end{tabular}
\end{center}
\end{table*}

\clearpage

\subsection{Case III: intra-pixel effects}
\label{sec:intraPSF02}

For simulations with $\sigma_{PSF} =$1 p.u., the effect of intra-pixel sensitivity variations is negligible, i.e. $\sim$10$^{-5}$, unless we consider unphysical or very unlikely cases, where the quantum efficiency can assume both positive and negative values in a pixel, or it is zero for a significant fraction of the area of the pixel. Intra-pixel effects become significant when the PSF is narrower, therefore we analyzed only the relevant simulations with $\sigma_{PSF} =$0.2 p.u.

\begin{figure*}
\epsscale{0.96}
\plotone{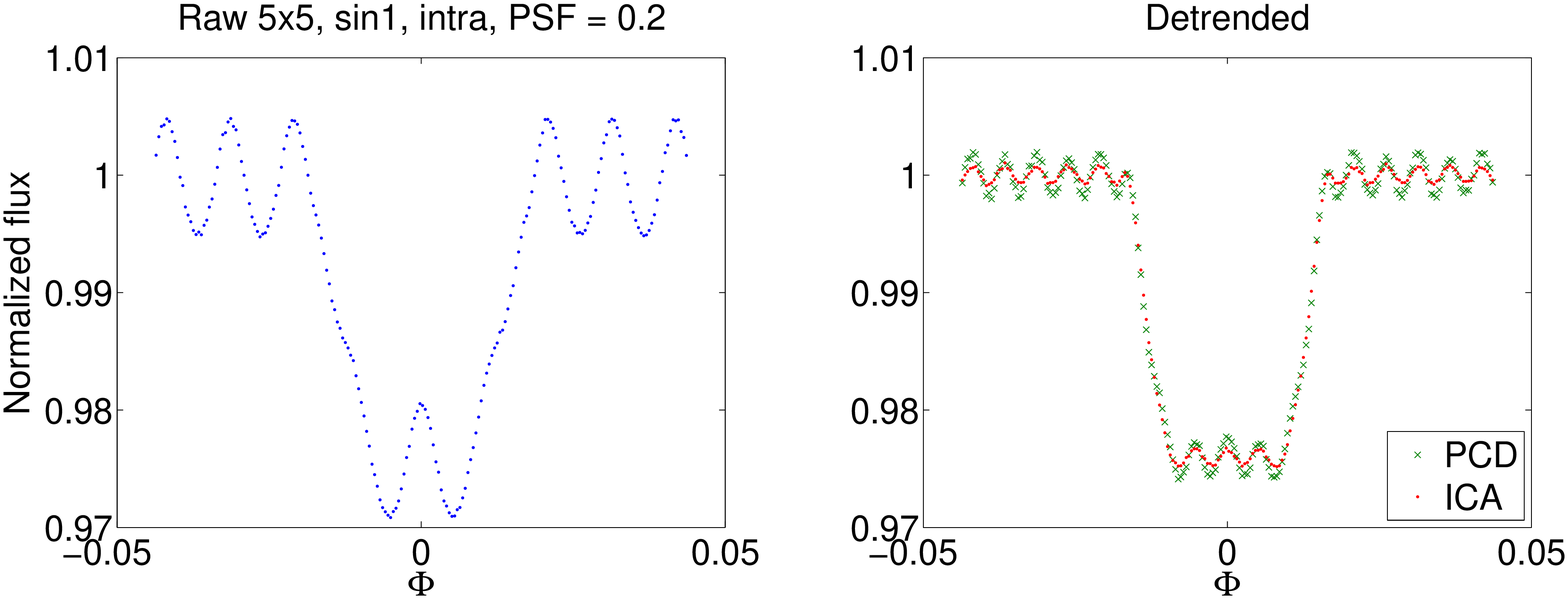}
\plotone{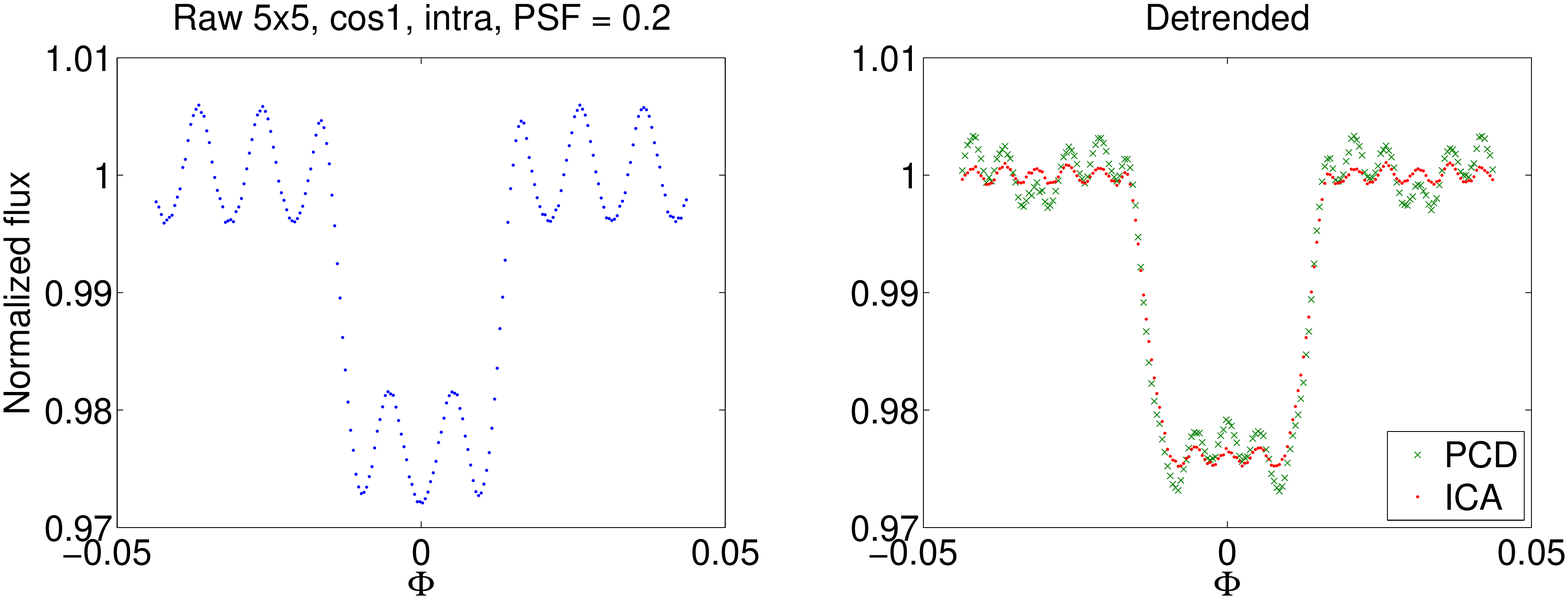}
\plotone{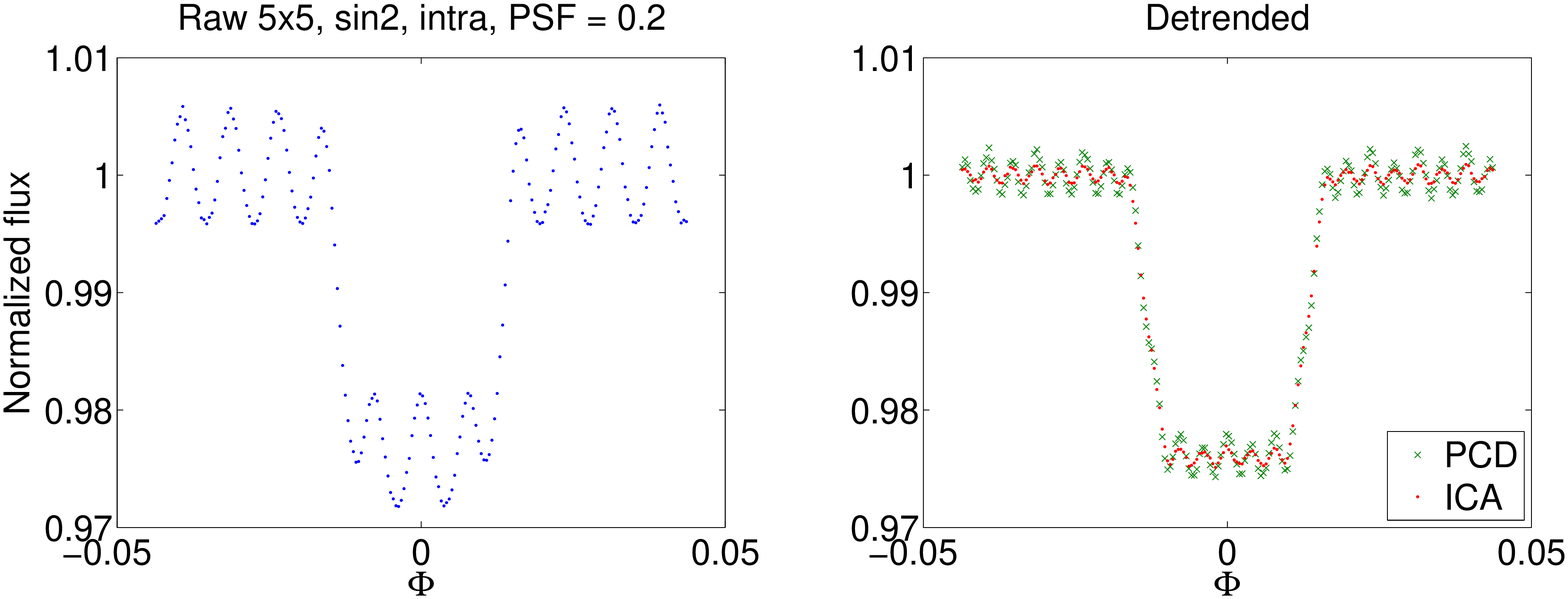}
\plotone{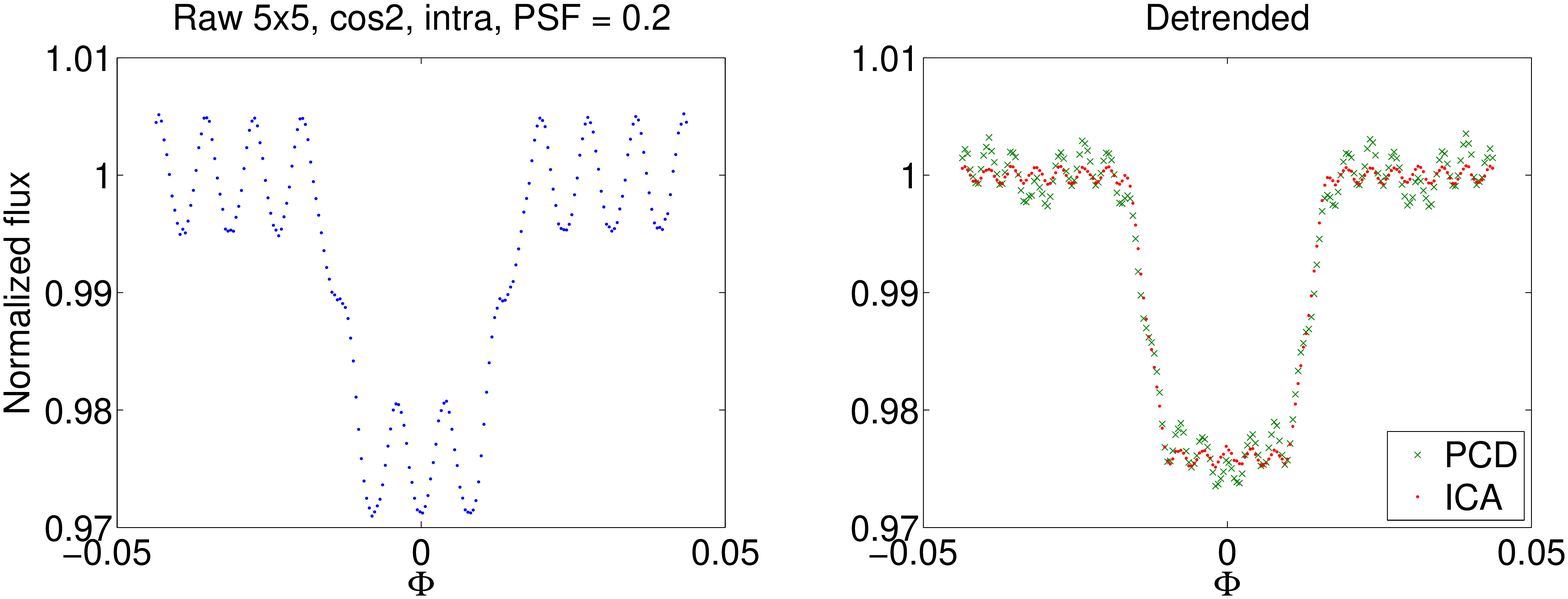}
\plotone{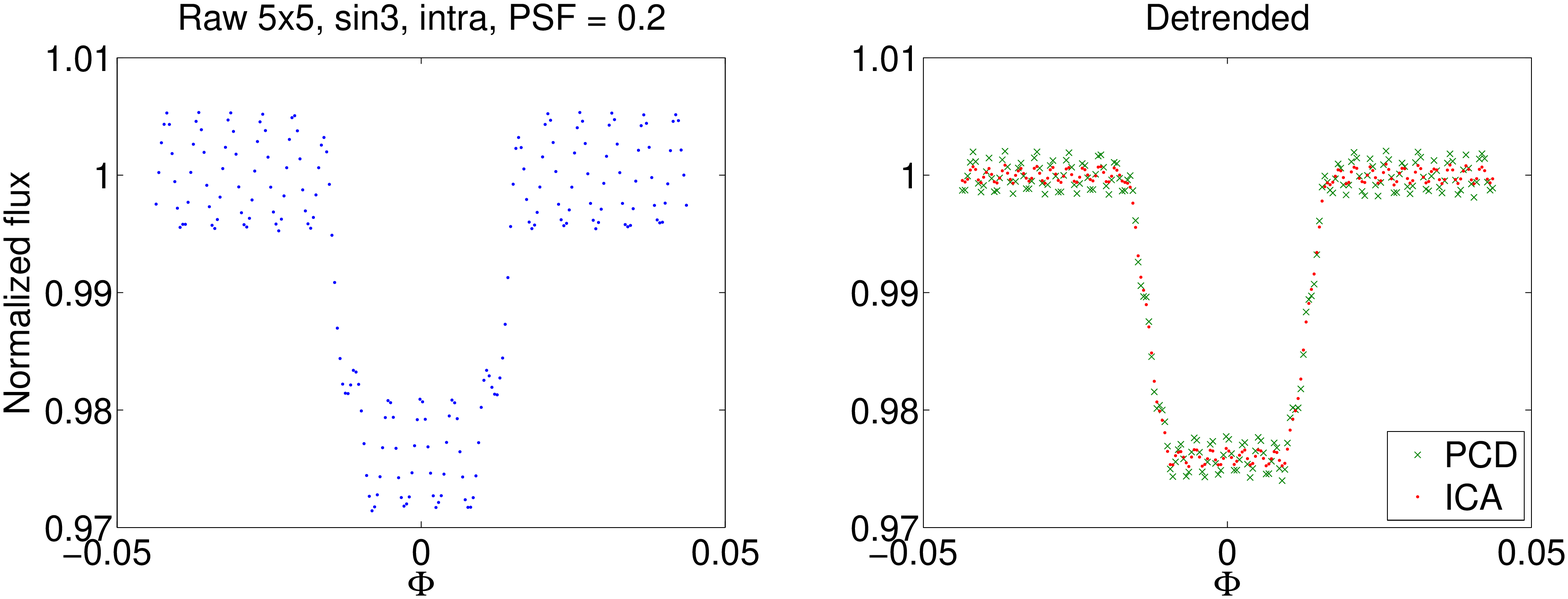}
\plotone{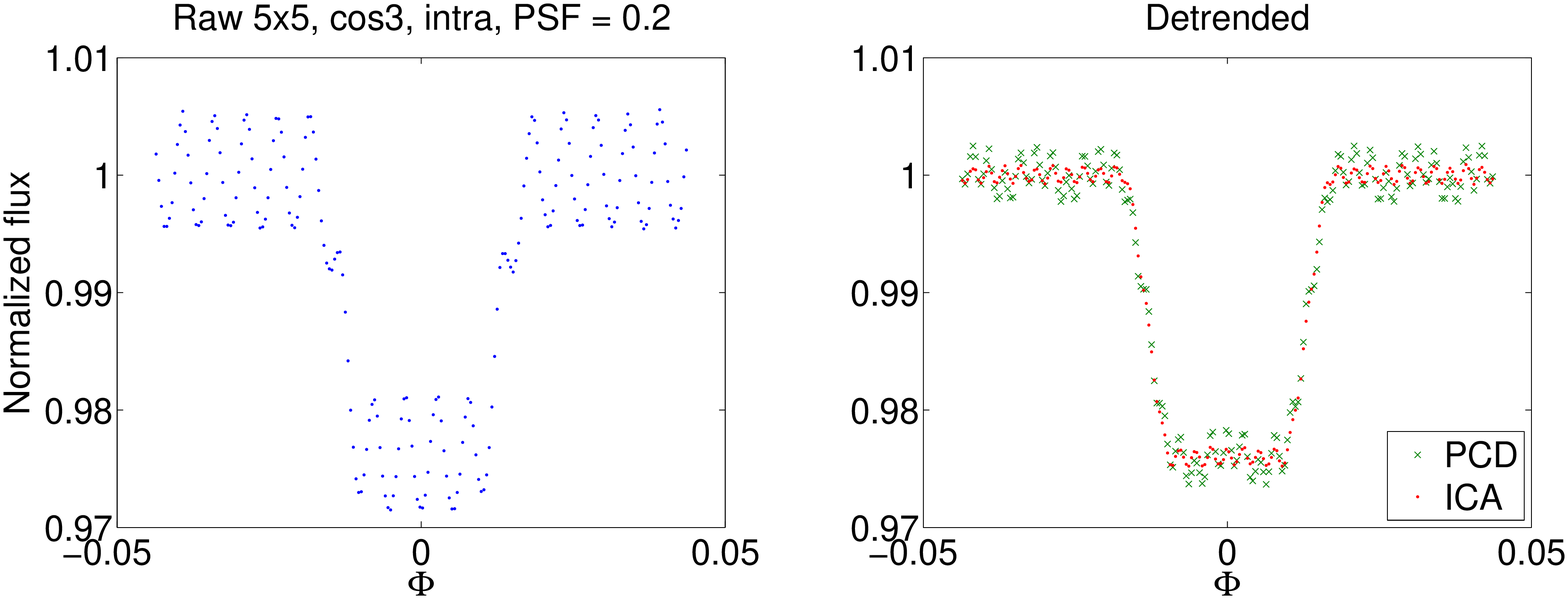}
\plotone{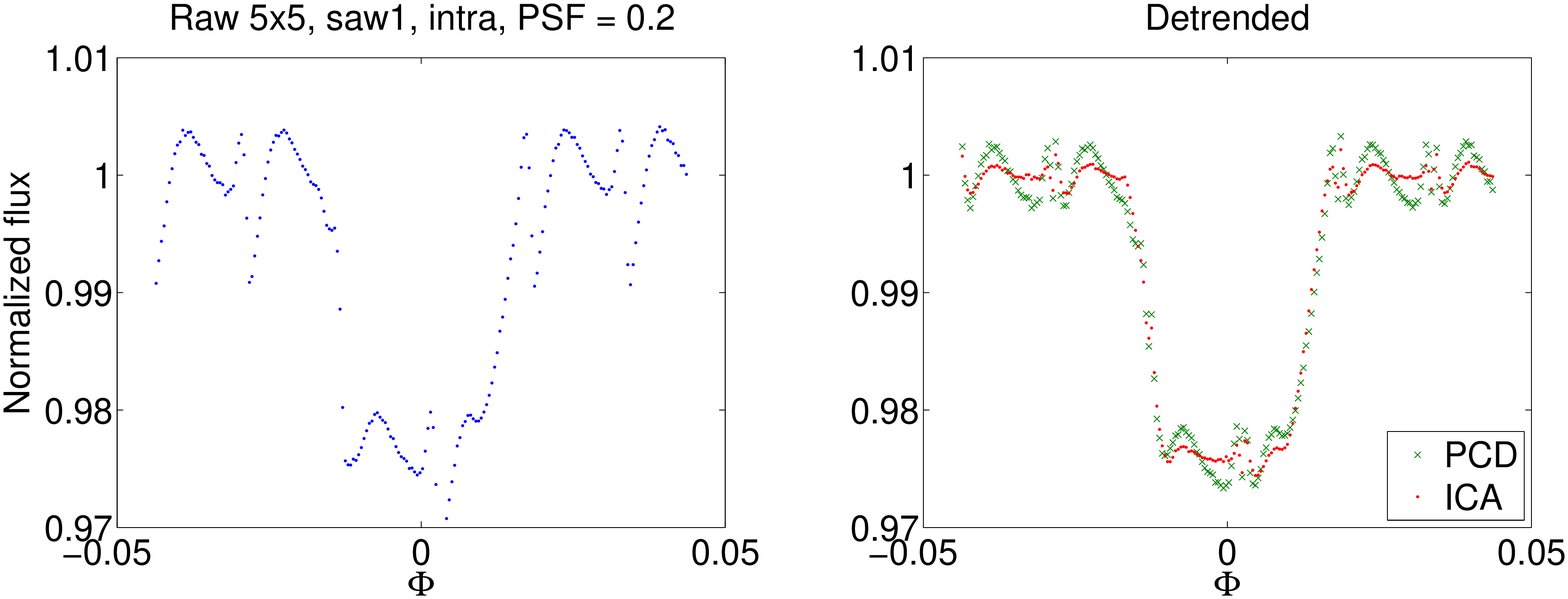}
\plotone{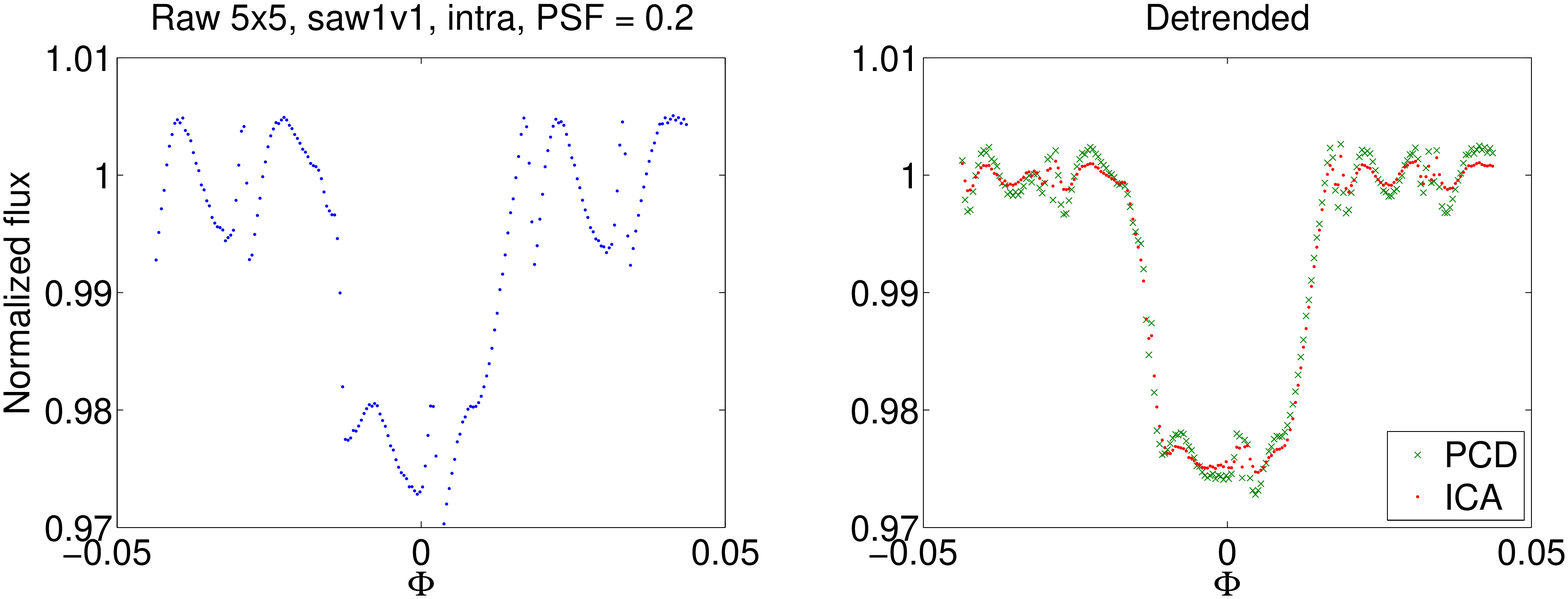}
\plotone{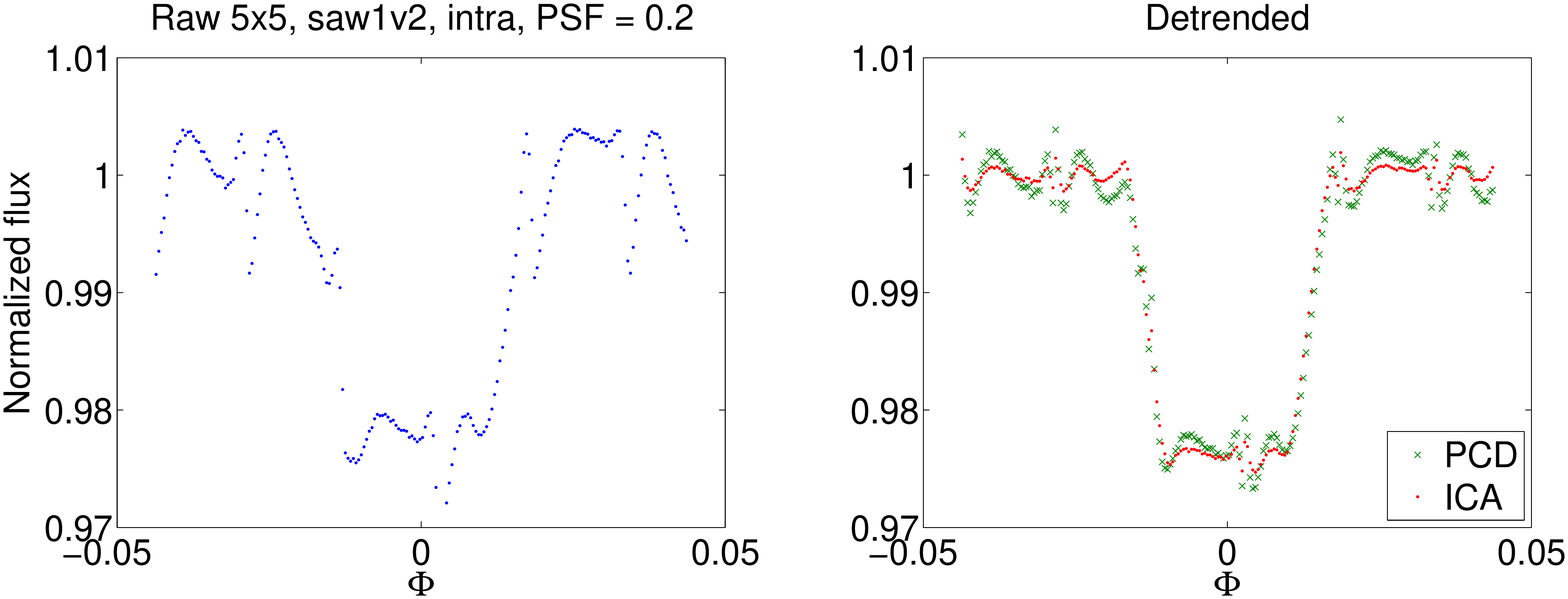}
\plotone{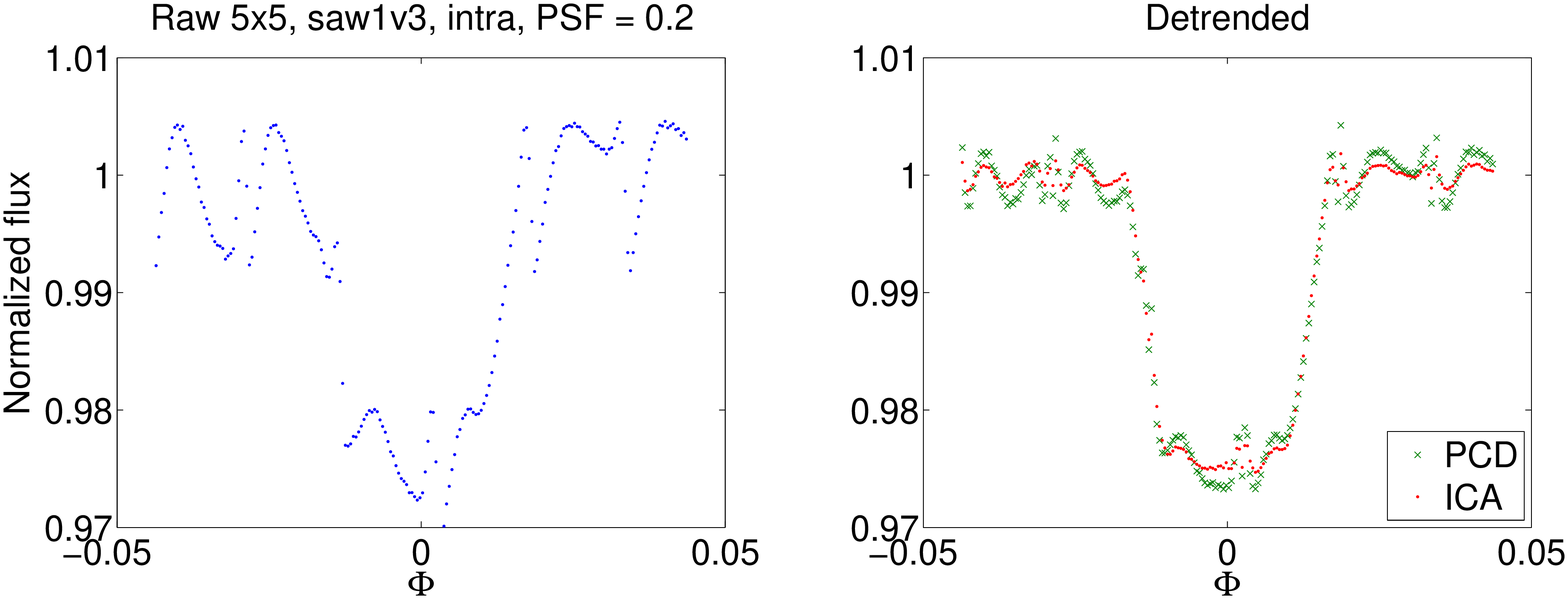}
\plotone{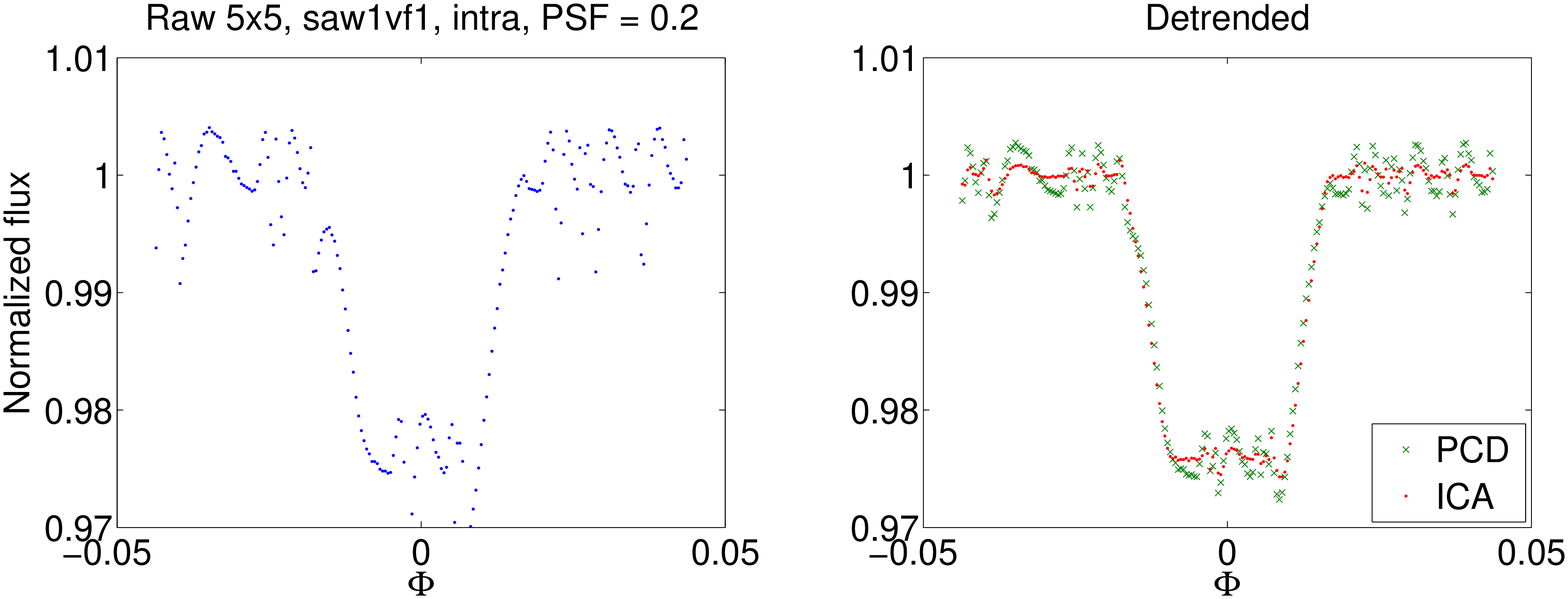}
\plotone{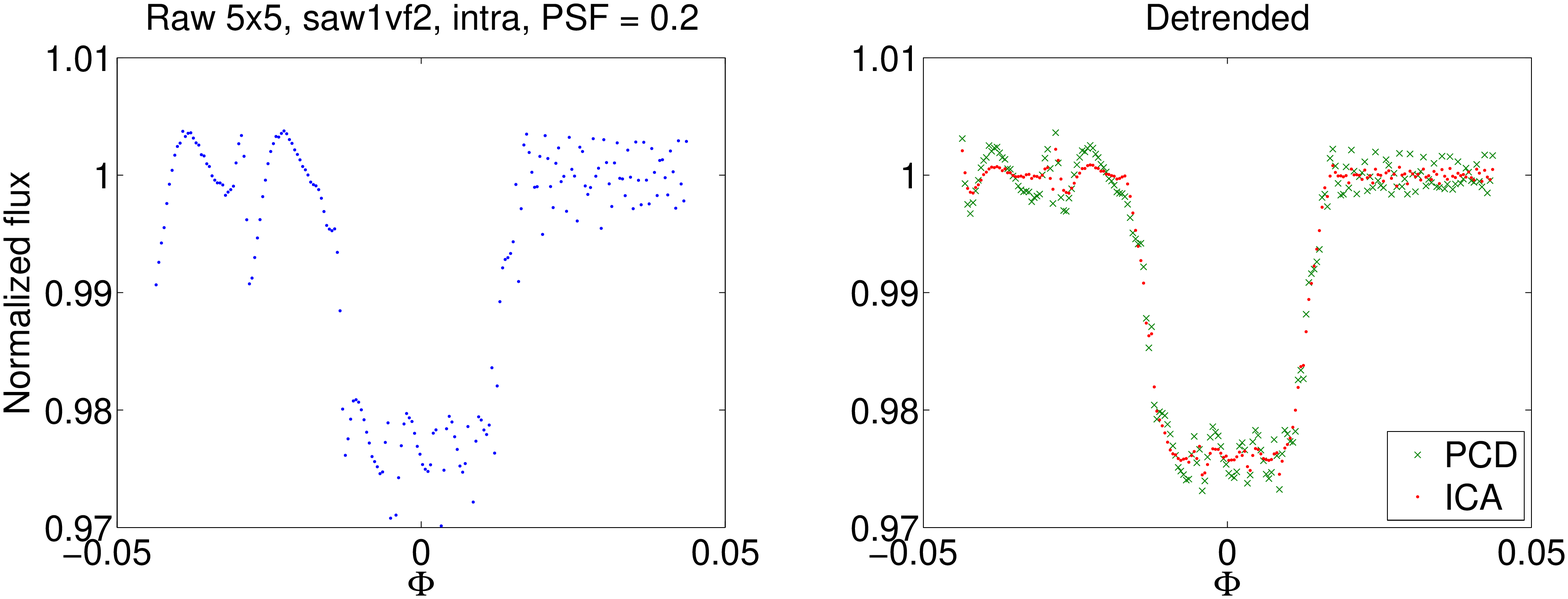}
\plotone{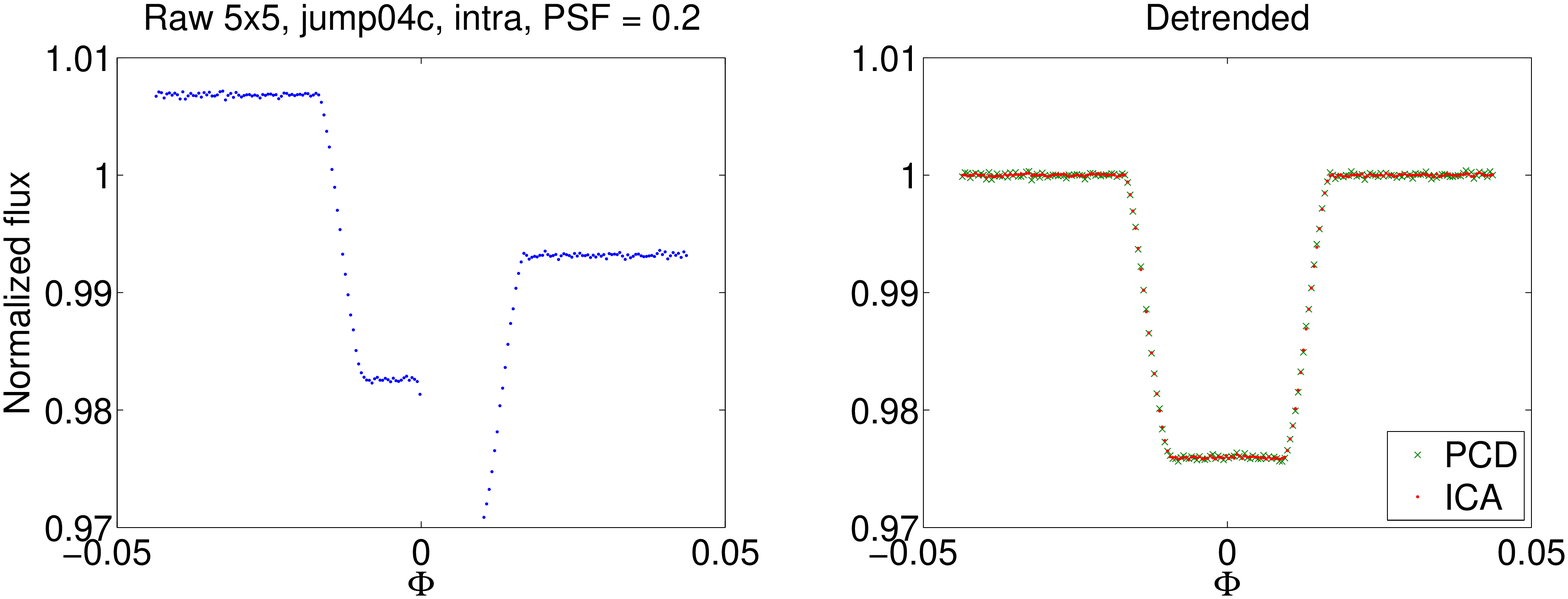}
\caption{Left panels: (blue) raw light-curves simulated with $\sigma_{PSF} = $0.2 p.u., and intra-pixel quantum efficiency variations over 5$\times$5 array of pixels. Right panels: detrended transit light-curves obtained with (green `x') polynomial centroid fitting method, and (red dots) pixel-ICA method. All the light-curves are binned over 10 points to make clearer visualization of the systematic effects. \label{fig13}}
\end{figure*}

\begin{figure*}
\epsscale{1.60}
\plotone{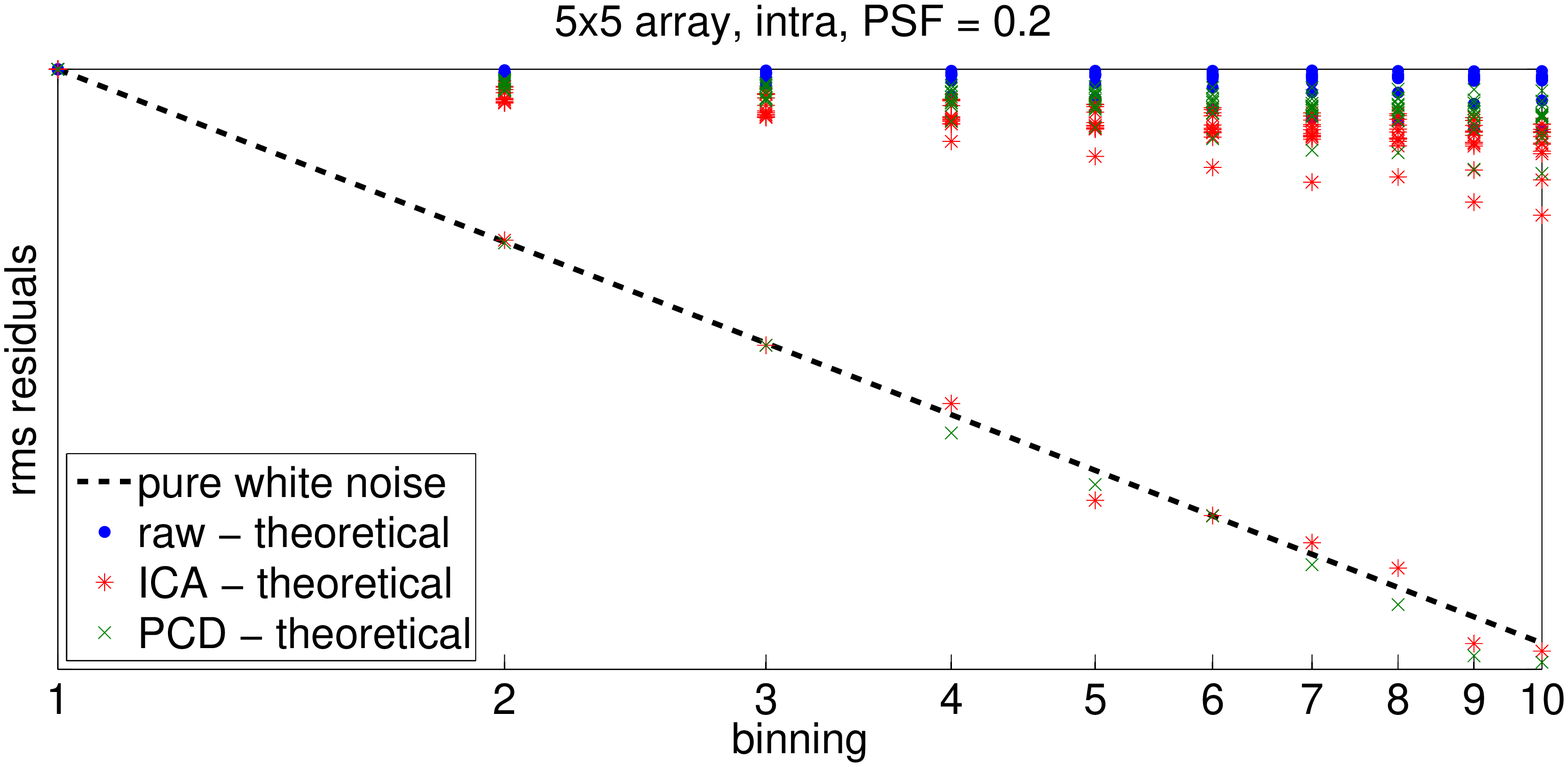}
\caption{Root mean square of residuals for light-curves binned over 1 to 10 points, scaled to their non-binned values. The simulations were obtained with $\sigma_{PSF} =$0.2 p.u., 5$\times$5 array, and intra-pixel effects. The dashed black line indicates the expected trend for white residuals, blue dots are for normalized raw light-curves, red `$\ast$' are for pixel-ICA detrendend light-curves, and green `x' for PCD detrended light-curves. \label{fig14}}
\end{figure*}

\begin{figure*}
\epsscale{1.60}
\plotone{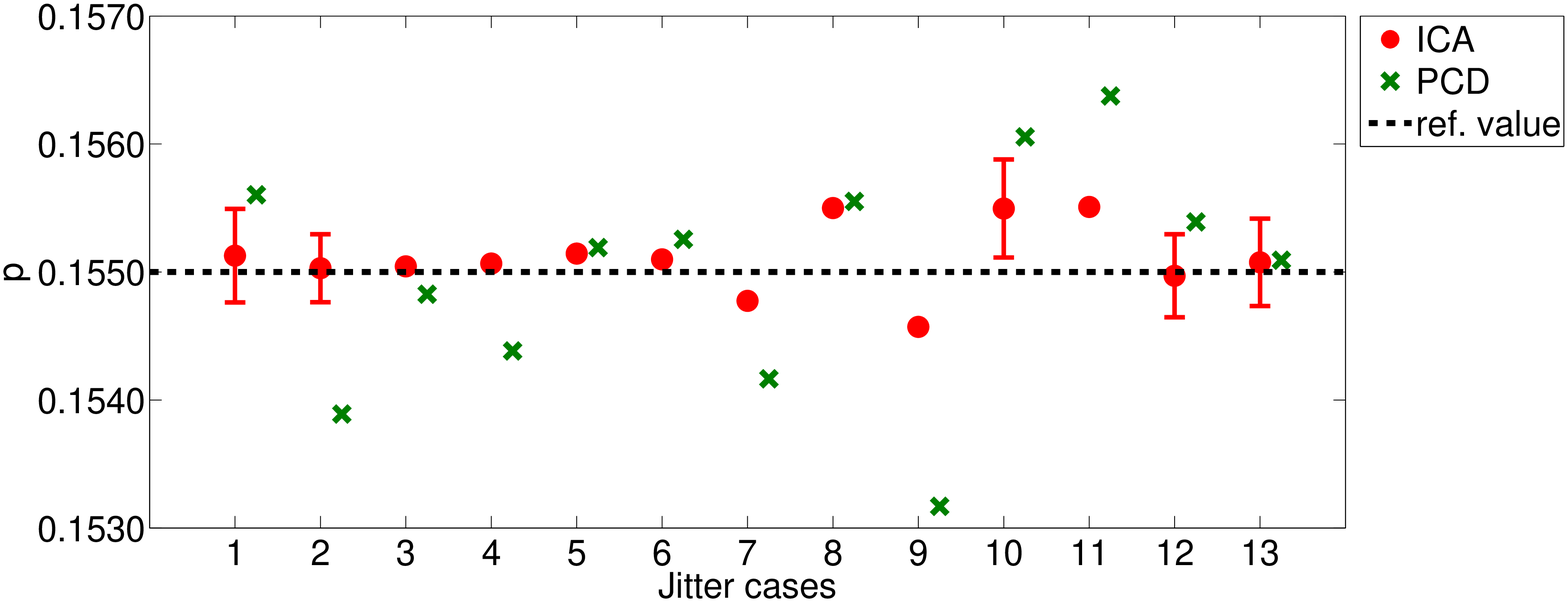}
\plotone{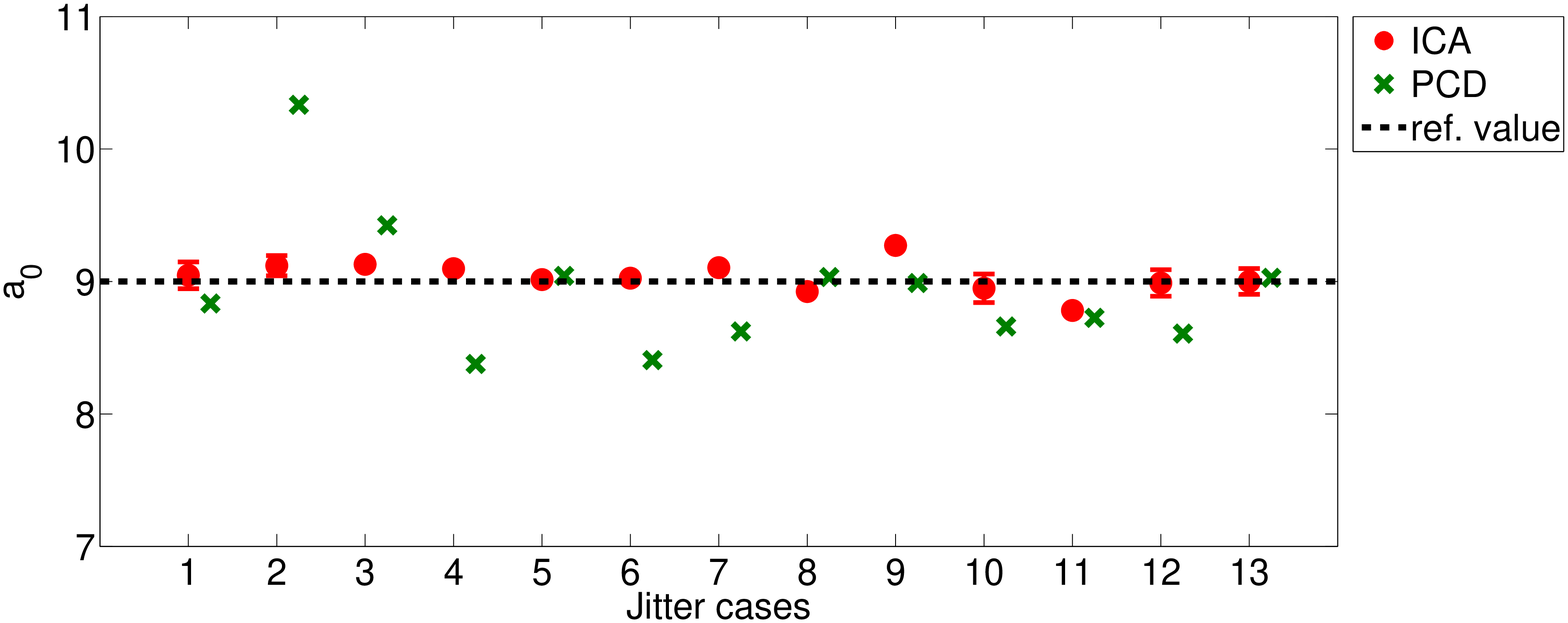}
\plotone{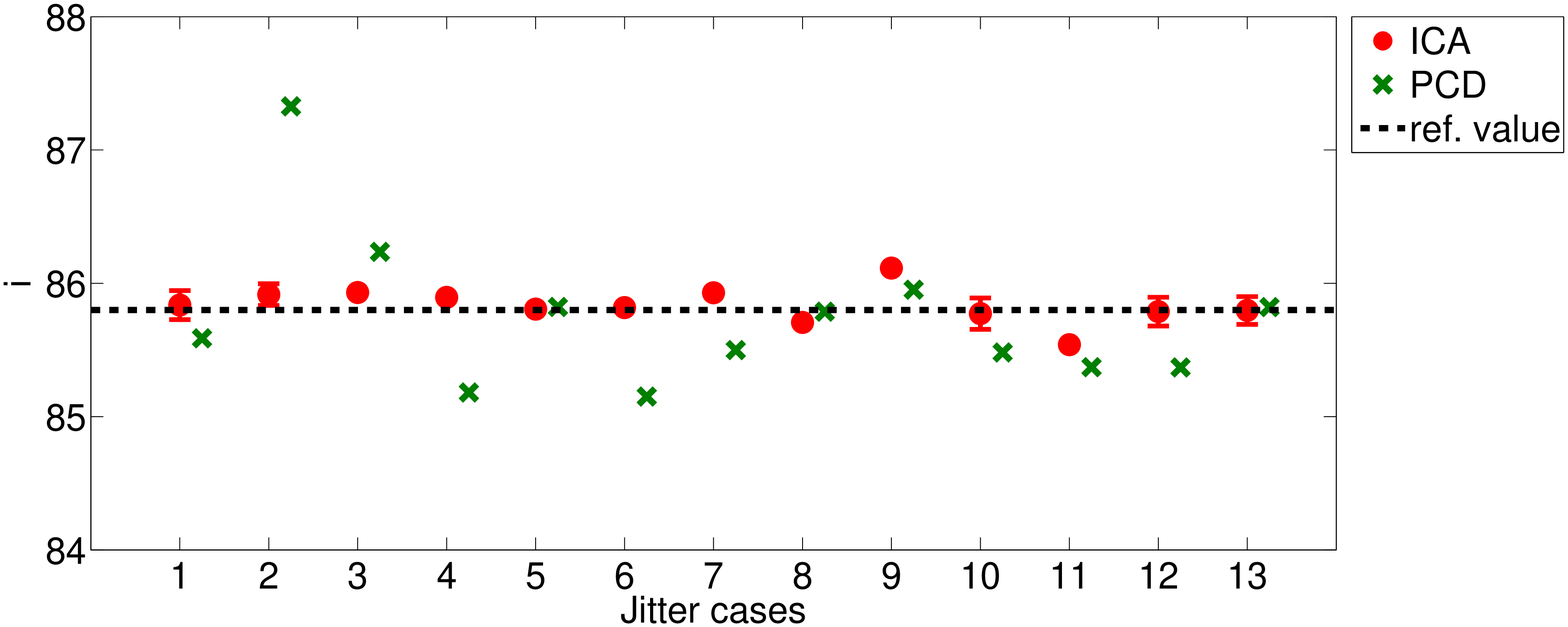}
\caption{Top panel: best estimates of the planet-to-star radii ratio, $p = r_p/R_s$, for detrended light-curves with (red dots) pixel-ICA, and (green `x') PCD method ($\sigma_{PSF} = $0.2 p.u., intra-pixel effects over 5$\times$5 array). Error bars are reported for representative cases of jitter signal, i.e. sin1, cos1, saw1v3, saw1vf2, and jump04c. Middle panel: the same for the orbital semimajor axis in units of the stellar radius, $a_0 = a/R_s$. Bottom panel: the same for the orbital inclination, $i$.\label{fig15}}
\end{figure*}

Fig. \ref{fig13} shows the raw light-curves simulated with $\sigma_{PSF} = $0.2 p.u., and intra-pixel quantum efficiency variations over 5$\times$5 array of pixels, and the correspondent detrended light-curves. The array is large enough that the observed modulations are only due to the pixel effects. Tab. \ref{tab6} reports the discrepancies between those light-curves and the theoretical model. The pixel-ICA technique reduces the dicrepancies by a factor 4-8 (for the selected binning) for the first 12 jitter series, and by a  factor 83 for the case `jump04c', outperforming the parametric method by a factor 2-4. Fig. \ref{fig14} shows how the residuals scale for binning over $n$ points, with $1 \le n \le 10$. In this case, the temporal structure is preserved in all detrended light-curves, except for the case `jump04c', which means that both methods have some troubles to decorrelate intra-pixel effects. Fig. \ref{fig15} shows the transit parameters retrieved from detrended light-curves; in representative cases, we calculated the error bars. Detailed numerical results are reported in Tab. \ref{tab10}. While in some cases the parametric method may perform better than pixel-ICA, if adopting higher order polynomials, in some other cases higher order polynomials lead to worse results than lower order polynomials. Higher order polynomials might approximate better the out-of-transit baseline, but the residuals to the theoretical light-curve are not necessarily improved, hence the correction can bias the astrophysical signal. The pixel-ICA method is less case dependent. We do not suggest that the same conclusions are necessarily valid for Spitzer, since our simulations have stronger systematics and much narrower PSFs. It is quite remarkable that, though the systematics are not well decorrelated, the parameter retrieval gives the correct results within the error bars.
\begin{table*}
\begin{center}
\caption{Root mean square of residuals between the light-curves and the theoretical model for simulations with $\sigma_{PSF} = 0.2$ p.u., and intra-pixel quantum efficiency variations over 5$\times$5 array of pixels; in particular they are calculated for the raw light-curves, light-curves detrended with pixel-ICA, and PCD method, binned over 10 points. \label{tab6}}
\begin{tabular}{cccc}
\tableline\tableline
Jitter & rms (raw $-$ theoretical) & rms (ICA $-$ theoretical) & rms (PCD $-$ theoretical)\\
\tableline
sin1 & 3.5$\times$10$^{-3}$ & 5.1$\times$10$^{-4}$ & 1.2$\times$10$^{-3}$\\
cos1 & 3.5$\times$10$^{-3}$ & 5.1$\times$10$^{-4}$ & 1.8$\times$10$^{-3}$\\
sin2 & 3.4$\times$10$^{-3}$ & 4.9$\times$10$^{-4}$ & 1.2$\times$10$^{-3}$\\
cos2 & 3.5$\times$10$^{-3}$ & 5.0$\times$10$^{-4}$ & 1.5$\times$10$^{-3}$\\
sin3 & 3.4$\times$10$^{-3}$ & 5.0$\times$10$^{-4}$ & 1.2$\times$10$^{-3}$\\
cos3 & 3.4$\times$10$^{-3}$ & 5.0$\times$10$^{-4}$ & 1.4$\times$10$^{-3}$\\
saw1 & 3.4$\times$10$^{-3}$ & 7.4$\times$10$^{-4}$ & 1.8$\times$10$^{-3}$\\
saw1v1 & 3.8$\times$10$^{-3}$ & 7.3$\times$10$^{-4}$ & 1.7$\times$10$^{-3}$\\
saw1v2 & 3.6$\times$10$^{-3}$ & 6.8$\times$10$^{-4}$ & 1.7$\times$10$^{-3}$\\
saw1v3 & 3.7$\times$10$^{-3}$ & 7.3$\times$10$^{-4}$ & 1.7$\times$10$^{-3}$\\
saw1vf1 & 3.2$\times$10$^{-3}$ & 6.6$\times$10$^{-4}$ & 1.7$\times$10$^{-3}$\\
saw1vf2 & 3.0$\times$10$^{-3}$ & 6.1$\times$10$^{-4}$ & 1.5$\times$10$^{-3}$\\
jump04c & 6.8$\times$10$^{-3}$ & 8.2$\times$10$^{-5}$ & 1.6$\times$10$^{-4}$\\
\tableline
\end{tabular}
\end{center}
\end{table*}

\clearpage

\subsection{The effect of astrophysical source Poisson noise}
\label{sec:poisson}

\begin{figure*}
\epsscale{0.96}
\plotone{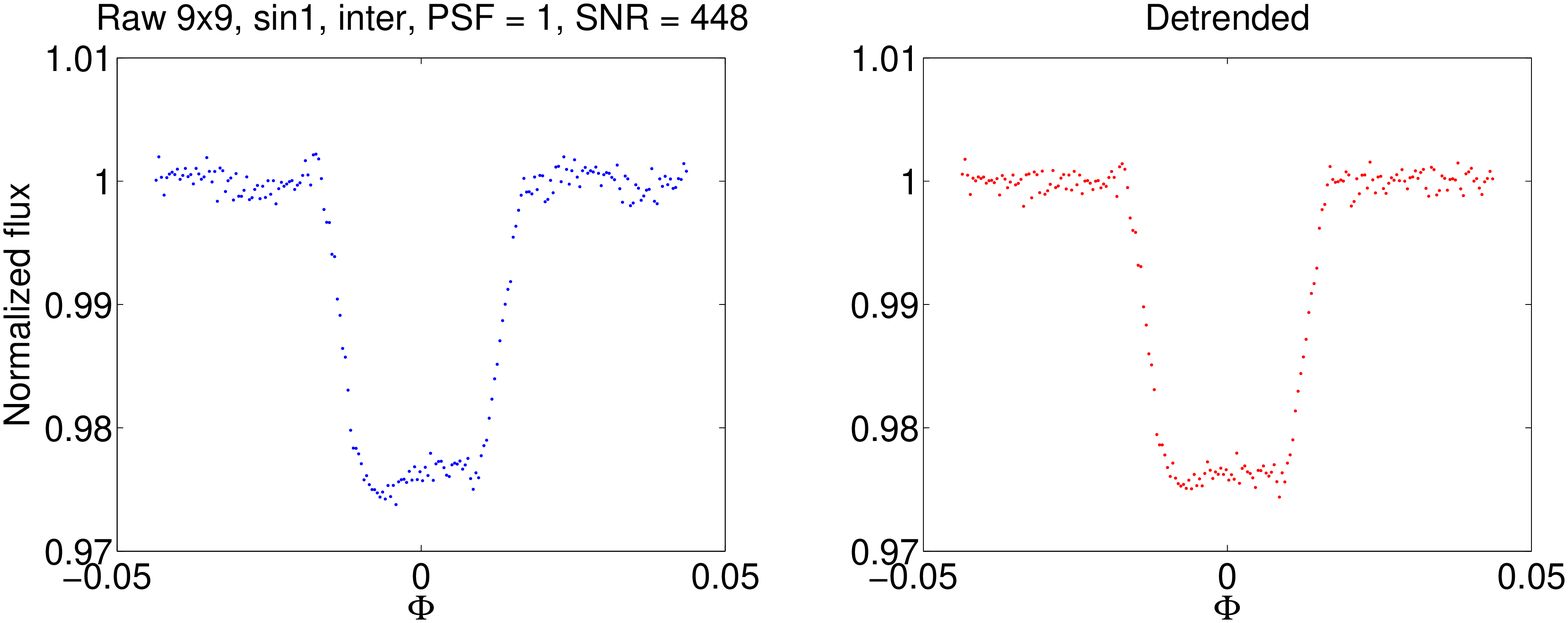}
\plotone{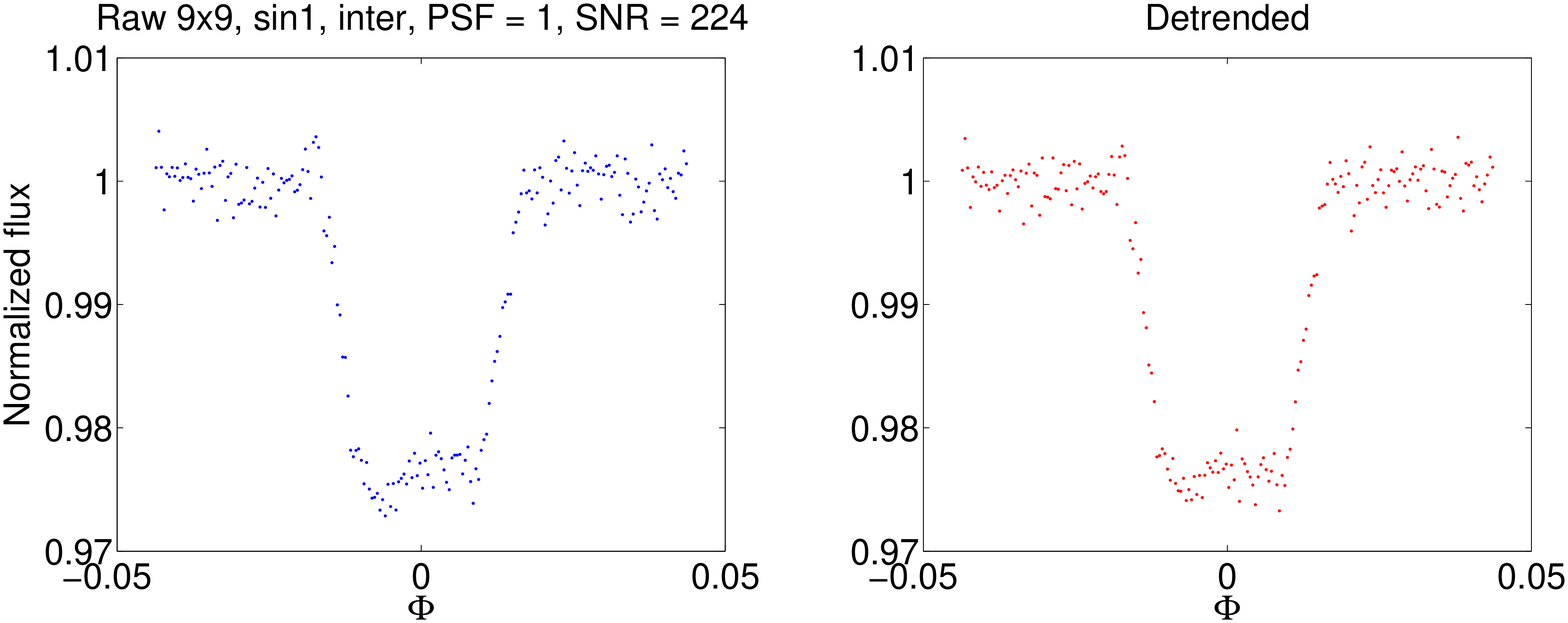}
\plotone{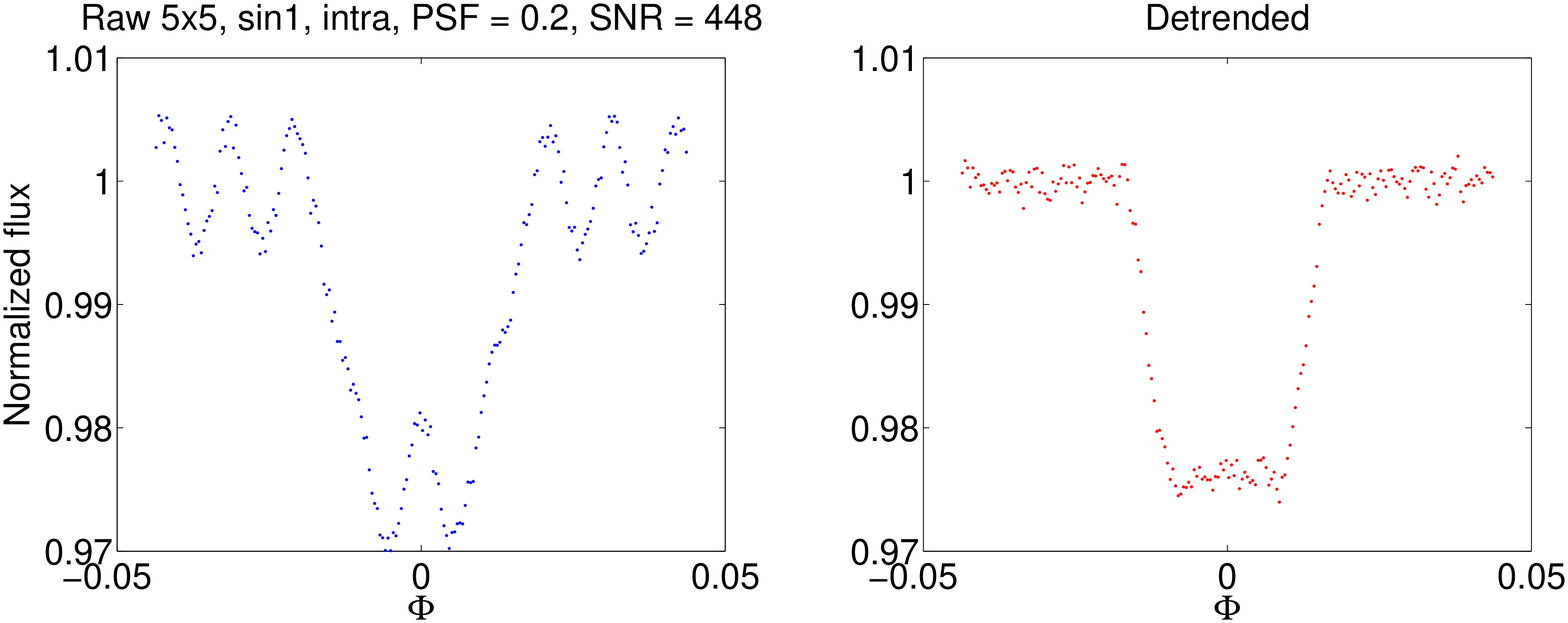}
\plotone{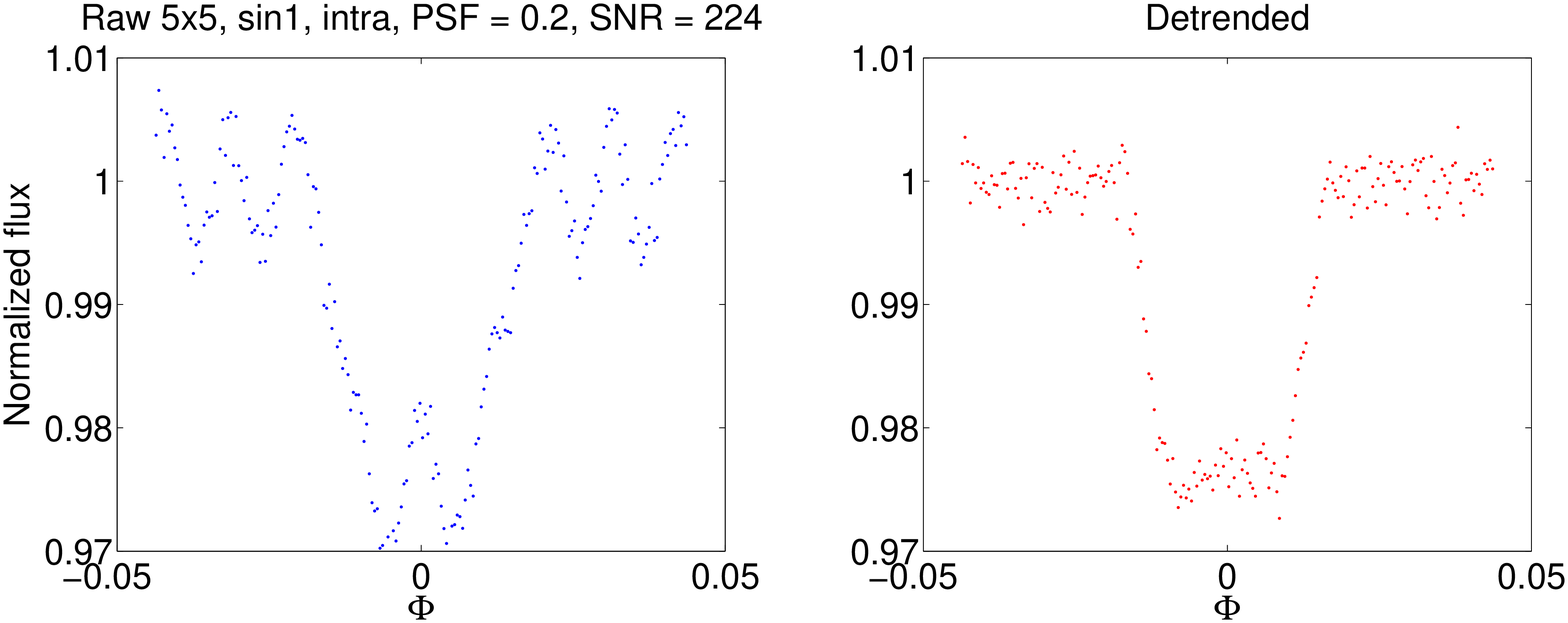}
\plotone{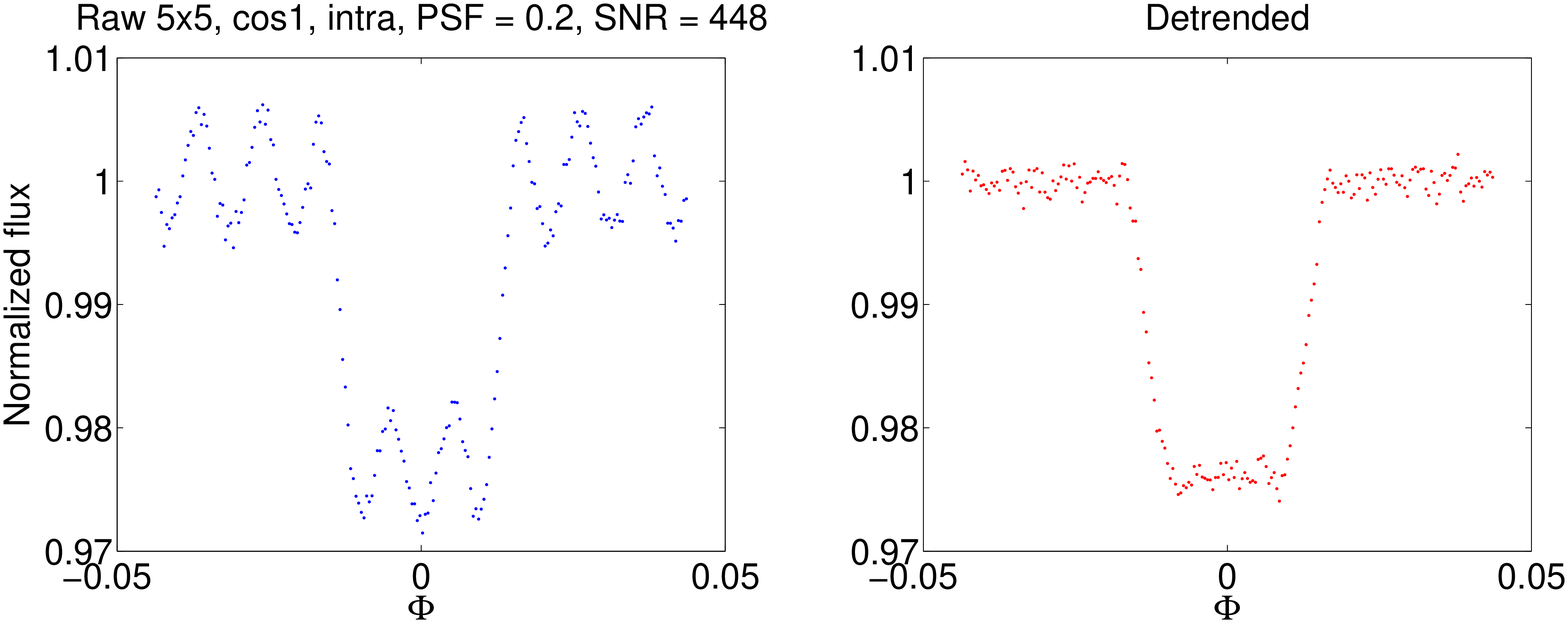}
\plotone{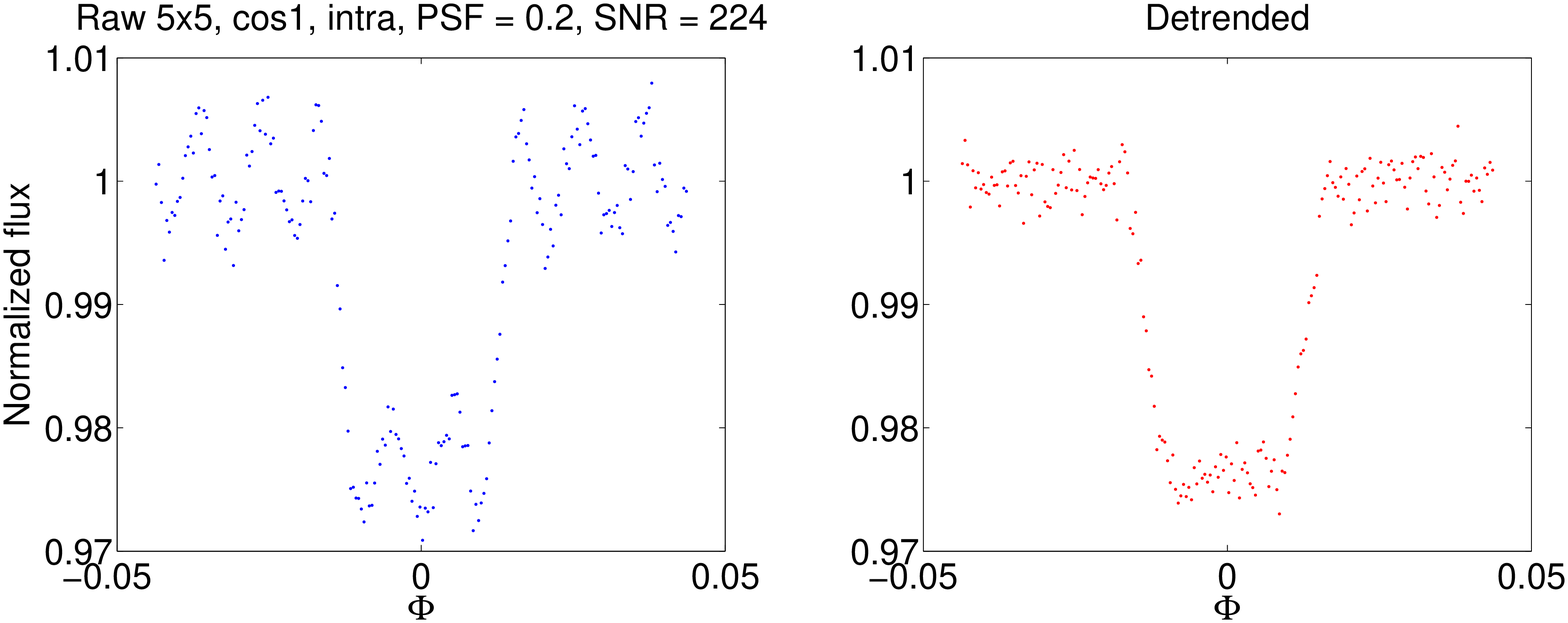}
\caption{Top panels:  (blue) raw light-curves simulated with $\sigma_{PSF} = $1 p.u., inter-pixel quantum efficiency variations over 9$\times$9 array of pixels, jitter `sin1', and poissonian noise with (left) SNR = 447, and (right) SNR = 224; (red) correspondent detrended light-curves with pixel-ICA. Middle panels: the same for raw light-curves simulated with $\sigma_{PSF} = $0.2 p.u., intra-pixel quantum efficiency variations over 5$\times$5 array of pixels, and jitter `sin1'. Bottom panels: the same for light-curves simulated with $\sigma_{PSF} = $0.2 p.u., intra-pixel quantum efficiency variations over 5$\times$5 array of pixels, and jitter `cos1'. All the light-curves are binned over 10 points, as in previous figures.  \label{fig16}}
\end{figure*}

\begin{figure*}
\epsscale{0.96}
\plotone{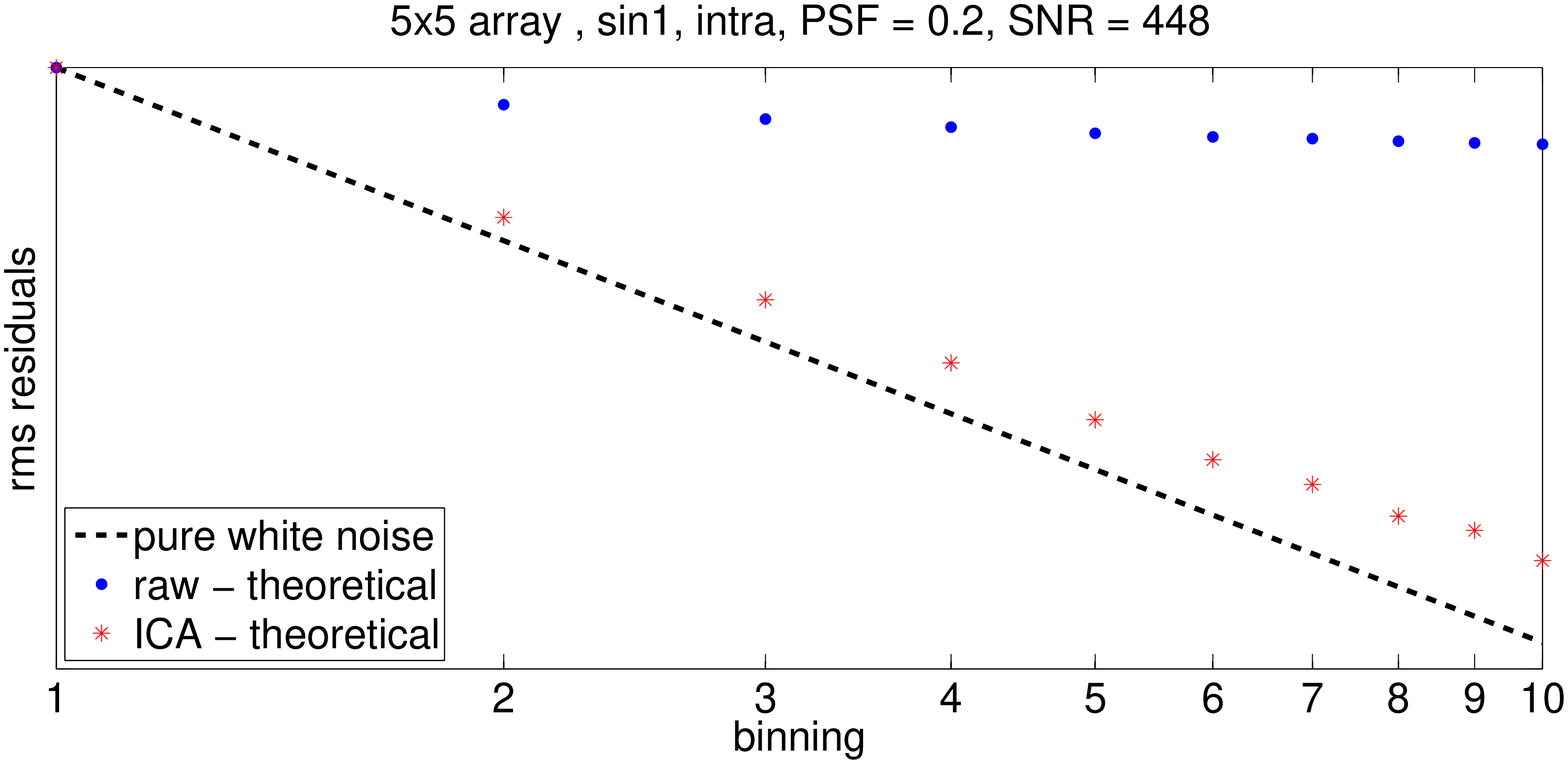}
\plotone{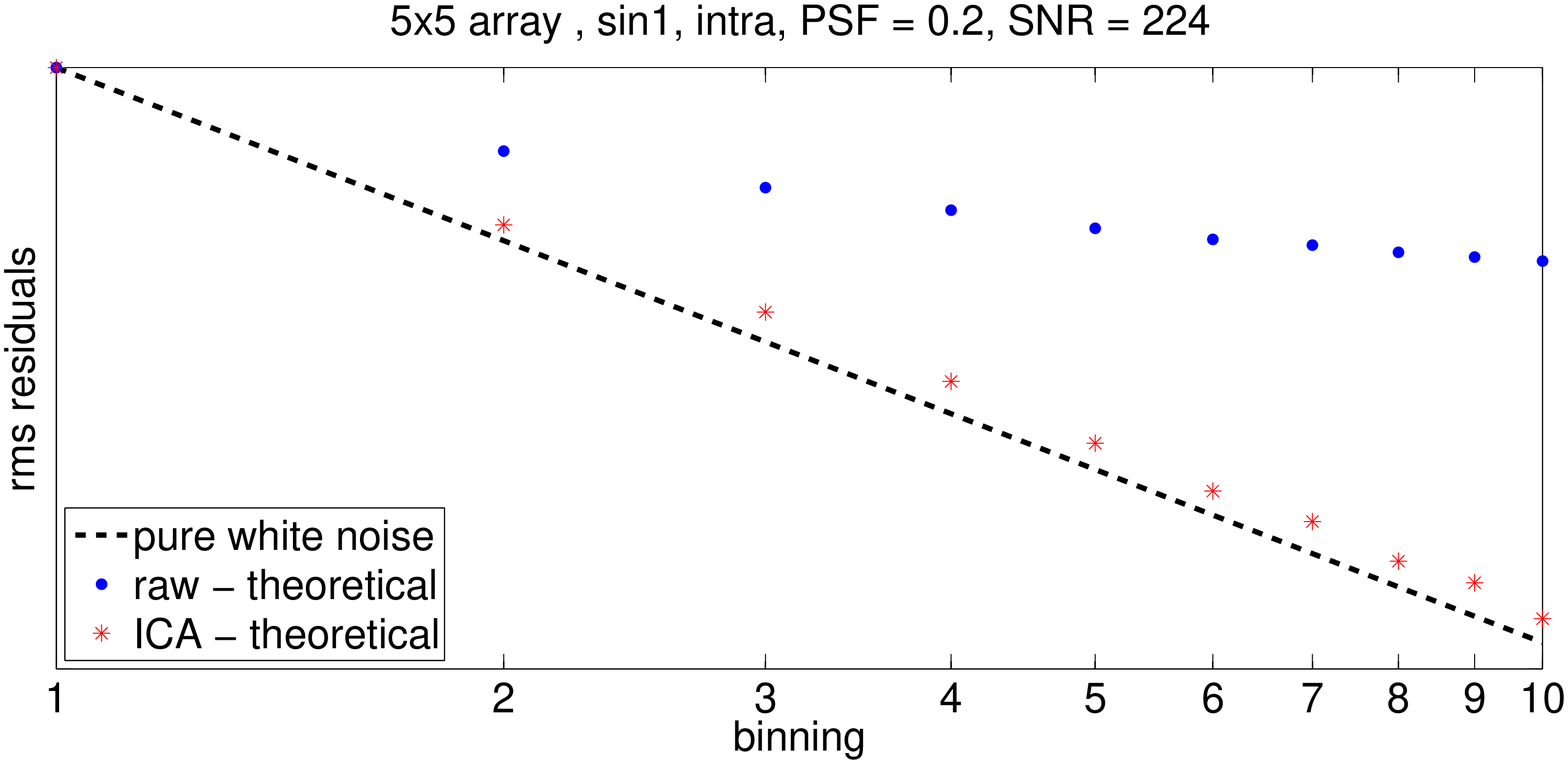}
\caption{Left panel: Root mean square of residuals for binning over 1 to 10 points, scaled to their non-binned values, obtained for simulations with $\sigma_{PSF} =$0.2 p.u., intra-pixel effects over a 5$\times$5 array, jitter `sin1', and poissonian noise with SNR = 447. The dashed black line indicates the expected trend for white residuals, blue dots are for normalized raw light-curves, and red `$\ast$' are for pixel-ICA detrendend light-curves. Right panel: the same, but with SNR = 224. \label{fig17}}
\end{figure*}

\begin{figure*}
\epsscale{1.60}
\plotone{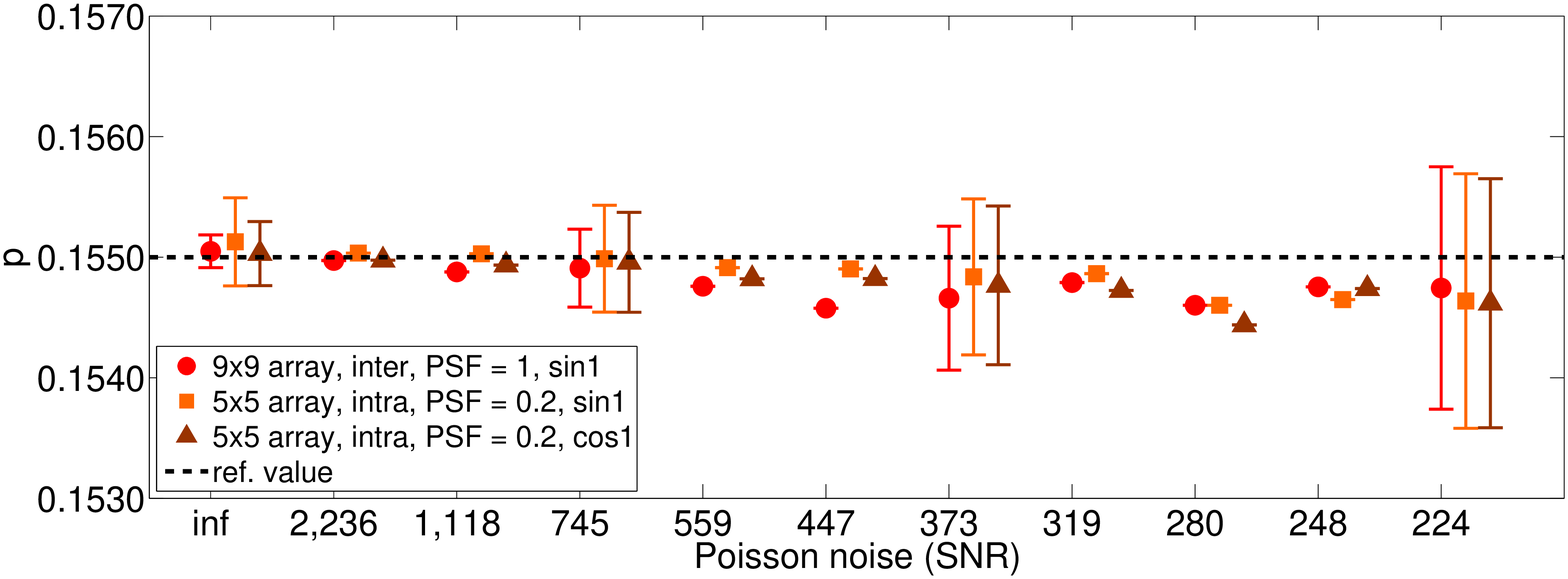}
\plotone{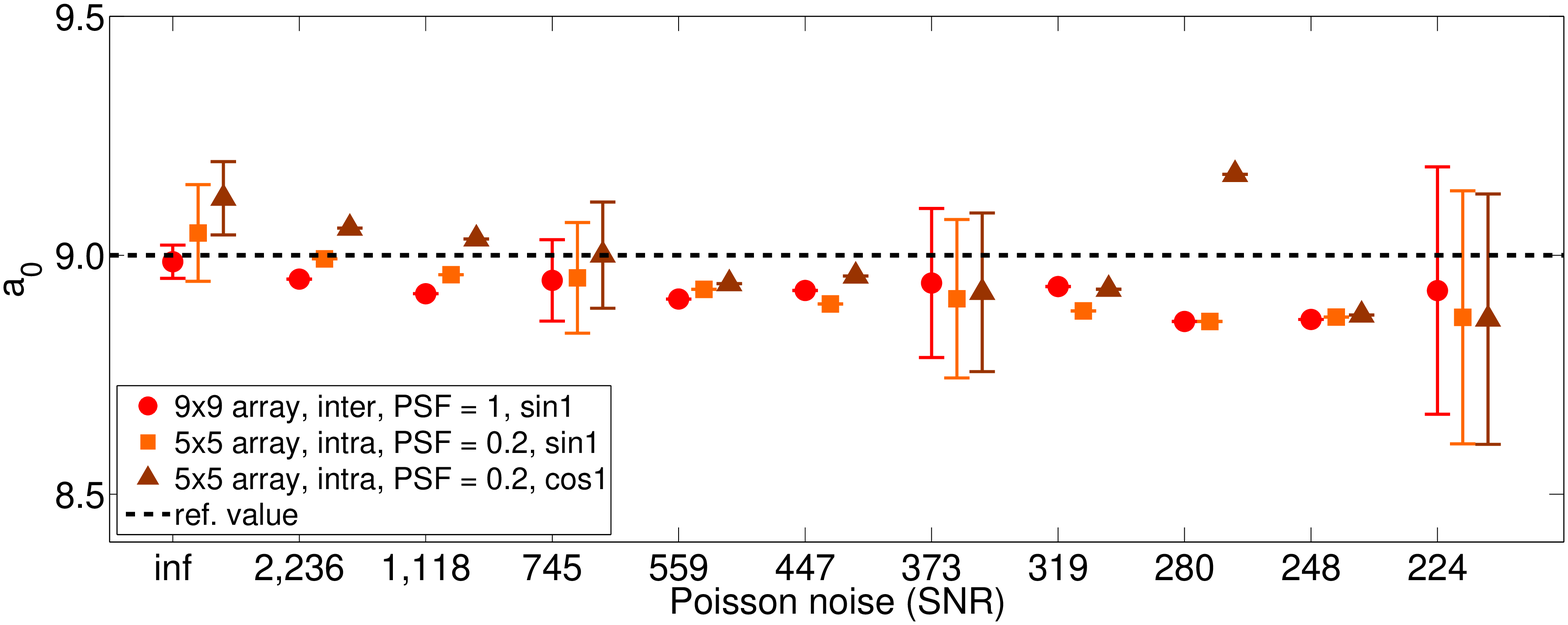}
\plotone{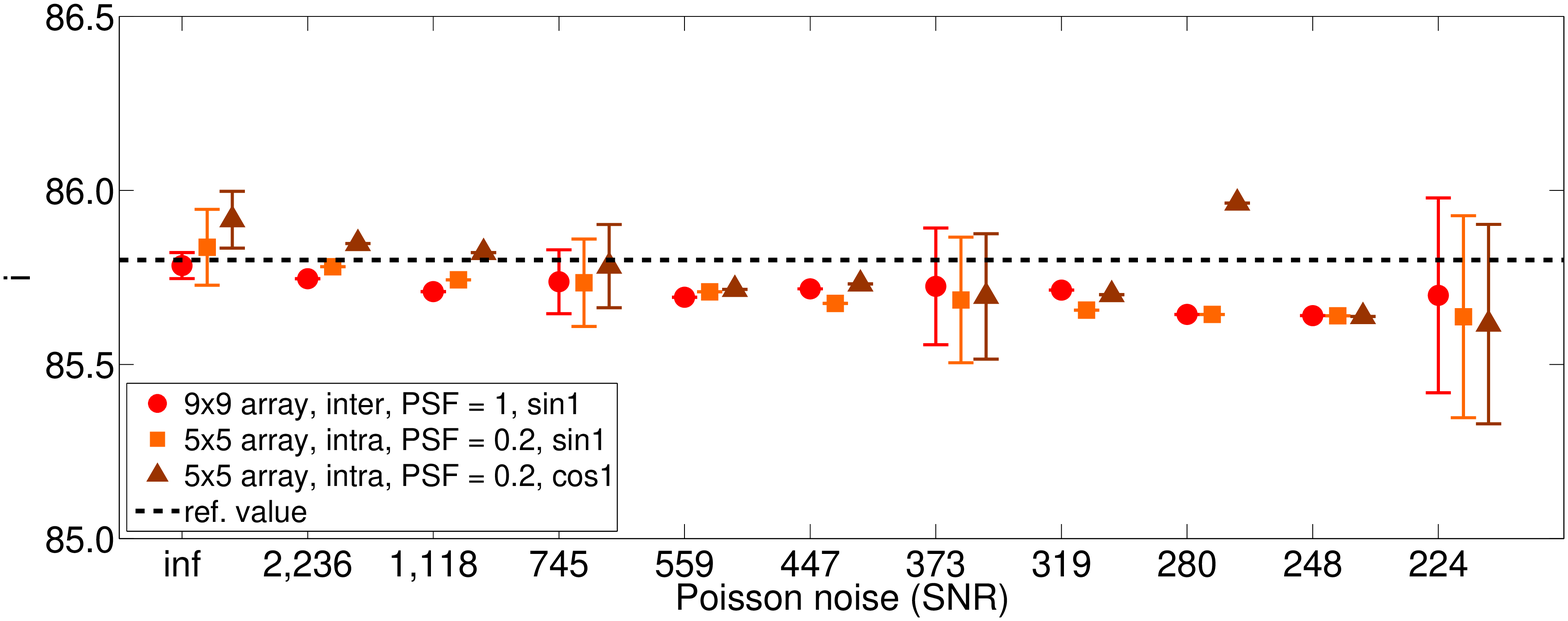}
\caption{Top panel: best estimates of the planet-to-star radii ratio, $p = r_p/R_s$, for some detrended light-curves with Poissonian noise at different levels. Error bars are reported for representative cases. Middle panel: the same for the orbital semimajor axis in units of the stellar radius, $a_0 = a/R_s$. Bottom panel: the same for the orbital inclination, $i$. \label{fig18}}
\end{figure*}

We tested the effect of astrophysical source Poisson noise at different levels. It is well explained by the results obtained with:
\begin{enumerate}
\item $\sigma_{PSF} = $1 p.u., inter-pixel variations over 9$\times$9 array, jitter `sin1';
\item $\sigma_{PSF} = $0.2 p.u., intra-pixel variations over 5$\times$5 array, jitter `sin1' and `cos1';
\end{enumerate}
Fig. \ref{fig16} shows the raw and detrended light-curves obtained for the listed cases with two finite values of Signal-to-Noise Ratio, SNR = 447 (intermediate) and SNR = 224 (lowest). Fig. \ref{fig17} shows how the residuals scale for binning over $n$ points, with $1 \le n \le 10$, in one of the most problematic case of intra-pixel variations (jitter `sin1'). As expected, binning properties depend on the amplitude of white noise relative to systematic noise; therefore, for cases with lower SNR it may superficially appear that systematics are better removed in the detrending process. Fig. \ref{fig18} reports the transit parameter retrieved from detrended light-curves; in representative cases, we calculated the error bars.

\begin{figure*}
\epsscale{1.60}
\plotone{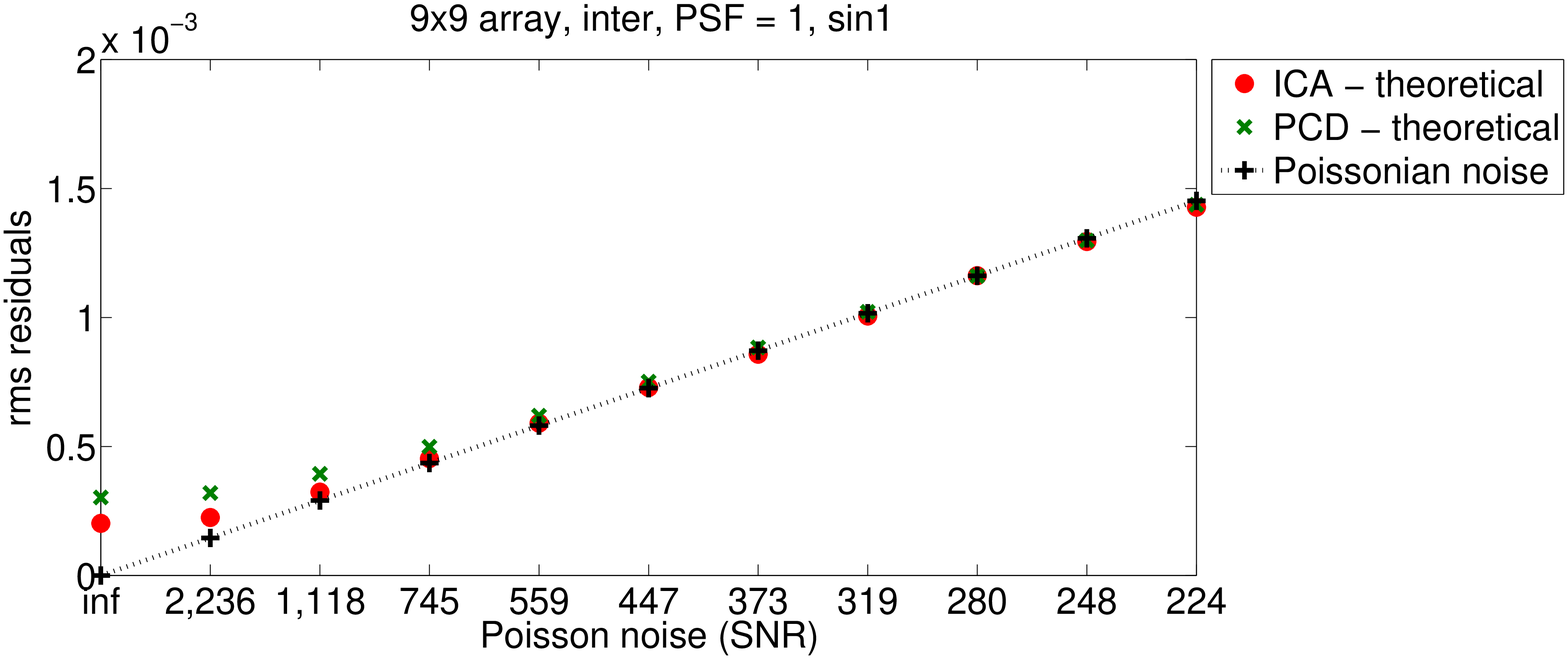}
\plotone{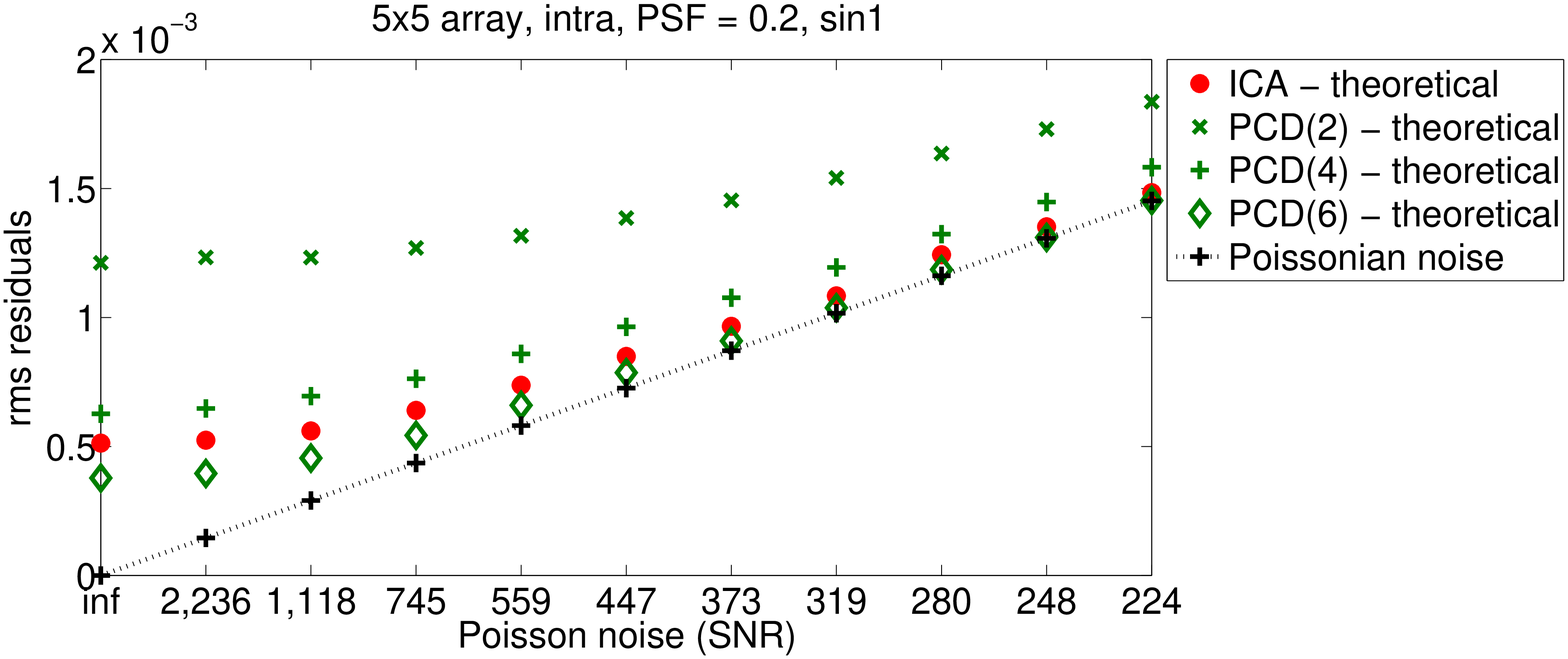}
\plotone{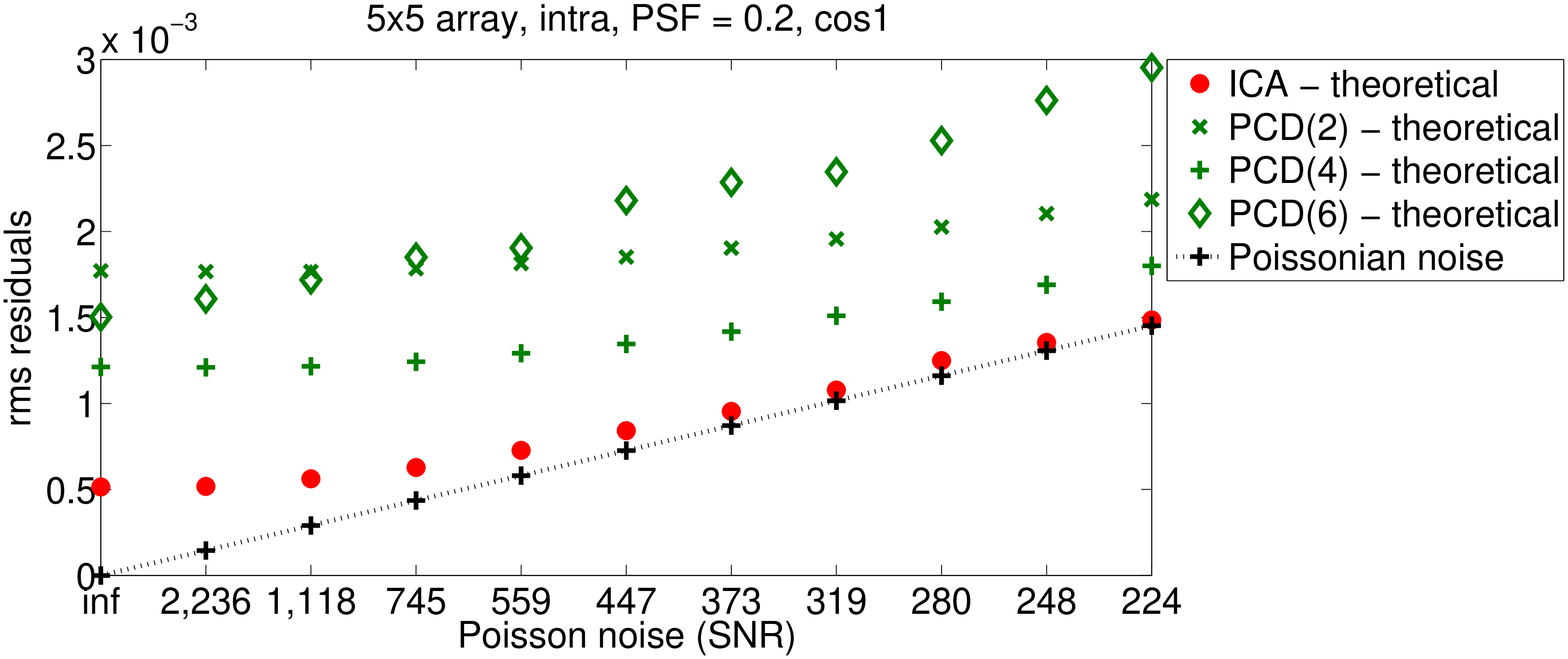}
\caption{Top panel:  Root mean square of discrepancies between detrended light-curves with (red circles) pixel-ICA, and (green) PCD method, and the theoretical model, obtained for simulations with $\sigma_{PSF} =$1 p.u., inter-pixel effects over a 9$\times$9 array, jitter `sin1', and poissonian noise at different levels. Black `+' along dashed line indicate the root mean square of the poissonian signal. Middle panel: The same for simulations with $\sigma_{PSF} =$0.2 p.u., intra-pixel effects over a 5$\times$5 array, jitter `sin1'. Green `+' refer to light-curves detrendend with a 4$^{th}$ order polynomial, and green rhomboids with a 6$^{th}$ order one. Bottom panel: The same for simulations with $\sigma_{PSF} =$0.2 p.u., intra-pixel effects over a 5$\times$5 array, jitter `cos1'. \label{fig19}}
\end{figure*}

Fig. \ref{fig19} shows the amplitude of discrepancies between the detrended light-curves and the theoretical model as a function of astrophysical Poisson noise, for pixel-ICA and PCD method with different polynomial orders:
\begin{enumerate}
\item For the inter-pixel effects the efficiency of the two methods is limited by the astrophysical Poisson noise level, except when it is smaller than the instrument HFPN, in which cases pixel-ICA slightly outperforms PCD.
\item For the intra-pixel effects, residual systematics are clearly present in high SNR cases, while for lower SNR the limit is the Poissonian threshold. Second order PCD method is far from optimal, while higher order polynomials in some cases can do better than pixel-ICA, but they do not always improve the results, as already mentioned in Sec. \ref{sec:intraPSF02}.
\end{enumerate}

\clearpage

\subsection{Inter-pixel effects without noise}

\begin{figure*}
\epsscale{0.96}
\plotone{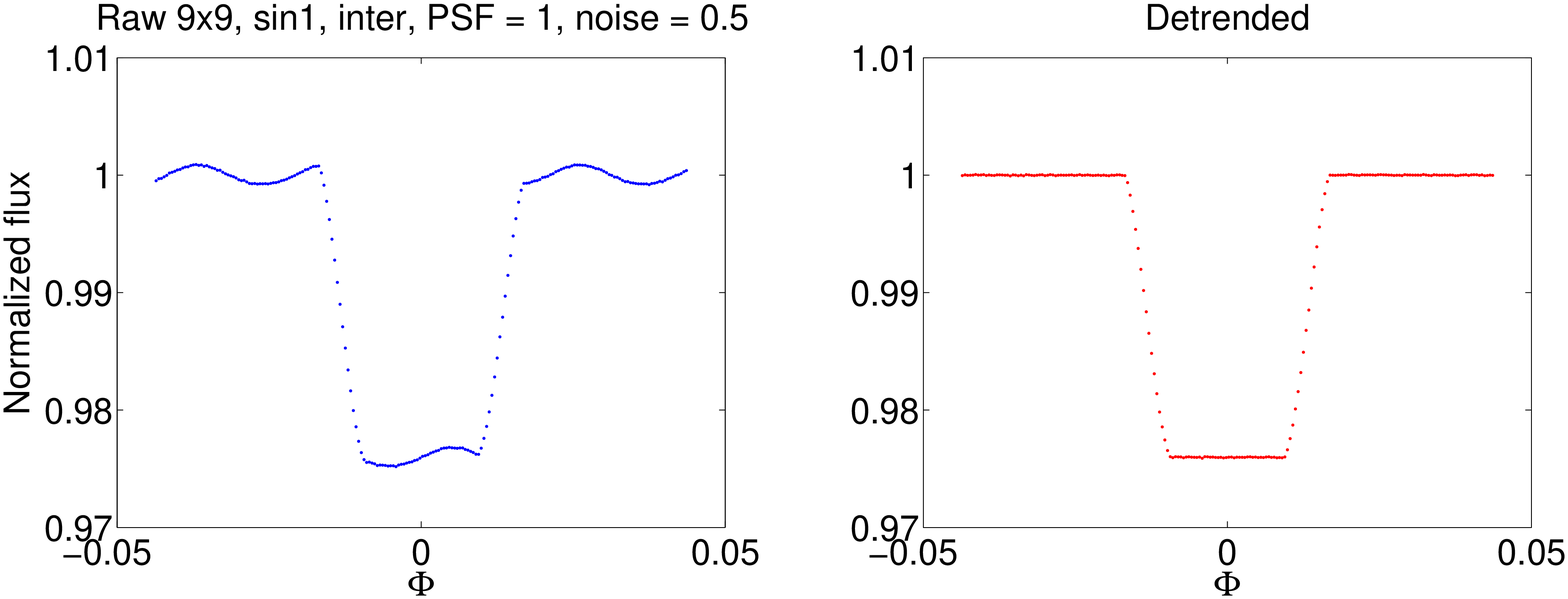}
\plotone{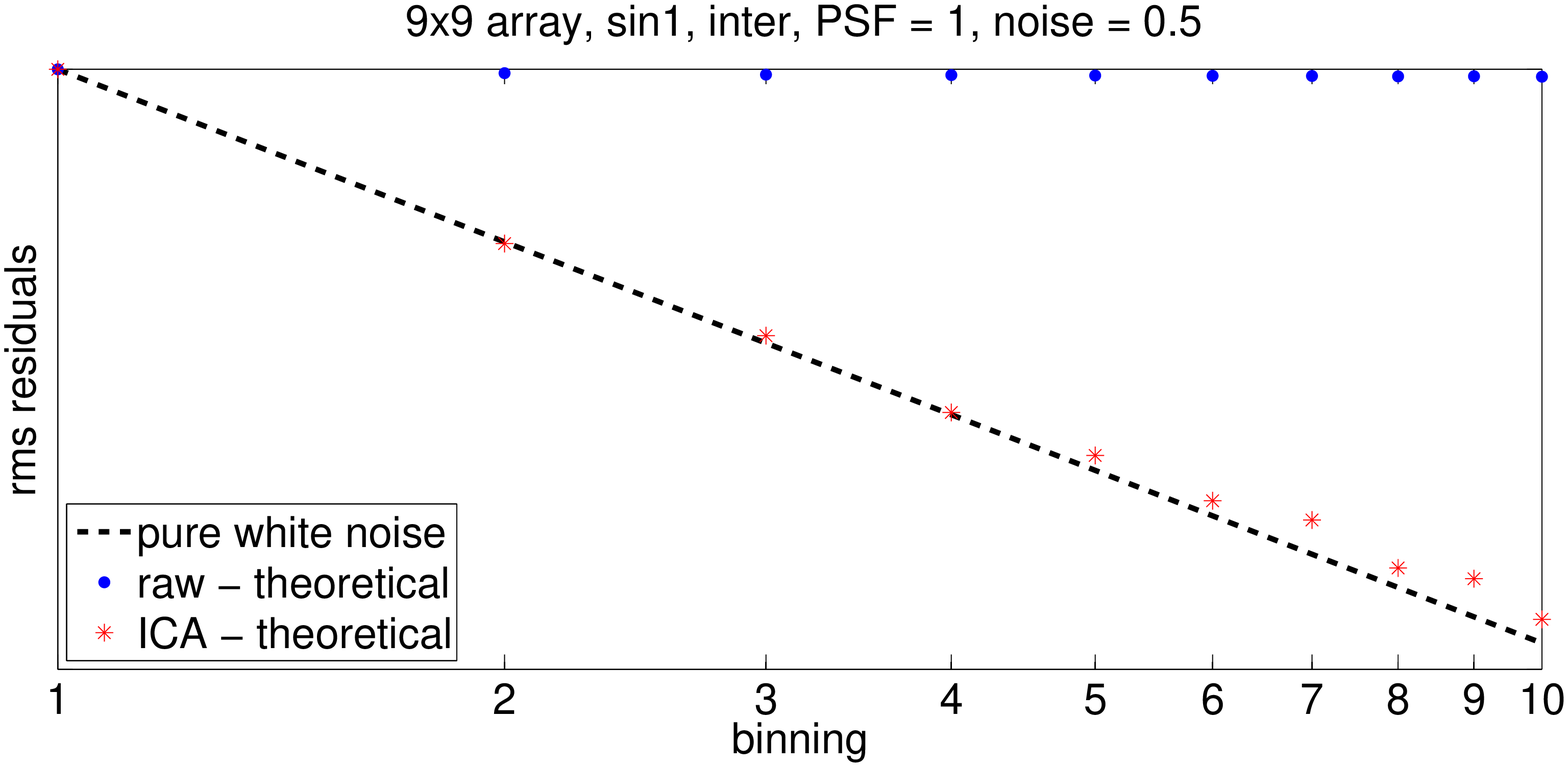}
\caption{Left panels: (blue) raw light-curve simulated with $\sigma_{PSF} = $1 p.u., inter-pixel sensitivity variations over 9$\times$9 array of pixels, jitter `sin1', and white noise at 0.5 photon counts/pixel/frame; (red) correspondent detrended light-curve with pixel-ICA. Right panel: Root mean square of residuals for binning over 1 to 10 points, scaled to their non-binned values. \label{fig20}}
\end{figure*}

In Sec. \ref{sec:interPSF1} we state that inter-pixel effects are well detrended with pixel-ICA method, based on the binning properties of residuals (and consistent results). Given that binning properties can only prove that systematics are negligible compared to the actual white noise level, we performed a last test for a simulation with inter-pixel effects ($\sigma_{PSF}=$1, 9$\times$9 array, jitter sin1) and a reduced white noise level, by a factor of 10. Note that it is an extreme low value of 0.5 photon counts/pixel/data point\footnote{We do not set this noise exactly to 0, because pixel time series with many zeroes would negatively affect the ICA detrending.}. Fig. \ref{fig20} shows the raw and detrended light-curves for this simulation, and the binning properties of their residuals. Time structure is very high for the raw light-curve, but it is again well detrended by pixel-ICA. We also checked that all the retrieved parameters are consistent with the original values within 1 $\sigma$.

\clearpage

\subsection{Testing a variant of pixel-ICA algorithm}
\label{sec:pixel-ICAvariant}

In sections \ref{sec:interPSF1}, \ref{sec:interPSF02}, and \ref{sec:intraPSF02} data are detrended before fitting the astrophysical model. In the previous literature, there are two schools of thought: (1) to use the out-of-transit data only to calibrate the systematics model (e.g. \cite{bea11, bal10}), (2) fitting for systematics and astrophysical models simultaneously (e.g. \cite{knu11, ste12, dem14}). A major advantage of the second approach is that correlations between transit and decorrelating parameters can be investigated. However, there are two arguments in favour of the first approach:
\begin{enumerate}
\item the out-of-transit can be used to calibrate the instrumental modulations, minimizing the risk of confounding them with the astrophysical signal. The risk of confusion is higher if the timescale of a systematic signal is similar to the transit duration; 
\item possible bumps due to occulted star spots during the transit would be reduced or canceled by an erronoeous correction of the instrument systematics.
\end{enumerate}
The second argument does not apply to our simulations, given that the only astrophysical signal is the planetary transit. In general, the best option is to check whether the two approaches lead to the same results, and if not, to investigate the origin of the discrepancy.

A variant of the original pixel-ICA algorithm consists in fitting the independent components plus a transit model to the raw light-curves (see Appendix C.5 in \cite{mor15}). We tested that, in our simulations, this alternative algorithm essentially leads to the same results obtained with the original algorithm. A few exceptions occur for some of the configurations obtained with `cos1' jitter (unfortunate timescale and phasing), for which results obtained with the two approaches are not identical, but consistent within 1$\sigma$. In all tested cases, the (partial) parameter error bars obtained by the MCMC chains are very similar for the two algorithms, which indicates the absence of strong correlations between decorrelating and transit parameters. However, we suggest that the final error bars should be computed according to Eq. \ref{eqn:sigmapar} to account for intrinsic errors on the components extracted.

\subsection{Comparison between pixel-ICA and PLD algorithms}
\label{sec:PLDdeming}

Recently, \cite{dem14} proposed a pixel-level detrending method. The underlying model to adapt to the data is the product of an astrophysical factor, a generalized function of the pixel intensities, and a temporal term:
\begin{equation}
\label{eqn:PLDmodel}
S^t = S_*(1-DE(t)) \times \mathcal{F}(P_1^t, P_2^t, ..., P_n^t) \times \mathcal{G}(t)
\end{equation}
where $S_*$ is the (constant) stellar luminosity, $D$ is the transit/eclipse depth, $E(t)$ is the eclipse shape normalized to unit amplitude, $P_i^t$ is the number of photon counts read in pixel $i$ at time $t$.
First order expansions of the the three factors in Eq. \ref{eqn:PLDmodel} lead to the following formula:
\begin{equation}
\label{eqn:PLDeclipse}
\Delta S^t = \sum_{i=1}^{n} c_i \delta P_i^t + DE(t) + ft + gt^2 + h
\end{equation}
where $\Delta$ indicates the total fluctuations from all sources, $c_i$ are partial derivatives from the Taylor expansion, $(ft + gt^2)$ represent the temporal variations (it is possible to adopt other functional forms). \cite{dem14} recommend the following replacement:
\begin{equation}
\delta P_i^t = \hat{P}_i^t = \frac{P_i^t}{ \sum_{i=1}^{n}P_i^t}
\end{equation}
Fitting Eq. \ref{eqn:PLDeclipse} to our simulated light-curves, the algorithm did not converge. A possible explanation is that the cross-terms between pixel fluctuations and transit are not negligible. Therefore, we readjusted Eq. \ref{eqn:PLDeclipse} to include those terms:
\begin{equation}
\Delta S^t = \left ( \sum_{i=1}^{n} c_i \hat{P}_i^t + h \right ) ( 1 - DE(t) )
\end{equation}
The explicit temporal dependence is ignored, given that it is not included in our simulations.

We fitted for three transit parameters ($p$, $a_0$ and $i$), and $c_i$ and $h$ coefficients  simultaneously. Results obtained with this method are practically indistinguishable from the ones obtained with the latest variant of pixel-ICA.

\section{Conclusions}
\label{sec:end}

We have tested the pixel-ICA algorithm, i.e. a non-parametric method proposed by \cite{mor14, mor15} to detrend Spitzer/IRAC primary transit observations, on simulated datasets. Systematics similar to the ones present in Spitzer/IRAC datasets are obtained by combining instrumental jitter with inter- or intra-pixel sensitivity variations. A variety of jitter time series is used to test the pixel-ICA method with:
\begin{enumerate}
\item periodic signals with different frequencies, phasing, and shape;
\item non-stationary signals with varying amplitudes or frequencies;
\item sudden change.
\end{enumerate}
The detrending performances of pixel-ICA method have been compared with division by a polynomial function of the centroid, in this paper PCD method, and PLD method \citep{dem14}. Here we summarize the main results found:
\begin{enumerate}
\item Pixel-ICA algorithm can detrend non-stationary signals and sudden changes, as well as periodic signals with different frequencies and phasing, relative to the transit.
\item Inter-pixel effects are well-detrended with pixel-ICA method.
\item Even if the instrument PSF is not entirely within the array of pixels, pixel-ICA leads to results which are consistent at $\sim$1$\sigma$ with the input parameters.
\item In most cases, pixel-ICA outperforms PCD method, especially if the instrument PSF is narrow, or it is not entirely within the photometric aperture.
\item Intra-pixel effects are only detectable if the PSF is relatively small.
\item Intra-pixel effects cannot be totally detrended by any of the three methods, but pixel-ICA, in most cases, outperforms PCD method, which is largely case-dependent. Also, pixel-ICA method provides consistent results within the error bars.
\item It is possible to fit the astrophysical signal after detrending or together with the other components. The only differences are registered if at least one of the non-transit components has a similar shape at the time of transit, in which case the first approach is preferable, but the two results were consistent within 1$\sigma$.
\item The PLD method, updated to include cross-term between pixel fluctuations and the astrophysical signals, lead to very similar results than pixel-ICA, particularly if the astrophysical signal is fitted together with the other components. 
\end{enumerate}
In conclusion, we have found in a variety of simulated cases that pixel-ICA performs as well or better than other methods used in the literature, in particular polynomial centroid corrections and pixel-level decorrelation \citep{dem14}. The main advantage of pixel-ICA over other methods relies on its purely statistical foundation without the need of imposing prior knowledge on the instrument systematics, therefore avoiding a potential source of error. The results of this paper, together with previous analyses of real Spitzer/IRAC datasets \citep{mor14, mor15}, suggest that photometric precision and repeatability at the level of one part in 10$^4$ can be achieved with current infrared space instruments.

\acknowledgments

The author would like to thank Prof. G. Tinetti and Dr. I. P. Waldmann for useful comments. G. Morello is funded by UCL Perren/Impact scholarship (CJ4M/CJ0T). This work was partially supported by ERC project numbers 617119 (ExoLights).

\clearpage

\appendix

\section{\textit{Independent} vs \textit{Principal} Component Analysis}

Principal Component Analysis (PCA) is a statistical technique to separate various components in a set of observations/recordings. The main difference between PCA and ICA is that the former just assumes the components to be uncorrelated, while the criterion of separation  adopted in ICA is the maximum statistical independence between the components, which is a more constraining property.\\
Two scalar random variables $x$ and $y$ are uncorrelated if their covariance is zero, or equivalently:
\begin{equation}
E\{xy\} = E\{x\}E\{y\}
\end{equation}
where $E\{\cdot\}$ denotes the expected value.\\
Two variables are statistically independent if their joint probability density factorizes into the product of their marginal densities:
\begin{equation}
p_{x,y}(x,y) = p_{x}(x) p_{y}(y)
\end{equation}
Equivalently, they must satisfy the property:
\begin{equation}
E\{f(x)g(y)\} = E\{f(x)\}E\{g(y)\}
\end{equation}
being $f(x)$ and $g(x)$ any integrable functions of $x$ and $y$, respectively.\\
Therefore, independent variables are also uncorrelated, but not viceversa. Only if the variables have gaussian distributions, uncorrelatedness implies independence.\\
In practice:
\begin{itemize}
\item PCA uses up to second-order statistics. It can only deals with mixtures of gaussian signals, where uncorrelatedness equals independence.
\item ICA uses third and fourth order statistics, and it can separate mixtures of non-gaussian signals.
\end{itemize}
Many ICA algorithms ran PCA as a useful preprocessing step \citep{hyv01}.

\clearpage

\section{Additional tables}
\label{app3}

\begin{table}
\begin{center}
\caption{Retrieved transit parameters for simulations with $\sigma_{PSF} =$1, 9$\times$9 array, inter-pixel effects (see Sec. \ref{sec:interPSF1}). In representative cases, we report the partial error bars obtained by the residuals, the final error bars, and the worst case error bars (see Sec. \ref{sec:pixel-ICA}). \label{tab7}}
\resizebox{0.6 \textwidth}{!}{%
\begin{tabular}{cccccc}
\tableline\tableline
Jitter & Parameters & Best values & 1-$\sigma$ errors & 1-$\sigma$ errors & 1-$\sigma$ errors\\
 & & & (residual scatter only) & (ICA) & (ICA worst case)\\
\tableline
 & $p$ & 0.15505 & 9$\times$10$^{-5}$ & 1.4$\times$10$^{-4}$ & 1.4$\times$10$^{-4}$\\
sin1 & $a_0$ & 8.99 & 0.02 & 0.03 & 0.04\\
 & $i$ & 85.78 & 0.03 & 0.04 & 0.04\\
\tableline
 & $p$ & 0.15501 & 1.0$\times$10$^{-4}$ & 1.5$\times$10$^{-4}$ & 1.5$\times$10$^{-4}$\\
cos1 & $a_0$ & 9.05 & 0.03 & 0.04 & 0.04\\
 & $i$ & 85.85 & 0.03 & 0.04 & 0.05\\
\tableline
 & $p$ & 0.15507 &  &  & \\
sin2 & $a_0$ & 8.98 &  &  & \\
 & $i$ & 85.78 &  &  & \\
\tableline
 & $p$ & 0.15510 &  &  & \\
cos2 & $a_0$ & 8.94 &  &  & \\
 & $i$ & 85.74 &  &  & \\
\tableline
 & $p$ & 0.15505 &  &  & \\
sin3 & $a_0$ & 8.99 &  &  & \\
 & $i$ & 85.79 &  &  & \\
\tableline
 & $p$ & 0.15503 &  &  & \\
cos3 & $a_0$ & 9.02 &  &  & \\
 & $i$ & 85.82 &  &  & \\
\tableline
 & $p$ & 0.15506 &  &  & \\
saw1 & $a_0$ & 8.98 &  &  & \\
 & $i$ & 85.78 &  &  & \\
\tableline
 & $p$ & 0.15506 &  &  & \\
saw1v1 & $a_0$ & 8.99 &  &  & \\
 & $i$ & 85.79 &  &  & \\
\tableline
 & $p$ & 0.15511 &  &  & \\
saw1v2 & $a_0$ & 9.02 &  &  & \\
 & $i$ & 85.81 &  &  & \\
\tableline
 & $p$ & 0.15508 & 1.0$\times$10$^{-4}$ & 1.5$\times$10$^{-4}$ & 1.5$\times$10$^{-4}$\\
saw1v3 & $a_0$ & 8.99 & 0.03 & 0.04 & 0.04\\
 & $i$ & 85.79 & 0.03 & 0.05 & 0.05\\
\tableline
 & $p$ & 0.15509 &  &  & \\
saw1vf1 & $a_0$ & 8.98 &  &  & \\
 & $i$ & 85.77 &  &  & \\
\tableline
 & $p$ & 0.15506 & 1.0$\times$10$^{-4}$ & 1.5$\times$10$^{-4}$ & 1.6$\times$10$^{-4}$\\
saw1vf2 & $a_0$ & 8.99 & 0.03 & 0.04 & 0.05\\
 & $i$ & 85.79 & 0.03 & 0.05 & 0.05\\
\tableline
 & $p$ & 0.15503 & 1.0$\times$10$^{-4}$ & 1.4$\times$10$^{-4}$ & 1.5$\times$10$^{-4}$\\
jump04c & $a_0$ & 9.00 & 0.03 & 0.04 & 0.04\\
 & $i$ & 85.80 & 0.03 & 0.04 & 0.05\\
\tableline
\end{tabular}}
\end{center}
\end{table}

\begin{table}
\begin{center}
\caption{Retrieved transit parameters for simulations with $\sigma_{PSF} =$1, 5$\times$5 array, inter-pixel effects (see Sec. \ref{sec:interPSF1}). In representative cases, we report the partial error bars obtained by the residuals, the final error bars, and the worst case error bars (see Sec. \ref{sec:pixel-ICA}). \label{tab8}}
\resizebox{0.6 \textwidth}{!}{%
\begin{tabular}{cccccc}
\tableline\tableline
Jitter & Parameters & Best values & 1-$\sigma$ errors & 1-$\sigma$ errors & 1-$\sigma$ errors\\
 & & & (residual scatter only) & (ICA) & (ICA worst case)\\
\tableline
 & $p$ & 0.15524 & 1.5$\times$10$^{-4}$ & 2.2$\times$10$^{-4}$ & 2.3$\times$10$^{-4}$\\
sin1 & $a_0$ & 8.97 & 0.04 & 0.06 & 0.07\\
 & $i$ & 85.76 & 0.05 & 0.07 & 0.07\\
\tableline
 & $p$ & 0.15505 & 1.6$\times$10$^{-4}$ & 2.2$\times$10$^{-4}$ & 2.3$\times$10$^{-4}$\\
cos1 & $a_0$ & 9.18 & 0.05 & 0.07 & 0.07\\
 & $i$ & 85.99 & 0.05 & 0.07 & 0.07\\
\tableline
 & $p$ & 0.15510 &  &  & \\
sin2 & $a_0$ & 9.08 &  &  & \\
 & $i$ & 85.88 &  &  & \\
\tableline
 & $p$ & 0.15536 &  &  & \\
cos2 & $a_0$ & 8.89 &  &  & \\
 & $i$ & 85.66 &  &  & \\
\tableline
 & $p$ & 0.15528 &  &  & \\
sin3 & $a_0$ & 8.99 &  &  & \\
 & $i$ & 85.79 &  &  & \\
\tableline
 & $p$ & 0.15516 &  &  & \\
cos3 & $a_0$ & 9.07 &  &  & \\
 & $i$ & 85.86 &  &  & \\
\tableline
 & $p$ & 0.15519 &  &  & \\
saw1 & $a_0$ & 8.97 &  &  & \\
 & $i$ & 85.77 &  &  & \\
\tableline
 & $p$ & 0.15517 &  &  & \\
saw1v1 & $a_0$ & 9.02 &  &  & \\
 & $i$ & 85.81 &  &  & \\
\tableline
 & $p$ & 0.15494 &  &  & \\
saw1v2 & $a_0$ & 8.92 &  &  & \\
 & $i$ & 85.74 &  &  & \\
\tableline
 & $p$ & 0.15527 & 1.5$\times$10$^{-4}$ & 2.4$\times$10$^{-4}$ & 2.6$\times$10$^{-4}$\\
saw1v3 & $a_0$ & 8.98 & 0.04 & 0.07 & 0.07\\
 & $i$ & 85.77 & 0.05 & 0.07 & 0.08\\
\tableline
 & $p$ & 0.15524 &  &  & \\
saw1vf1 & $a_0$ & 9.00 &  &  & \\
 & $i$ & 85.78 &  &  & \\
\tableline
 & $p$ & 0.15531 & 1.6$\times$10$^{-4}$ & 2.2$\times$10$^{-4}$ & 2.4$\times$10$^{-4}$\\
saw1vf2 & $a_0$ & 8.99 & 0.04 & 0.06 & 0.07\\
 & $i$ & 85.79 & 0.05 & 0.07 & 0.07\\
\tableline
 & $p$ & 0.15508 & 1.3$\times$10$^{-4}$ & 2.6$\times$10$^{-4}$ & 2.8$\times$10$^{-4}$\\
jump04c & $a_0$ & 9.04 & 0.04 & 0.07 & 0.08\\
 & $i$ & 85.83 & 0.04 & 0.08 & 0.09\\
\tableline
\end{tabular}}
\end{center}
\end{table}

\begin{table}
\begin{center}
\caption{Retrieved transit parameters for simulations with $\sigma_{PSF} =$0.2, 5$\times$5 array, inter-pixel effects (see Sec. \ref{sec:interPSF02}). In representative cases, we report the partial error bars obtained by the residuals, the final error bars, and the worst case error bars (see Sec. \ref{sec:pixel-ICA}). \label{tab9}}
\resizebox{0.6 \textwidth}{!}{%
\begin{tabular}{cccccc}
\tableline\tableline
Jitter & Parameters & Best values & 1-$\sigma$ errors & 1-$\sigma$ errors & 1-$\sigma$ errors\\
 & & & (residual scatter only) & (ICA) & (ICA worst case)\\
\tableline
 & $p$ & 0.15507 & 3$\times$10$^{-5}$ & 6$\times$10$^{-5}$ & 7$\times$10$^{-5}$\\
sin1 & $a_0$ & 9.010 & 0.010 & 0.018 & 0.020\\
 & $i$ & 85.808 & 0.010 & 0.020 & 0.022\\
\tableline
 & $p$ & 0.15505 & 4$\times$10$^{-5}$ & 6$\times$10$^{-5}$ & 7$\times$10$^{-5}$\\
cos1 & $a_0$ & 9.018 & 0.010 & 0.016 & 0.019\\
 & $i$ & 85.815 & 0.011 & 0.017 & 0.020\\
\tableline
 & $p$ & 0.15508 &  &  & \\
sin2 & $a_0$ & 9.009 &  &  & \\
 & $i$ & 85.806 &  &  & \\
\tableline
 & $p$ & 0.15505 &  &  & \\
cos2 & $a_0$ & 9.004 &  &  & \\
 & $i$ & 85.801 &  &  & \\
\tableline
 & $p$ & 0.15510 &  &  & \\
sin3 & $a_0$ & 9.001 &  &  & \\
 & $i$ & 85.798 &  &  & \\
\tableline
 & $p$ & 0.15507 &  &  & \\
cos3 & $a_0$ & 9.015 &  &  & \\
 & $i$ & 85.812 &  &  & \\
\tableline
 & $p$ & 0.15506 &  &  & \\
saw1 & $a_0$ & 8.993 &  &  & \\
 & $i$ & 85.791 &  &  & \\
\tableline
 & $p$ & 0.15506 &  &  & \\
saw1v1 & $a_0$ & 9.000 &  &  & \\
 & $i$ & 85.798 &  &  & \\
\tableline
 & $p$ & 0.15506 &  &  & \\
saw1v2 & $a_0$ & 8.997 &  &  & \\
 & $i$ & 85.795 &  &  & \\
\tableline
 & $p$ & 0.15506 & 3$\times$10$^{-5}$ & 5$\times$10$^{-5}$ & 6$\times$10$^{-5}$\\
saw1v3 & $a_0$ & 8.998 & 0.009 & 0.015 & 0.018\\
 & $i$ & 85.796 & 0.010 & 0.016 & 0.019\\
\tableline
 & $p$ & 0.15508 &  &  & \\
saw1vf1 & $a_0$ & 8.999 &  &  & \\
 & $i$ & 85.797 &  &  & \\
\tableline
 & $p$ & 0.15506 & 3$\times$10$^{-5}$ & 5$\times$10$^{-5}$ & 6$\times$10$^{-5}$\\
saw1vf2 & $a_0$ & 9.001 & 0.009 & 0.015 & 0.018\\
 & $i$ & 85.798 & 0.010 & 0.016 & 0.019\\
\tableline
 & $p$ & 0.15508 & 4$\times$10$^{-5}$ & 6$\times$10$^{-5}$ & 6$\times$10$^{-5}$\\
jump04c & $a_0$ & 9.00 & 0.011 & 0.017 & 0.019\\
 & $i$ & 85.80 & 0.011 & 0.018 & 0.020\\
\tableline
\end{tabular}}
\end{center}
\end{table}

\begin{table}
\begin{center}
\caption{Retrieved transit parameters for simulations with $\sigma_{PSF} =$0.2, 5$\times$5 array, intra-pixel effects (see Sec. \ref{sec:intraPSF02}). In representative cases, we report the partial error bars obtained by the residuals, the final error bars, and the worst case error bars (see Sec. \ref{sec:pixel-ICA}). \label{tab10}}
\resizebox{0.6 \textwidth}{!}{%
\begin{tabular}{cccccc}
\tableline\tableline
Jitter & Parameters & Best values & 1-$\sigma$ errors & 1-$\sigma$ errors & 1-$\sigma$ errors\\
 & & & (residual scatter only) & (ICA) & (ICA worst case)\\
\tableline
 & $p$ & 0.1551 & 3$\times$10$^{-4}$ & 4$\times$10$^{-4}$ & 5$\times$10$^{-4}$\\
sin1 & $a_0$ & 9.05 & 0.07 & 0.10 & 0.13\\
 & $i$ & 85.84 & 0.08 & 0.11 & 0.14\\
\tableline
 & $p$ & 0.1550 & 2$\times$10$^{-4}$ & 3$\times$10$^{-4}$ & 3$\times$10$^{-4}$\\
cos1 & $a_0$ & 9.12 & 0.07 & 0.08 & 0.09\\
 & $i$ & 85.92 & 0.08 & 0.08 & 0.09\\
\tableline
 & $p$ & 0.1550 &  &  & \\
sin2 & $a_0$ & 9.13 &  &  & \\
 & $i$ & 85.93 &  &  & \\
\tableline
 & $p$ & 0.1551 &  &  & \\
cos2 & $a_0$ & 9.10 &  &  & \\
 & $i$ & 85.89 &  &  & \\
\tableline
 & $p$ & 0.1551 &  &  & \\
sin3 & $a_0$ & 9.01 &  &  & \\
 & $i$ & 85.81 &  &  & \\
\tableline
 & $p$ & 0.1551 &  &  & \\
cos3 & $a_0$ & 9.02 &  &  & \\
 & $i$ & 85.82 &  &  & \\
\tableline
 & $p$ & 0.1548 &  &  & \\
saw1 & $a_0$ & 9.10 &  &  & \\
 & $i$ & 85.93 &  &  & \\
\tableline
 & $p$ & 0.1555 &  &  & \\
saw1v1 & $a_0$ & 8.92 &  &  & \\
 & $i$ & 85.71 &  &  & \\
\tableline
 & $p$ & 0.1546 &  &  & \\
saw1v2 & $a_0$ & 9.27 &  &  & \\
 & $i$ & 86.11 &  &  & \\
\tableline
 & $p$ & 0.1555 & 4$\times$10$^{-4}$ & 4$\times$10$^{-4}$ & 5$\times$10$^{-4}$\\
saw1v3 & $a_0$ & 8.95 & 0.10 & 0.11 & 0.13\\
 & $i$ & 85.77 & 0.11 & 0.12 & 0.15\\
\tableline
 & $p$ & 0.1555 &  &  & \\
saw1vf1 & $a_0$ & 8.78 &  &  & \\
 & $i$ & 85.54 &  &  & \\
\tableline
 & $p$ & 0.1550 & 3$\times$10$^{-4}$ & 3$\times$10$^{-4}$ & 3$\times$10$^{-4}$\\
saw1vf2 & $a_0$ & 8.99 & 0.09 & 0.10 & 0.10\\
 & $i$ & 85.79 & 0.10 & 0.11 & 0.11\\
\tableline
 & $p$ & 0.15508 & 4$\times$10$^{-5}$ & 3$\times$10$^{-4}$ & 4$\times$10$^{-4}$\\
jump04c & $a_0$ & 9.001 & 0.011 & 0.10 & 0.12\\
 & $i$ & 85.796 & 0.012 & 0.10 & 0.13\\
\tableline
\end{tabular}}
\end{center}
\end{table}

\clearpage

\section{List of acronyms}
\label{app4}

HFPN = High-Frequency Pixel Noise \\
ICA = Independent Component Analysis \\
ISR = Interference-to-Signal Ratio \\
MCMC = Markov Chain MonteCarlo \\
PCA = Principal Component Analysis \\
PCD = Polynomial Centroid Division \\
PLD = Pixel-Level Decorrelation \\
PSF = Point Spread Function \\
RMS = Root Mean Square \\
SNR = Signal-to-Noise Ratio \\

\clearpage

\clearpage

\end{document}